%% file: main.tex
\documentclass[conference,10pt]{IEEEtran}
\IEEEoverridecommandlockouts
\usepackage{cite}
\usepackage{amsmath,amssymb,amsfonts}
\usepackage{algorithmic}
\usepackage{graphicx}
\usepackage{textcomp}
\usepackage{xcolor}
\usepackage{enumitem}
\usepackage{comment}
\setlist[itemize]{leftmargin=*}
\def\BibTeX{{\rm B\kern-.05em{\sc i\kern-.025em b}\kern-.08em
    T\kern-.1667em\lower.7ex\hbox{E}\kern-.125emX}}

\input{packages.tex}
\input{macro.tex}
\input{performance.tex}

\begin{document}

\title{The Case for Strong Scaling in Deep Learning: Training Large 3D CNNs with Hybrid Parallelism}

\author{
  \IEEEauthorblockN{
    Yosuke Oyama\IEEEauthorrefmark{1}\,\IEEEauthorrefmark{2},
    Naoya Maruyama\IEEEauthorrefmark{2},
    Nikoli Dryden\IEEEauthorrefmark{3}\IEEEauthorrefmark{2},
    Erin McCarthy\IEEEauthorrefmark{4}\IEEEauthorrefmark{2},
    Peter Harrington\IEEEauthorrefmark{5},\\
    Jan Balewski\IEEEauthorrefmark{5},
    Satoshi Matsuoka\IEEEauthorrefmark{6}\,\IEEEauthorrefmark{1},
    Peter Nugent\IEEEauthorrefmark{5} and
    Brian Van Essen\IEEEauthorrefmark{2}
  }

  \IEEEauthorblockA{
    \small{
      \IEEEauthorrefmark{1}
      Tokyo Institute of Technology,
      \texttt{oyama.y.aa@m.titech.ac.jp}
    }
  }
  \IEEEauthorblockA{
    \small{
      \IEEEauthorrefmark{2}
      Lawrence Livermore National Laboratory,
      \texttt{\{maruyama3,vanessen1\}@llnl.gov}
    }
  }
  \IEEEauthorblockA{
    \small{
      \IEEEauthorrefmark{3}
      ETH Z\"{u}rich,
      \texttt{ndryden@ethz.ch}
    }
  }
  \IEEEauthorblockA{
    \small{
      \IEEEauthorrefmark{4}
      University of Oregon,
      \texttt{emccarth@cs.uoregon.edu}
    }
  }
  \IEEEauthorblockA{
    \small{
      \IEEEauthorrefmark{5}
      Lawrence Berkeley National Laboratory,
      \texttt{\{PHarrington,balewski,penugent\}@lbl.gov}
    }
  }
  \IEEEauthorblockA{
    \small{
      \IEEEauthorrefmark{6}
      RIKEN Center for Computational Science,
      \texttt{matsu@acm.org}
    }
  }
}

\maketitle
\thispagestyle{plain}
\pagestyle{plain}

\input{abstract.tex}


\input{1_introduction.tex}

\input{2_background.tex}
\input{3_scaling.tex}

\input{3_perfmodel.tex}
\input{3_model.tex}
\input{4_evaluation.tex}
\input{5_related.tex}
\input{6_conclusion.tex}

\input{acknowledgement.tex} 

\bibliographystyle{src/IEEEtran}
\bibliography{IEEEabrv,src/references}

\end{document}

%% file: packages.tex
\makeatletter
\@ifundefined{notodos}{
\usepackage[colorinlistoftodos,prependcaption]{todonotes}
}{}
\makeatother

\usepackage{multirow}
\usepackage{multicol}
\usepackage{ifthen}
\usepackage{pifont}
\usepackage{subfig}
\usepackage{xcolor}
\usepackage[group-separator={,}]{siunitx}
\usepackage{hyperref}
\usepackage{booktabs}

\usepackage{tikz}
\usetikzlibrary{shapes}
\usetikzlibrary{patterns}
\usetikzlibrary{decorations.pathreplacing}

%% file: macro.tex
\newcommand{\cosmoflow}{CosmoFlow}
\newcommand{\unet}{U-Net}
\newcommand{\distconv}{Distconv}

\newcommand{\conduit}{Conduit}
\newcommand{\lassen}{Lassen}
\newcommand{\llnl}{Lawrence Livermore National Laboratory}
\newcommand{\cudnn}{cuDNN}

\newcommand{\cubic}[1]{#1^3}

\makeatletter
\newcommand{\perf}[1]{\@nameuse{#1}}
\newcommand{\perfx}[1]{\@nameuse{#1}}
\makeatother

\makeatletter
\@ifundefined{figref}{\newcommand{\figref}[1]{Figure~\ref{#1}}}{}
\@ifundefined{tabref}{\newcommand{\tabref}[1]{Table~\ref{#1}}}{}
\@ifundefined{secref}{\newcommand{\secref}[1]{Section~\ref{#1}}}{}

\@ifundefined{toprule}{\newcommand{\toprule}{\hline}}{}
\@ifundefined{midrule}{\newcommand{\midrule}{\hline}}{}
\@ifundefined{bottomrule}{\newcommand{\bottomrule}{\hline}}{}

\@ifundefined{text}{\newcommand{\text}[1]{\mathrm{#1}}}{}
\makeatother

%% file: abstract.tex
\begin{abstract}

We present scalable hybrid-parallel algorithms for training large-scale 3D convolutional neural networks. Deep learning-based emerging scientific workflows often require model training with large, high-dimensional samples, which can make training much more costly and even infeasible due to excessive memory usage. We solve these challenges by extensively applying hybrid parallelism throughout the end-to-end training pipeline, including both computations and I/O. Our hybrid-parallel algorithm extends the standard data parallelism with spatial parallelism, which partitions a single sample in the spatial domain, realizing strong scaling beyond the mini-batch dimension with a larger aggregated memory capacity. We evaluate our proposed training algorithms with two challenging 3D CNNs, CosmoFlow and 3D U-Net. Our comprehensive performance studies show that good weak and strong scaling can be achieved for both networks using up to 2K GPUs. More importantly, we enable training of CosmoFlow with much larger samples than previously possible, realizing an order-of-magnitude improvement in prediction accuracy.

\end{abstract}

%% file: 1_introduction.tex
\section{Introduction} \label{section:introduction}
Recent advances in deep learning, especially convolutional neural networks (CNNs), have become a subject of significant research interest in emerging scientific workflows in research fields such as cosmology~\cite{Mathuriya:2018:CUD:3291656.3291743}, medical image analysis~\cite{CABR16}, climate analysis~\cite{kurth2018exascale}, and turbulent flow simulations~\cite{duraisamy15}. This is thanks to its potential for extraordinary impact, as demonstrated in image and speech classification~\cite{Krizhevsky2012a, pmlr-v48-amodei16}, playing games~\cite{44806}, and translating natural languages~\cite{DBLP:journals/corr/abs-1810-04805}, among others. While early successful results have been reported in applying deep CNNs to scientific problems, training highly robust and accurate scientific models can be severely constrained as they tend to need to use much larger samples, such as 3D medical images or simulation outputs. Larger, high-dimensional samples can make deep CNNs even deeper with each layer becoming more compute intensive, making already long training times even longer. Furthermore, increased sample sizes directly increase the memory usage of model training, which is especially problematic in accelerators with limited memory capacity such as GPUs. 3D CNNs can consume tens to hundreds of gigabytes of memory, as exemplified in the cosmology and medical models evaluated in this work (see \secref{section:evaluation}). This easily exceeds the available memory capacity of typical high-end GPUs. These two major problems, high computational cost and memory usage, can severely reduce the effectiveness of deep learning, especially in scientific domains.

Parallelizing training can alleviate these problems by employing more compute and memory resources. However, the most commonly-used approach, data-parallelism, fails to adequately address the challenges posed by extreme scientific problems due to its limited parallelism. While training samples are distributed over multiple processing elements (PEs) such as GPUs, each PE must still process at least one complete sample, resulting in limited reduction of per-PE memory pressure. Reducing the size of training samples by, e.g., lowering resolution is a common workaround. However, it inevitably loses information in the original data, which can be critical to train highly accurate models. Even if memory is not a problem, the degree of parallelism is still limited by the number of samples per mini-batch, which cannot be arbitrarily increased without adversely affecting the model accuracy~\cite{keskar2017large}.

To address these issues in large-scale 3D CNNs, \textbf{strong scaling is required}. We present an end-to-end training framework that extends the parallelism beyond the current state of practice for strong scaling. First, we develop a highly efficient hybrid-parallel algorithm for 3D convolutions that exploits parallelism both in the spatial and mini-batch dimensions, allowing one sample to be distributed over multiple PEs. This provides improved performance and is indispensable for breaking the memory barrier in 3D CNNs. Second, to achieve scalable end-to-end performance, \textbf{the performance of I/O pipeline must also strong scale} for a fixed number of samples, which may be gigabytes in size.
We apply the same hybrid-parallel techniques to I/O, maximizing parallelism, increasing throughput, and minimizing scaling bottlenecks.

We demonstrate end-to-end scalable training by extending an existing DNN framework with our hybrid-parallel compute and I/O algorithms. In our experimental studies, we use CosmoFlow, a regression model for cosmology~\cite{Mathuriya:2018:CUD:3291656.3291743}, and the 3D U-Net, a segmentation model for medical images~\cite{CABR16}, as representative large-sample 3D CNNs. We present comprehensive performance analyses of our algorithms and show that they collectively enable efficient strong scaling for both problems, i.e., parallel speedups without increasing the mini-batch size. Even more importantly,
we demonstrate that by training CosmoFlow with much larger samples than previously possible, it is possible to realize \textbf{an order-of-magnitude improvement in prediction accuracy}, which can have a tremendous impact in accelerating scientific discoveries. This new capability enabled by our training pipeline is not limited to the particular problem but can be a critical tool for broader ML-enhanced scientific applications.

We summarize our contributions as follows:
\begin{itemize}
\item We present an end-to-end approach for strong-scaling training large-sample 3D CNNs. We address the compute, memory, and I/O challenges using hybrid-parallelism.
\item We present a prototype implementation of our proposed approach by extending the LBANN framework~\cite{VanEssen:2015:LLB:2834892.2834897} and demonstrate training on full-resolution samples for CosmoFlow ($512^3$) and the 3D U-Net ($256^3$). Our performance results show show good strong and weak scaling on up to 2048 GPUs. Training the CosmoFlow model is 1.77x faster when using 2048 GPUs over 512 GPUs, both using the same mini-batch size of 64. Similarly, a 1.42x speedup is achieved for the 3D U-Net when using 512 GPUs over 256 GPUs.
\item We provide detailed model-based performance analyses of both problems in order to give comprehensive understanding of their scaling efficiencies.
\item We demonstrate a significant improvement in prediction accuracy by using full-resolution data. The CosmoFlow model trained with $512^3$ samples, while requiring at least eight V100 GPUs per sample, realizes ten times lower mean squared error than when trained with $128^3$ samples, which was the largest size reported previously.
\end{itemize}

%% file: 2_background.tex
\section{Background} \label{section:background}

Here, we first describe two fundamental methods to train a single DNN in parallel: data- and model-parallelism. We also describe hybrid-parallel training, a combination of both which couples spatial partitioning and data-parallelism for improved scalability. Then we introduce the CosmoFlow and the 3D U-Net models in more detail.

\subsection{Data-, model-, and hybrid-parallelism} \label{subsection:parallelisms}
Mini-batch Stochastic Gradient Descent (SGD) is the most widely used technique to
optimize the parameters of a given deep neural network, and has the form:
\begin{displaymath}
  W^{(t+1)} = W^{(t)} - \eta^{(t)} \sum_{n=1}^{N} \nabla L \left( x_n; W^{(t)} \right) ,
\end{displaymath}
where
$W^{(t)}$ are the network parameters at step $t$,
$\eta^{(t)}$ is the learning rate at step $t$,
$N$ is the mini-batch size,
$L$ is the loss function,
and $x_n$ is the $n$th sample.

\subsubsection{Data-parallelism}
Data-parallel training, which partitions samples to compute $\nabla L$ in parallel,
is the most widely-used training technique
(\figref{figure:parallelism}, top left). It takes advantage of the fact that
$N$ is typically large enough to efficiently parallelize training on up to thousands of GPUs~\cite{2017arXiv170602677G,2018arXiv180900778A,2019arXiv190312650Y} and that
communication requirements are relatively small compared
to the compute requirements for CNNs.

However, data-parallelism is limited by the number of samples in a mini-batch,
the memory requirements of training, and how well the model learns.

\subsubsection{Model-parallelism}

\newcommand\zzz{{\setlength{\unitlength}{1mm}%
    \begin{picture}(5,3)
      \put(0,1){\line(5,0){5}}
    \end{picture}}}

\begin{figure}[t]
  \begin{center}
    \input{figures/parallelism.tikz}
  \end{center}
  \caption{
    Three different parallel strategies for deep neural networks.
    $N$ denotes the mini-batch size.
    ``Data-parallel'' and ``Hybrid-parallel'' compute a mini-batch of two samples
    (\protect\input{figures/sample_1.tikz}
    and \protect\input{figures/sample_2.tikz}),
    while ``Model-parallel'' splits the spatial dimension of one sample
    (\protect\input{figures/sample_1.tikz}) into two GPUs.
    Note that data movement within a single process is typically cheap.
  }
  \label{figure:parallelism}
\end{figure}
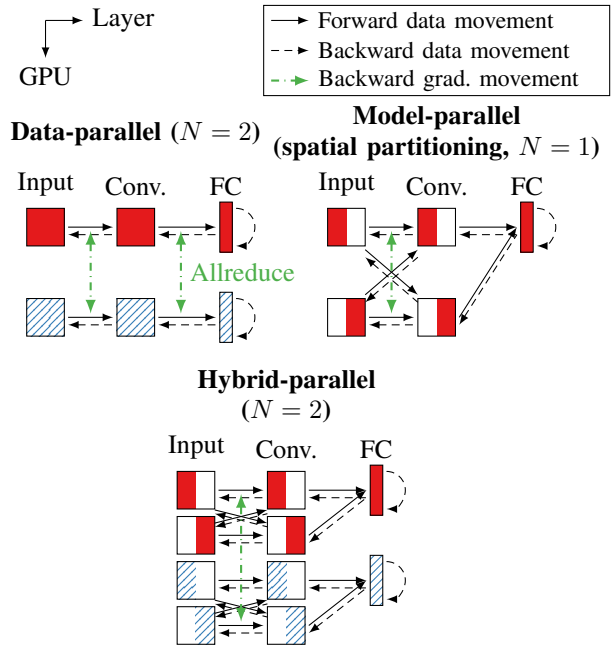

On the other hand, model-parallel training (\figref{figure:parallelism}, top right)
parallelizes the computation of $\nabla L$ for each data sample across multiple GPUs.
This enables additional parallelism to be exploited, and partitions data across
multiple GPUs to reduce memory requirements. In data-parallel training, if the
memory requirements (including necessary intermediate activations) exceed the memory
capacity of a GPU, training is infeasible. In contrast, with model-parallelism,
the memory requirements are roughly inversely proportional to the number of partitions.
This strong-scaling advantage is especially attractive for high-dimensional CNNs, since the large
input data results in huge intermediate activation tensors during training.

At the same time, model-parallel training requires careful framework design
to mitigate overheads. While data-parallel
training requires only a single global allreduce per layer to aggregate parameter
updates, model-parallel training can involve many fine-grained communication
operations and synchronizations in each layer.

Many strategies exist to exploit model-parallelism in CNNs, including spatial
partitioning, channel/filter partitioning, and layer-wise pipelining (see \secref{section:related}). In this
paper, due to the high-resolution and 3-dimensional input data, we focus on
spatial partitioning as a natural way to decompose the computation.
Spatial partitioning distributes input and activation tensors
by partitioning the height, width, and depth dimensions, similarly to traditional
stencil computations. This introduces a halo exchange (an exchange of spatial boundary data between neighbor processes)
within convolutional and pooling layers to maintain correctness,
although it can be overlapped
to hide communication overhead. For 3D data, the improved surface-to-volume ratio
of the problem mitigates these communication overheads even further.
Existing work on spatial partitioning has been limited to 2D data; here,
we extend spatial partitioning to support 3D data.

\subsubsection{Hybrid-parallelism}
``Hybrid-parallelism'' is the combination of data-parallelism and model-parallelism (\figref{figure:parallelism}, bottom).
Hybrid parallelism takes advantage of both the low overhead of data-parallelism to weak scale and of model-parallelism to strong-scale onto more compute resources. It requires careful selection of the relative balance of parallelism to achieve good scalability, as demonstrated in \secref{subsection:strong_scale}.

\subsection{\cosmoflow{}}
\cosmoflow{} \cite{Mathuriya:2018:CUD:3291656.3291743} is a project to use deep
learning to estimate the values of important cosmological parameters from 3D
cosmological simulations.
One of the goals in cosmology is to understand and control the underlying
systematics in a cosmological survey. As there is only one universe to observe,
and the entire universe is needed to make these measurements, constraining the
effects of systematics falls to computationally expensive simulations.
The \cosmoflow{} network aims to replicate both the systematics from survey operations as well
as those nature forces upon us. Creating surrogates for these simulations is
necessary to generate the sheer statistical numbers needed to control the
systematics.

Mathuriya et al. conducted thousands of independent N-body dark matter simulations with varied initial cosmological parameters, and constructed a dataset from them. The task is to predict the initial parameters from 3D mass distributions. The original spatial dimensions, $512^3$ voxels, required too much memory to train on, so each sample was split into $128^3$ voxel sub-volumes which are used as different data samples.
As a result, the \cosmoflow{} network was trained with 99,456 training samples, each a $128^3$-voxel 3D histogram of particle counts. It was reported that training with large mini-batches containing 8192 samples did not converge to comparable accuracy as smaller, 2048-sample mini-batches. These two problems are easily solved by our approach, because each sample is distributed among multiple GPUs, avoiding memory limits and allowing scaling without large mini-batches.

The latest \cosmoflow{} dataset is the
``2019\_05\_4parE'' dataset~\cite{cosmoflow-dataset}. It contains \num{10017}
simulated universes, each of which is composed of four channels (redshifts)
and is $512^3$ voxels, stored as 16-bit integers, along with four
cosmological parameters that were used to generate the universe. These
are $\Omega_M$, the proportion of matter in the universe; $\sigma_8$, the
amplitude of mass fluctuations at a distance scale of 8 Mpc/h; $n_s$, the scalar
spectral index of the spatial curvature of a comoving slice of space-time; and
$H_0$, Hubble's constant.
The dataset is about 9.77 TiB in size.
We normalize the parameters to be in $[-1, 1]$ when training, in line with prior work.

In this paper, we distinguish the datasets with the spatial input sizes, \perf{dim128} to \perf{dim512}.
We split each dataset into 80\%, 10\%, and 10\% as training, validation, and test datasets respectively.
Additionally in this work, we will test the hypothesis that by allowing neural networks to
observe entire data samples during training it is possible to learn longer range properties
in the data and improve the quality of this type of regression model.





\subsection{The 3D \unet{}} \label{subsection:background_unet}
The 3D \unet{}~\cite{CABR16} is a 3D version of the \unet{}~\cite{DBLP:journals/corr/RonnebergerFB15}, a 2D CNN for image segmentation. It replaces all 2D operations with 3D operations to perform volumetric segmentation on 3D data. \unet{}s have been applied to a wide range of 2D and 3D segmentation applications, such as biological image analysis~\cite{DBLP:journals/corr/RonnebergerFB15,CABR16} and CT image analysis~\cite{ohou2019high}.

In this paper, we apply the 3D \unet{} to the Liver Tumor Segmentation (LiTS) dataset~\cite{DBLP:journals/corr/abs-1901-04056}, where the task is to segment liver lesions in 3D CT scans. It consists of 131 CT scans for training and 70 for testing. Each is composed of a variable number of $512 \times 512$ slices and per-voxel ground-truth labels. To use a consistent input size, we down-sample the non-slice dimensions and up- or down-sample the slice dimension so each sample is $256^3$ voxels.
We convert the original dataset to equivalent HDF5 files with 16-bit integers.

The most significant difference between the 3D \unet{} and \cosmoflow{} networks is that it uses deconvolution layers
to upsample activations to their original size.
As the memory requirements for activations is cubic in the layer's spatial dimensions,
the 3D \unet{} requires a huge amount of memory near both the input and output layers,
compared to the \cosmoflow{} network with the same input size.
Furthermore, the CT image and labeled segmentation of each sample in the LiTS dataset are both the same size.
Since the labels are not small (e.g. a class label), we must consider the I/O performance of reading them in addition to the inputs.
Thus, the 3D \unet{} helps demonstrate performance in different regimes than \cosmoflow{}.


%% file: figures/parallelism.tikz
\begin{tikzpicture}
  \def\tensorh{0.5};
  \def\aroffset{0.05};
  \def\ardist{0.1};
  \def\secondx{1.2};
  \def\thirdx{2.4};
  \def\nodehcm{1.2cm};
  \def\nodeh{1.2};
  \def\modelxcm{4cm};
  \def\modelycm{3.5cm};

  \tikzstyle{bp}=[densely dashed];
  \definecolor{gradcolor}{RGB}{77,175,74}
  \tikzstyle{grad}=[dashdotted,color=gradcolor,thick];

  \def\fcy{0};

  \def\axisx{0};
  \def\axisy{2.75};
  \draw[-latex] ({\axisx},{\axisy}) -- ({\axisx+\tensorh},{\axisy}) node[anchor=west]{Layer};
  \draw[-latex] ({\axisx},{\axisy}) -- ({\axisx},{\axisy-\tensorh}) node[anchor=north]{GPU};

  \def\legendx{3};
  \def\legendh{\tensorh*0.8};
  \draw ({\legendx-\aroffset*2},{\axisy+\legendh/2}) rectangle ({\legendx+4.5},{\axisy-\legendh*2.5});
  \draw[-latex] ({\axisx+\legendx},{\axisy}) -- ({\axisx+\legendx+\tensorh},{\axisy}) node[anchor=west]{\small Forward data movement};
  \draw[-latex,bp] ({\axisx+\legendx},{\axisy-\legendh}) -- ({\axisx+\legendx+\tensorh},{\axisy-\legendh}) node[anchor=west]{\small Backward data movement};
  \draw[-latex,grad] ({\axisx+\legendx},{\axisy-\legendh*2}) -- ({\axisx+\legendx+\tensorh},{\axisy-\legendh*2}) node[anchor=west,black]{\small Backward grad. movement};

  \node[align=center] at({\secondx},{\tensorh*2.5}) {\textbf{Data-parallel ($N=2$)}};
  \foreach \n in {0,1} {
    \ifthenelse{\n=0}
    {
      \definecolor{fillcolor}{RGB}{228,26,28};
      \tikzstyle{myfill}=[fill=fillcolor];
    }
    {
      \definecolor{fillcolor}{RGB}{55,126,184};
      \tikzstyle{myfill}=[pattern color=fillcolor, pattern=north east lines];
    }
    \begin{scope}[yshift={-\nodehcm*\n}]
      \draw[myfill] ({-\tensorh/2},{-\tensorh/2}) rectangle ({\tensorh/2},{\tensorh/2});
      \draw[-latex] ({\tensorh/2+\aroffset},0) -- ({\secondx-\tensorh/2-\aroffset},0);
      \draw[latex-,bp] ({\tensorh/2+\aroffset},{-\ardist}) -- ({\secondx-\tensorh/2-\aroffset},{-\ardist});

      \draw[myfill] ({\secondx-\tensorh/2},{-\tensorh/2}) rectangle ({\secondx+\tensorh/2},{\tensorh/2});
      \draw[-latex] ({\secondx+\tensorh/2+\aroffset},0) -- ({\thirdx-\tensorh/6-\aroffset},0);
      \draw[latex-,bp] ({\secondx+\tensorh/2+\aroffset},{-\ardist}) -- ({\thirdx-\tensorh/6-\aroffset},{-\ardist});

      \draw[myfill] ({\thirdx-\tensorh/6},{-\tensorh/1.5}) rectangle ({\thirdx+\tensorh/6},{\tensorh/1.5});
      \draw[-latex,bp] ({\thirdx+\tensorh/6+\aroffset},{\tensorh/2}) arc[start angle=90, end angle=-90, radius={\tensorh/2}];
    \end{scope}
  }

  \draw[latex-latex,grad] ({\secondx/2},{-\aroffset}) -- ({\secondx/2},{-\nodeh+\aroffset});
  \draw[latex-latex,grad] ({(\secondx+\thirdx)/2},{-\aroffset}) -- ({(\secondx+\thirdx)/2},{-\nodeh+\aroffset});
  \node[anchor=west,gradcolor] at({(\secondx+\thirdx)/2},{-\nodeh/2}) {Allreduce};

  \node    [anchor=south] at({0},       {\tensorh/2+\aroffset}) {Input};
  \node[anchor=south] at({\secondx},{\tensorh/2+\aroffset}) {Conv.};
  \node[anchor=south] at({\thirdx}, {\tensorh/2+\aroffset})  {FC};

  \definecolor{fillcolor}{RGB}{228,26,28};
  \begin{scope}[xshift={\modelxcm}]

    \node[align=center] at({\secondx},{\tensorh*2.5+0.2}) {\textbf{Model-parallel}};
    \node[align=center] at({\secondx},{\tensorh*2.5-0.2}) {\textbf{(spatial partitioning, $N=1$)}};
    \foreach \n in {0,1} {
      \begin{scope}[yshift={-\nodehcm*\n}]
        \fill[fill=fillcolor] ({\tensorh/2*\n-\tensorh/2},{-\tensorh/2}) rectangle ({\tensorh/2*\n},{\tensorh/2});
        \draw ({-\tensorh/2},{-\tensorh/2}) rectangle ({\tensorh/2},{\tensorh/2});

        \fill[fill=fillcolor] ({\secondx+\tensorh/2*\n-\tensorh/2},{-\tensorh/2}) rectangle ({\secondx+\tensorh/2*\n},{\tensorh/2});
        \draw ({\secondx-\tensorh/2},{-\tensorh/2}) rectangle ({\secondx+\tensorh/2},{\tensorh/2});

        \draw[-latex] ({\tensorh/2+\aroffset},0) -- ({\secondx-\tensorh/2-\aroffset},0);
        \draw[latex-,bp] ({\tensorh/2+\aroffset},{-\ardist}) -- ({\secondx-\tensorh/2-\aroffset},{-\ardist});
        \draw[-latex] ({\secondx+\tensorh/2+\aroffset},0) -- ({\thirdx-\tensorh/6-\aroffset},{\fcy+\nodeh*\n});
        \draw[latex-,bp] ({\secondx+\tensorh/2+\aroffset},{-\ardist}) -- ({\thirdx-\tensorh/6-\aroffset},{\fcy+\nodeh*\n-\ardist});
      \end{scope}
    }
    \draw[fill=fillcolor] ({\thirdx-\tensorh/6},{\fcy-\tensorh/1.5}) rectangle ({\thirdx+\tensorh/6},{\fcy+\tensorh/1.5});
    \draw[-latex,bp] ({\thirdx+\tensorh/6+\aroffset},{\fcy+\tensorh/2}) arc[start angle=90, end angle=-90, radius={\tensorh/2}];

    \draw[-latex] ({\tensorh/2},{-\tensorh/2-\aroffset}) -- ({\secondx-\tensorh/2},{-\nodeh+\tensorh/2+\aroffset});
    \draw[-latex] ({\tensorh/2},{-\nodeh+\tensorh/2+\aroffset}) -- ({\secondx-\tensorh/2},{-\tensorh/2-\aroffset});
    \draw[latex-,bp] ({\tensorh/2},{-\tensorh/2-\aroffset-\ardist}) -- ({\secondx-\tensorh/2},{-\nodeh+\tensorh/2+\aroffset-\ardist});
    \draw[latex-,bp] ({\tensorh/2},{-\nodeh+\tensorh/2+\aroffset-\ardist}) -- ({\secondx-\tensorh/2},{-\tensorh/2-\aroffset-\ardist});

    \draw[latex-latex,grad] ({\secondx/2},{-\aroffset}) -- ({\secondx/2},{-\nodeh+\aroffset});

    \node[anchor=south] at({0},       {\tensorh/2+\aroffset}) {Input};
    \node[anchor=south] at({\secondx},{\tensorh/2+\aroffset}) {Conv.};
    \node[anchor=south] at({\thirdx}, {\tensorh/2+\aroffset+\fcy}) {FC};
  \end{scope}

  \def\hybridx{\modelxcm/2};
  \def\hybridy{-\modelycm};

  \begin{scope}[xshift={\hybridx},yshift={\hybridy}]
    \node[align=center] at({\secondx},{\tensorh*2.5+0.2}) {\textbf{Hybrid-parallel}};
    \node[align=center] at({\secondx},{\tensorh*2.5-0.2}) {\textbf{($N=2$)}};
    \foreach \nd in {0,1} {
      \ifthenelse{\nd=0}
      {
        \definecolor{fillcolor}{RGB}{228,26,28};
        \tikzstyle{myfill}=[fill=fillcolor];
      }
      {
        \definecolor{fillcolor}{RGB}{55,126,184};
        \tikzstyle{myfill}=[pattern color=fillcolor, pattern=north east lines];
      }
      \foreach \nm in {0,1} {
        \begin{scope}[yshift={-\nodehcm/2*(\nm+\nd*2)}]
          \fill[myfill] ({\tensorh/2*\nm-\tensorh/2},{-\tensorh/2}) rectangle ({\tensorh/2*\nm},{\tensorh/2});
          \draw ({-\tensorh/2},{-\tensorh/2}) rectangle ({\tensorh/2},{\tensorh/2});
          \draw[-latex] ({\tensorh/2+\aroffset},0) -- ({\secondx-\tensorh/2-\aroffset},0);
          \draw[latex-,bp] ({\tensorh/2+\aroffset},{-\ardist}) -- ({\secondx-\tensorh/2-\aroffset},{-\ardist});
          \fill[myfill] ({\secondx+\tensorh/2*\nm-\tensorh/2},{-\tensorh/2}) rectangle ({\secondx+\tensorh/2*\nm},{\tensorh/2});
          \draw ({\secondx-\tensorh/2},{-\tensorh/2}) rectangle ({\secondx+\tensorh/2},{\tensorh/2});
          \draw[-latex] ({\secondx+\tensorh/2+\aroffset},0) -- ({\thirdx-\tensorh/6-\aroffset},{\fcy+\nodeh/2*\nm});
          \draw[latex-,bp] ({\secondx+\tensorh/2+\aroffset},{-\ardist}) -- ({\thirdx-\tensorh/6-\aroffset},{\fcy+\nodeh/2*\nm-\ardist});
        \end{scope}
      }
      \begin{scope}[yshift={-\nodehcm*\nd}]
        \draw[myfill] ({\thirdx-\tensorh/6},{\fcy-\tensorh/1.5}) rectangle ({\thirdx+\tensorh/6},{\fcy+\tensorh/1.5});
        \draw[-latex,bp] ({\thirdx+\tensorh/6+\aroffset},{\fcy+\tensorh/2}) arc[start angle=90, end angle=-90, radius={\tensorh/2}];
        \def\dy{0.1};
        \draw[-latex] ({\tensorh/2},{-\tensorh/2-\aroffset/4}) -- ({\secondx-\tensorh/2},{-\nodeh/2+\tensorh/2+\aroffset/4-\dy});
        \draw[-latex] ({\tensorh/2},{-\nodeh/2+\tensorh/2+\aroffset/4-\dy}) -- ({\secondx-\tensorh/2},{-\tensorh/2-\aroffset/4});
        \draw[latex-,bp] ({\tensorh/2},{-\tensorh/2-\aroffset/4-\ardist/2}) -- ({\secondx-\tensorh/2},{-\nodeh/2+\tensorh/2+\aroffset/4-\ardist/2-\dy});
        \draw[latex-,bp] ({\tensorh/2},{-\nodeh/2+\tensorh/2+\aroffset/4-\ardist/2-\dy}) -- ({\secondx-\tensorh/2},{-\tensorh/2-\aroffset/4-\ardist/2});
      \end{scope}
    }
    \begin{scope}[yshift={-\nodehcm/4*2}]
    \end{scope}
    \draw[latex-latex,grad] ({\secondx/2},{-\aroffset}) -- ({\secondx/2},{-\nodeh*1.5+\aroffset});

    \node[anchor=south] at({0},       {\tensorh/2+\aroffset}) {Input};
    \node[anchor=south] at({\secondx},{\tensorh/2+\aroffset}) {Conv.};
    \node[anchor=south] at({\thirdx}, {\tensorh/2+\aroffset+\fcy})  {FC};
  \end{scope}

\end{tikzpicture}

%% file: figures/sample_1.tikz
\begin{tikzpicture}
  \definecolor{fillcolor}{RGB}{228,26,28};
  \def\tensorh{0.25};
  \def\offset{0.1};
  \draw[white] (0,0) rectangle ({-\offset},{\tensorh});
  \draw[fill=fillcolor] (0,0) rectangle ({\tensorh},{\tensorh});
\end{tikzpicture}

%% file: figures/sample_2.tikz
\begin{tikzpicture}
  \definecolor{fillcolor}{RGB}{55,126,184};
  \def\tensorh{0.25};
  \draw[pattern color=fillcolor, pattern=north east lines] (0,0) rectangle ({\tensorh},{\tensorh});
\end{tikzpicture}

%% file: 3_scaling.tex
\begin{figure*}[t]
  \begin{center}
    \input{figures/overview.tikz}
  \end{center}
  \caption{Overview of hybrid-parallel training. Each node contains four processes which partition a single data sample.}
  \label{figure:overview}
\end{figure*}
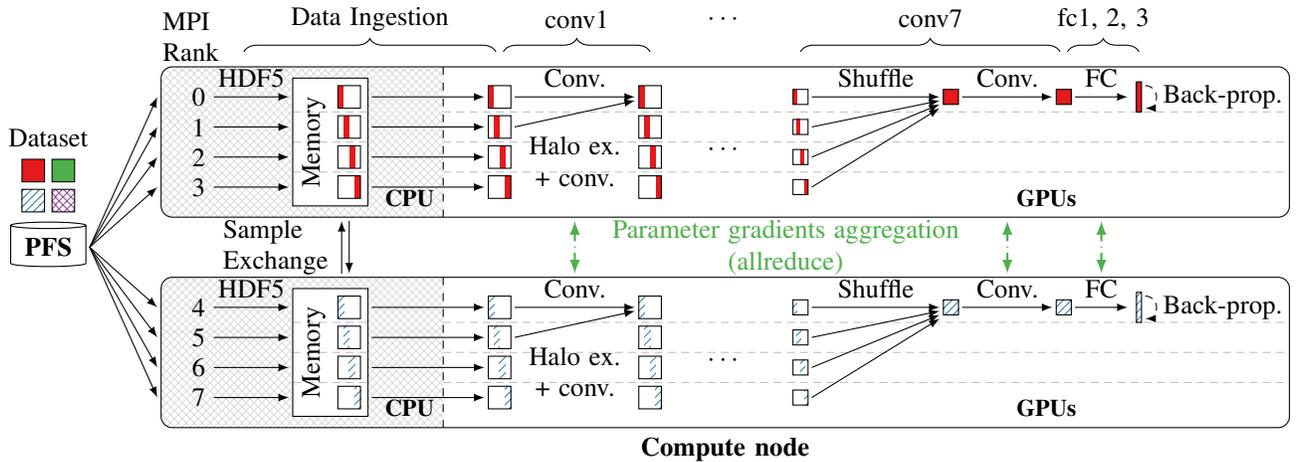

\section{Scalable training of 3D CNNs}
Scaling up the training of neural networks for large 3D data cubes requires innovation in both spatially distributed convolution/deconvolution,
and parallel data ingestion, reuse, and movement. In this section, we discuss each of these in turn.
We also propose a performance model to predict layer-wise computation and communication time for a given network and runtime configuration to validate our computational efficiency. We use the Livermore Big Artificial Neural Network Toolkit (LBANN)~\cite{VanEssen:2015:LLB:2834892.2834897} to implement our approach,
as it has already demonstrated good scalability for 2D spatial partitioning~\cite{distconv}. Our training pipeline is summarized in \figref{figure:overview}.

Part of the contributions described here derive from traditional solutions in HPC systems and software architecture, but are extended to tackle the complexities of training deep neural networks.


\emph{Notation.} We adopt cuDNN's notation for tensor dimensions, using $N$, $C$, $D$, $H$, and $W$ to
refer to samples, channels, depth, height, and width, respectively. Unless explicitly
mentioned, we assume all tensors are fully packed. We use ``$D$-way'', ``$D \times H$-way'', or ``$D \times H \times W$-way''
to refer to how many ways the depth, height, and/or width are partitioned. We
omit $N$ when the remaining GPUs are used for data-parallelism in a hybrid-parallel manner.
For example, if the total number of GPUs is 16, $2$-way means there are $16/ (2 \times 1 \times 1) = 8$ groups each of which split a data sample onto two GPUs in the depth dimension.

\subsection{Hybrid-parallel Implementation of 3D CNNs} \label{subsection:proposal-scaling-extending}

The hybrid-parallel training requires partitioning activations in their spatial and sample dimensions over distributed PEs. Once they are partitioned, each layer computation is done locally except for layers that involve data dependencies across partitions, which include convolutions, pooling, and batch normalization. Convolution and pooling have a spatial dependency that can be resolved with halo exchanges.
For batch normalization, partial statistics over partitions need to be aggregated with allreduces to correctly compute per-channel statistics for samples.

In this work, we extend an existing library~\cite{distconv} that provides the basic infrastructure for implementing hybrid-parallel 2D CNNs.
This extension requires significant additional work to both support 3D data and achieve good performance. We began by adding support for 5D tensors (required for 3D data) to both the library and underlying distributed linear algebra backend.

However, we found several performance issues and missing support for realizing real end-to-end training of 3D CNNs, which had not been reported before. Several operations were designed for 2D data and not well-optimized for large, 3D tensor shapes. For example, we identified that the existing packing and unpacking CUDA kernels for the neighbor communication of boundary regions were sub optimal for our target problems. We developed a suite of new optimized packing/unpacking kernels for common convolutional filters (e.g., $3^3$ and $5^3$). Such optimized kernels are automatically picked when possible. Similarly, we extend other network components, such as distributed batch-normalization and the cross-entropy loss, with optimized versions for large 3D problems. In general, we find that due to the large data sizes, operations that are normally considered cheap can in fact dominate runtime if not well implemented.

The library also lacked support for many layers necessary for more general CNN architectures, as it was designed primarily for sequential networks. The 3D U-Net, which contains both down- and upsampling branches with residual connections between, required additional features. We developed distributed, hybrid-parallel implementations of deconvolution and support for more flexible distributed tensor manipulations for the residual connections.

Unlike the activations, the layer parameters are relatively small in our networks (e.g., 9.5M for CosmoFlow, see \tabref{table:network_architecture}). Thus, we do not expect a significant benefit from partitioning them (e.g., with channel/filter parallelism~\cite{dryden2019channel}) as we do the activations. We use standard data-parallel techniques to aggregate parameter updates with allreduces in backpropagation (green arrows in \figref{figure:overview}).

Note that these extensions do not change the fundamental architecture of the underlying framework, but extend some base classes such as the tensor class and convolutional layer classes to support hybrid-parallelism. Thus, our techniques can be applied to any other deep learning framework.

\subsection{I/O performance optimization} \label{subsection:proposal-scaling-io}


Training the \cosmoflow{} and the 3D \unet{} networks requires the ingestion and shuffling of many huge samples,
each of which is 1 GiB and 64 MiB in size respectively, and is accessed once per epoch in random order.
A key challenge is that in the steady state, ingesting training data from the PFS quickly becomes the dominant portion of the runtime.
Furthermore, the 10 TB \cosmoflow{} dataset is too large to cache in local storage on our compute nodes.
For example, our typical configuration for the \cosmoflow{} network uses a mini-batch of size 64 and our system has 240 GB/s of PFS bandwidth. Thus, loading each mini-batch requires at least 256 ms, which is prohibitively slow (\figref{figure:strong_scale_breakdown}).
Handling this workload efficiently requires solving three tasks: maximizing the utilized bandwidth to the parallel file system (PFS), caching the data set efficiently in distributed memory to avoid subsequent access to the PFS, and efficient shuffling of samples from the data cache during each epoch.

However, when training with hybrid-parallelism,
even if in-memory caching is enabled, we found that
we require that data samples be spatially partitioned and mini-batches are typically small. Traditional sample-parallel I/O approaches have limited parallelism in this regime, and would require data be redistributed to match the spatial parallelism, limiting strong scaling and opportunities for hiding I/O overhead.
In \figref{figure:strong_scale_breakdown_conduit}, we demonstrate that without this spatial-parallel I/O technique training of the \cosmoflow{} network does not scale at all with any number of GPUs, even if the entire dataset is distributed among the host memory of the computing nodes.
This problem is even more acute for the 3D \unet{}, where we also spatially distribute the ground-truth segmentation.

To address this, we develop a new parallel I/O pipeline where each process fetches its local \emph{hyperslab}, or contiguous 3D fragment, of a data sample. This incorporates spatial parallelism into the I/O process to enable strong scaling PFS bandwidth and minimize data shuffling, redistribution, and memory footprints.
We build on existing infrastructure in LBANN, including its C++ data readers and distributed, in-memory data cache~\cite{jacobs2019} to reduce PFS accesses.


\begin{figure}
  \centering
  \begin{minipage}{0.49\textwidth}\centering
    \subfloat[Epoch 0]{
      \def\overviewpartone{1}
      \resizebox{0.65\width}{!}{\input{figures/overview.tikz}}
      \label{figure:data_store_e0}
    }
    \subfloat[Epoch 1+]{
      \def\overviewparttwo{1}
      \resizebox{0.65\width}{!}{\input{figures/overview.tikz}}
      \label{figure:data_store_e1}
    }
  \end{minipage}
  \caption{Data movement during both the initial epoch 0 (left) and steady state epoch 1+ (right).  During epoch 0, HDF5 ingests hyperslabs in parallel into the data store.  During epoch 1+, the data store distributes the hyperslabs for each sample in the mini-batch that is about to be trained on.
  }
  \vspace{-1em}
  \label{figure:data_store}
\end{figure}
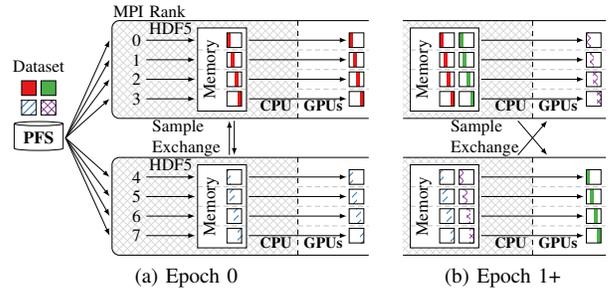

Our data reader uses \conduit{}~\cite{conduit} as both an in-memory data structure and an interface to an I/O backend, such as HDF5 (\figref{figure:data_store}).
\conduit{} is an open source data exchange library that provides efficient ways of exchanging scientific data.  In prior work, LBANN has been optimized to provide parallel I/O using both MPI and multi-threading, but was limited to a single MPI rank per sample.
To overcome this performance bottleneck and support spatially parallel I/O, we rearchitected the data ingestion pipeline to use parallel HDF5 with MPI-IO. This allows multiple ranks which each require one hyperslab of a sample to coordinate their activity when ingesting large samples. Using this, data loading can now track the strong-scaling of our hybrid-parallelism while minimizing data redistribution, as each rank reads only the data it needs.

However, as GPU performance continues to outstrip I/O bandwidth, it is also necessary to minimize PFS accesses. To do this, when samples are loaded (\figref{figure:data_store_e0}), they are placed into \conduit{} nodes and then into LBANN's distributed, in-memory data store to cache the samples for the duration of training. We extended the data store to hold a sample as a collection of hyperslabs. This aligns the spatially parallel I/O, training, and data caching for best performance with hybrid-parallel convolution.

After the first epoch is complete, the data store has cached the entire data set, which it will distribute on subsequent epochs.
Before each epoch, the data store computes a global owner map and a schedule mapping samples to SGD iterations.
This allows the data store to redistribute hyperslabs of samples as needed for the upcoming mini-batch (see \figref{figure:data_store_e1}).

As we strong scale, the capacity of the data store increases in proportion to the compute resources, allowing increasingly large datasets to be cached. This is also well-positioned to take advantage of node-local storage and non-volatile memories on future systems.

%% file: figures/overview.tikz
\begin{tikzpicture}
  \def\numprocs{4}; 

  \ifthenelse{\isundefined{\overviewpartone}}{}{\def\overviewpart{1}}
  \ifthenelse{\isundefined{\overviewparttwo}}{}{\def\overviewpart{1}}

  \def\nodehcm{2.8cm};      
  \def\nodeh{2.8};          
  \def\nodepadh{0.2};       
  \def\proch{0.4};          
  \def\ssdw{1};             
  \def\pfsw{1};             
  \def\tensorh{0.3} ;       
  \def\tensorsh{0.2};       
  \def\arrowmargin{0.05};   
  \def\nodecorners{0.15cm}; 

  \def\pfsx{-4};          
  \def\rankx{-2};         
  \def\inputx{0};         
  \def\convx{4};          
  \def\convsx{6};         
  \def\convshufflex{8};   
  \def\convoutx{9.5};     
  \def\fcx{10.5};         
  \def\braceyshift{12.5}; 
  \def\pfsy{(\nodetopy+(-\nodeh+\nodebtmy))/2}; 
  \def\memoryleftx{(\inputx-0.75)};
  \def\inmemorymarginw{(\proch-\tensorh)};

  \ifthenelse{\isundefined{\overviewparttwo}}{
    \def\memoryrightx{(\inputx+\tensorh/2+0.1)};
  }{
    \def\memoryrightx{(\inputx+\tensorh/2+0.1+\inmemorymarginw+\tensorh)};
  }

  \ifthenelse{\isundefined{\overviewpart}}{
    \def\inputshufflex{2}; 
    \def\cpugpux{(\inputshufflex+\tensorh/2+\arrowmargin*2-1)}; 
  }{
    \ifthenelse{\isundefined{\overviewparttwo}}{
      \def\inputshufflex{2.5};
    }{
      \def\inputshufflex{3};
    }
    \def\cpugpux{(\memoryrightx+(\inputshufflex-\tensorh/2))/2};
  }

  \def\labely{(\nodetopy+0.2)}; 
  \def\nodeleftx{(\rankx-0.5)}; 
  \def\noderightx{(\fcx+2)};    
  \def\nodetopy{(\proch*0.5+\nodepadh)};              
  \def\nodebtmy{(-\proch*(\numprocs-0.5)-\nodepadh)}; 

  \ifthenelse{\isundefined{\overviewpart}}{
    \def\gpux{(\convshufflex+\convoutx)/2};
  }{
    \def\gpux{(\cpugpux)};
  }

  \definecolor{cpucolor}{RGB}{220,220,220};
  \definecolor{gradcolor}{RGB}{77,175,74};
  \tikzstyle{cpufill}=[pattern color=cpucolor, pattern=crosshatch];
  \tikzstyle{grad}=[dashdotted,color=gradcolor,thick];
  \definecolor{fillcolor}{RGB}{228,26,28};
  \definecolor{fillcolor2}{RGB}{55,126,184};
  \definecolor{fillcolor3}{RGB}{77,175,74};
  \definecolor{fillcolor4}{RGB}{152,78,163};

  \tikzstyle{fillstyle}=[fill=fillcolor];
  \tikzstyle{fillstyle2}=[pattern color=fillcolor2, pattern=north east lines];
  \tikzstyle{fillstyle3}=[fill=fillcolor3];
  \tikzstyle{fillstyle4}=[pattern color=fillcolor4, pattern=crosshatch];

  \ifthenelse{\isundefined{\overviewparttwo}}{
    \node[cylinder, shape border rotate=90, draw, minimum height=0.5cm, minimum width=1cm, aspect=0.1] at({\pfsx},{\pfsy}) {\textbf{PFS}};
    \foreach \n in {0,1} {
      \foreach \ip in {1,...,\numprocs} {
        \def\i{(\ip-1)};
        \def\y{-(\proch*\i+\nodeh*\n)};
        \ifthenelse{\isundefined{\overviewslidesampleparallel} \OR \ip=1}{
          \draw[-latex] ({\pfsx+\pfsw/2+\arrowmargin},{\pfsy}) -- ({\nodeleftx-\arrowmargin},{\y});
        }{}
      }
    }

    \def\y{(\pfsy+0.5+\proch*0.8)};
    \draw[fillstyle ] ({\pfsx-\proch/2-\tensorh/2},{\y+\proch/2-\tensorh/2}) rectangle ({\pfsx-\proch/2+\tensorh/2},{\y+\proch/2+\tensorh/2});
    \draw[fillstyle2] ({\pfsx-\proch/2-\tensorh/2},{\y-\proch/2-\tensorh/2}) rectangle ({\pfsx-\proch/2+\tensorh/2},{\y-\proch/2+\tensorh/2});
    \draw[fillstyle3] ({\pfsx+\proch/2-\tensorh/2},{\y+\proch/2-\tensorh/2}) rectangle ({\pfsx+\proch/2+\tensorh/2},{\y+\proch/2+\tensorh/2});
    \draw[fillstyle4] ({\pfsx+\proch/2-\tensorh/2},{\y-\proch/2-\tensorh/2}) rectangle ({\pfsx+\proch/2+\tensorh/2},{\y-\proch/2+\tensorh/2});
    \node[anchor=south] at({\pfsx},{\y+\proch}) {Dataset};
  }{
  }

  \foreach \n in {0,1} {
    \ifthenelse{\n=0} {
      \tikzstyle{myfill}=[fillstyle];
    }{
      \tikzstyle{myfill}=[fillstyle2];
    }

    \begin{scope}[yshift={-\nodehcm*\n}]
      \ifthenelse{\isundefined{\overviewparttwo}}{
        \fill[cpufill]
        ({\cpugpux},{\nodebtmy})
        [rounded corners={\nodecorners}]-- ({\rankx-0.5},{\nodebtmy})
        -- ({\rankx-0.5},{\nodetopy})
        -- ({\cpugpux},{\nodetopy});
      }{
        \fill[cpufill] ({\memoryleftx-\tensorh/2},{\nodebtmy}) rectangle ({\cpugpux},{\nodetopy});
      }

      \ifthenelse{\isundefined{\overviewpart}}{
        \draw[rounded corners={\nodecorners}] ({\nodeleftx},{\nodetopy}) rectangle ({\noderightx},{\nodebtmy});
      }{
        \def\x{(\inputshufflex+\tensorh/2+0.1)};
        \ifthenelse{\isundefined{\overviewparttwo}}{
          \draw
          ({\x},{\nodebtmy})
          [rounded corners={\nodecorners}] -- ({\rankx-0.5},{\nodebtmy})
          -- ({\rankx-0.5},{\nodetopy})
          -- ({\x},{\nodetopy});
        }{
          \draw ({\memoryleftx-\tensorh/2},{\nodetopy}) -- ({\x},{\nodetopy});
          \draw ({\memoryleftx-\tensorh/2},{\nodebtmy}) -- ({\x},{\nodebtmy});
        }
      }

      \draw[dashed] ({\cpugpux},{\nodetopy}) -- ({\cpugpux},{\nodebtmy});
      \node[align=center,anchor=east] at({\cpugpux},{-\proch*(\numprocs-0.65)}) {\small \textbf{CPU}};
      \node[align=center,anchor=west] at({\gpux},{-\proch*(\numprocs-0.65)}) {\small \textbf{GPUs}};



      \draw[fill=white] ({\memoryleftx},{\tensorh/2+0.1}) rectangle ({\memoryrightx},{-\proch*(\numprocs-1)-\tensorh/2-0.1});
      \node[anchor=north,align=center,rotate=90] at({\memoryleftx},{-\proch*(\numprocs-1)/2}) {Memory};

      \ifthenelse{\isundefined{\overviewparttwo}} {
        \ifthenelse{\isundefined{\overviewslidesampleparallel}}{
          \node[anchor=south] at({((\rankx+\tensorh/2)+\memoryleftx)/2},0) {HDF5};
        }{
          \node[anchor=south] at({((\rankx+\tensorh/2)+\memoryleftx)/2},{-\arrowmargin}) {I/O};
        }
      }{}

      \foreach \ip in {1,...,\numprocs} {
        \def\i{(\ip-1)};

        \ifthenelse{\isundefined{\overviewparttwo}}{
          \newcount\tmp
          \tmp=\ip
          \advance\tmp by -1
          \ifthenelse{\n=1}{\advance\tmp by \numprocs}{}
          \node at({\rankx},{-\proch*\i}) {\number\tmp};

          \ifthenelse{\isundefined{\overviewslidesampleparallel} \OR \ip=1}{
            \draw[-latex] ({\rankx+\tensorh/2+\arrowmargin},{-\proch*\i}) -- ({\memoryleftx-\arrowmargin},{-\proch*\i});
          }{
          }
        }{
        }

        \ifthenelse{\ip=1}{
        }{
          \def\y{-\proch*(\i-0.5)};
          \draw[densely dashed, opacity=0.25] ({\memoryleftx},{\y}) -- ({\memoryrightx},{\y});
        }

        \ifthenelse{\ip=1}{
        }{
          \ifthenelse{\isundefined{\overviewpart}}{
            \def\x{\noderightx}
          }{
            \def\x{(\inputshufflex+\tensorh/2+0.1)}
          }
          \def\y{-\proch*(\i-0.5)};
          \draw[densely dashed, opacity=0.25] ({\cpugpux},{\y}) -- ({\x},{\y});
        }

        \ifthenelse{\isundefined{\overviewslidesampleparallel}}{
          \def\x{\inputx-\tensorh/2+\tensorh/\numprocs*(\i+1)};
        }{
          \def\x{\inputx+\tensorh/2};
        }
        \ifthenelse{\isundefined{\overviewslidesampleparallel} \OR \ip=1}{
          \fill[myfill] ({\inputx-\tensorh/2+\tensorh/\numprocs*\i},{-\proch*\i-\tensorh/2}) rectangle ({\x},{-\proch*\i+\tensorh/2});
          \draw ({\inputx-\tensorh/2},{-\proch*\i-\tensorh/2}) rectangle ({\inputx+\tensorh/2},{-\proch*\i+\tensorh/2});
        }{}

        \ifthenelse{\isundefined{\overviewparttwo}}{
        }{
          \ifthenelse{\n=0} {
            \tikzstyle{myfill3}=[fillstyle3];
            \tikzstyle{myfill4}=[fillstyle4];
          }{
            \tikzstyle{myfill4}=[fillstyle3];
            \tikzstyle{myfill3}=[fillstyle4];
          }
          \def\x{(\inputx+\inmemorymarginw+\tensorh)};
          \fill[myfill3] ({\x-\tensorh/2+\tensorh/\numprocs*\i},{-\proch*\i-\tensorh/2}) rectangle ({\x-\tensorh/2+\tensorh/\numprocs*(\i+1)},{-\proch*\i+\tensorh/2});
          \draw ({\x-\tensorh/2},{-\proch*\i-\tensorh/2}) rectangle ({\x+\tensorh/2},{-\proch*\i+\tensorh/2});
        }

        \def\leftx{(\memoryrightx+\arrowmargin)};
        \def\rightx{(\inputshufflex-\tensorh/2-\arrowmargin)};
        \ifthenelse{\isundefined{\overviewslidesampleparallel}}{
          \def\y{-\proch*\i};
        }{
          \def\y{0};
          \ifthenelse{\ip=1}{
            \node[anchor=south] at({(\leftx+\rightx)/2},0) {Scatter};
          }{}
        }
        \draw[-latex] ({\leftx},{\y}) -- ({\rightx},{-\proch*\i});

        \fill[fill=white] ({\inputshufflex-\tensorh/2},{-\proch*\i-\tensorh/2}) rectangle ({\inputshufflex+\tensorh/2},{-\proch*\i+\tensorh/2});

        \ifthenelse{\isundefined{\overviewparttwo}}{
          \tikzstyle{myfillx}=[myfill];
        }{
          \tikzstyle{myfillx}=[myfill4];
        }

        \fill[myfillx] ({\inputshufflex-\tensorh/2+\tensorh/\numprocs*\i},{-\proch*\i-\tensorh/2}) rectangle ({\inputshufflex-\tensorh/2+\tensorh/\numprocs*(\i+1)},{-\proch*\i+\tensorh/2});
        \draw ({\inputshufflex-\tensorh/2},{-\proch*\i-\tensorh/2}) rectangle ({\inputshufflex+\tensorh/2},{-\proch*\i+\tensorh/2});

        \ifthenelse{\isundefined{\overviewpart}}{
          \fill[myfill] ({\convx-\tensorh/2+\tensorh/\numprocs*\i},{-\proch*\i-\tensorh/2}) rectangle ({\convx-\tensorh/2+\tensorh/\numprocs*(\i+1)},{-\proch*\i+\tensorh/2});
          \draw ({\convx-\tensorh/2},{-\proch*\i-\tensorh/2}) rectangle ({\convx+\tensorh/2},{-\proch*\i+\tensorh/2});

          \fill[myfill] ({\convsx-\tensorsh/2+\tensorsh/\numprocs*\i},{-\proch*\i-\tensorsh/2}) rectangle ({\convsx-\tensorsh/2+\tensorsh/\numprocs*(\i+1)},{-\proch*\i+\tensorsh/2});
          \draw ({\convsx-\tensorsh/2},{-\proch*\i-\tensorsh/2}) rectangle ({\convsx+\tensorsh/2},{-\proch*\i+\tensorsh/2});
          \draw[-latex] ({\convsx+\tensorsh/2+\arrowmargin},{-\proch*\i}) -- ({\convshufflex-\tensorsh/2-\arrowmargin},{-\arrowmargin*\i});
        }{
        }
      }

      \ifthenelse{\isundefined{\overviewpart}}{
        \node[anchor=south] at({(\inputshufflex+\convx)/2},0) {Conv.};
        \node[anchor=north, minimum width=2cm, text width=2cm, align=center] at({(\inputshufflex+\convx)/2},{-\proch}) {Halo ex. \\ + conv.};
        \draw[-latex] ({\inputshufflex+\tensorh/2+\arrowmargin},0) -- ({\convx-\tensorh/2-\arrowmargin},0);
        \draw[-latex] ({\inputshufflex+\tensorh/2+\arrowmargin},{-\proch}) -- ({\convx-\tensorh/2-\arrowmargin},{-\arrowmargin});

        \node[align=center] at({(\convx+\convsx)/2},{-\proch*(\numprocs-1)/2-\proch/4}) {$\cdots$};
        \node[anchor=south] at({(\convsx+\convshufflex)/2},0) {Shuffle};

        \draw[myfill] ({\convshufflex-\tensorsh/2},{-\tensorsh/2}) rectangle ({\convshufflex+\tensorsh/2},{\tensorsh/2});
        \node[anchor=south] at({(\convshufflex+\convoutx)/2},0) {Conv.};
        \draw[-latex] ({\convshufflex+\tensorsh/2+\arrowmargin},0) -- ({\convoutx-\tensorsh/2-\arrowmargin},0);
        \draw[myfill] ({\convoutx-\tensorsh/2},{-\tensorsh/2}) rectangle ({\convoutx+\tensorsh/2},{\tensorsh/2});
        \node[anchor=south] at({(\convoutx+\fcx)/2},0) {FC};
        \draw[-latex] ({\convoutx+\tensorsh/2+\arrowmargin},0) -- ({\fcx-\tensorsh/2-\arrowmargin},0);
        \draw[myfill] ({\fcx-\tensorsh/6},{-\tensorsh}) rectangle ({\fcx+\tensorsh/6},{\tensorsh});

        \draw[-latex, densely dashed] ({\fcx+\tensorsh/6+\arrowmargin},{\tensorsh-0.05}) arc[start angle=90, end angle=-90, radius={\tensorh/2}];
        \node[anchor=west,align=left] at({\fcx+\tensorh/2+\arrowmargin},0) {Back-prop.};
      }{
      }
    \end{scope}
  }

  \ifthenelse{\isundefined{\overviewslidesampleparallel}}{
    \ifthenelse{\isundefined{\overviewparttwo}}{
      \def\i{((\inputx+\tensorh/2+\inputx-\tensorh/2)/2)};
      \draw[latex-] ({\i-.1},{\nodebtmy-\arrowmargin}) -- ({\i-.1},{-\nodeh+\nodetopy+\arrowmargin});
      \draw[-latex] ({\inputx*\tensorh},{\nodebtmy-\arrowmargin}) -- ({\inputx*\tensorh},{-\nodeh+\nodetopy+\arrowmargin});
    }{
      \def\i{(\cpugpux)};
      \def\w{0.3};
      \draw[latex-] ({\i+\w},{\nodebtmy-\arrowmargin}) -- ({\i-\w},{-\nodeh+\nodetopy+\arrowmargin});
      \draw[-latex] ({\i-\w},{\nodebtmy-\arrowmargin}) -- ({\i+\w},{-\nodeh+\nodetopy+\arrowmargin});
    }
    \node[minimum width=1cm, text width=1cm, anchor=west] at({\i-1.8},{(\nodebtmy+(-\nodeh+\nodetopy))/2}) {Sample \\ Exchange};
  }{}

  \ifthenelse{\isundefined{\overviewpart}}{
    \foreach \i in {((\inputshufflex+\convx)/2),((\convshufflex+\convoutx)/2),((\convoutx+\fcx)/2)} {
      \draw[latex-latex,grad] ({\i},{\nodebtmy-\arrowmargin}) -- ({\i},{-\nodeh+\nodetopy+\arrowmargin});
    }
    \node[minimum width=10cm, text width=10cm, align=center, gradcolor] at({(\convx+\convsx)/2+0.8},{(\nodebtmy+(-\nodeh+\nodetopy))/2})
    {Parameter gradients aggregation \\ (allreduce)};
  }{
  }

  \ifthenelse{\isundefined{\overviewparttwo}}{
    \ifthenelse{\isundefined{\overviewpart}}{
      \node[minimum width=1cm, text width=1cm, anchor=west] at({\rankx-.6},{(\labely+.2}) {MPI \\ Rank};
    }{
      \node[minimum width=2cm, text width=2cm, anchor=west] at({\rankx-.6},{(\labely}) {MPI Rank};
    }
    \ifthenelse{\isundefined{\overviewpart}}{
      \draw[decorate,decoration={brace,amplitude=5}] ({\rankx+.6},{\labely}) -- ({\inputshufflex-\arrowmargin},{\labely}) node[midway,yshift={\braceyshift}] {Data Ingestion};
    }{}
  }{
  }
  \ifthenelse{\isundefined{\overviewpart}}{
    \draw[decorate,decoration={brace,amplitude=5}] ({\inputshufflex+\arrowmargin},{\labely}) -- ({\convx},{\labely}) node[midway,yshift={\braceyshift}] {conv1};
    \path ({\convx},{\labely}) -- ({\convsx},{\labely}) node[midway,yshift={\braceyshift}] {$\cdots$};
    \draw[decorate,decoration={brace,amplitude=5}] ({\convsx},{\labely}) -- ({\convoutx-\arrowmargin},{\labely}) node[midway,yshift={\braceyshift}] {conv7};
    \draw[decorate,decoration={brace,amplitude=5}] ({\convoutx+\arrowmargin},{\labely}) -- ({\fcx},{\labely}) node[midway,yshift={\braceyshift}] {fc1, 2, 3};
  }{
  }

  \ifthenelse{\isundefined{\overviewpart}}{
    \node[anchor=north] at({(\nodeleftx+\noderightx)/2},{-\nodeh+\nodebtmy}) {\textbf{Compute node}};
  }{}

  \ifthenelse{\isundefined{\overviewslide}}{
  }{
    \onslide<1> {
      \def\r{0.1};
      \def\y{(\labely+1)};
      \draw[very thick, -latex] ({\rankx},{\y+\r*2}) -- ({\noderightx-\r},{\y+\r*2});
      \draw[very thick, densely dashed] ({\noderightx-\r},{\y+\r*2}) arc[start angle=90, end angle=-90, radius={\r}];
      \draw[very thick, -latex, densely dashed] ({\noderightx-\r},{\y}) -- ({\inputshufflex},{\y});
      \node[anchor=south] at({(\nodeleftx+\noderightx)/2},{\y+\r*2}) {\textbf{Forward/backward computation}};
    }

    \onslide<2> {
      \def\leftx{(\pfsx-\pfsw/2-\arrowmargin*4)};
      \def\topy{(\labely+0.7)};
      \def\btmy{(\nodebtmy-\nodeh-0.5)};
      \tikzstyle{sliderect}=[fill=black, fill opacity=0, rounded corners={\nodecorners}, very thick];

      \fill[white, opacity=0.5] ({\leftx},{\topy}) rectangle ({\noderightx},{\btmy});

      \draw[sliderect]
      ({\leftx},{\topy}) rectangle ({\inputshufflex-\arrowmargin},{\btmy});
      \draw[sliderect]
      ({\inputshufflex+\arrowmargin},{\topy}) rectangle ({\convoutx},{\btmy});

      \node[anchor=south, minimum width=6cm, text width=6cm] at({(\leftx+\inputshufflex)/2},{\topy}) {\large{\textbf{1. \alertbf{Parallel data reader}: \\ Parallelize the I/O workload by using \alertbf{hyperslabs}}}};
      \node[anchor=south, minimum width=7cm, text width=7cm] at({(\inputshufflex+\convoutx)/2},{\topy}) {\large{\textbf{2. \alertbf{\distconv{}}: \\ Parallelize the computation by partial convolution and halo exchanges}}};
    }
  }

  \def\y{(\labely+\arrowmargin*4)};
  \ifthenelse{\isundefined{\overviewslidepartone}}{
  }{
    \node[anchor=south, align=center, text width=5cm, minimum width=5cm] at({\inputx},{\y}) {A hyperslab};
    \draw[-latex] ({\inputx},{\y}) -- ({\inputx},{\tensorh/2});
  }

  \ifthenelse{\isundefined{\overviewslideparttwo}}{
  }{
    \node[anchor=south, align=center, text width=5cm, minimum width=5cm] at({\inputx},{\y}) {The dataset is distributed among the host memory};
    \draw[-latex] ({\inputx},{\y}) -- ({\inputx},{\nodetopy});
  }

\end{tikzpicture}

%% file: 3_perfmodel.tex
\subsection{Performance Modeling} \label{section:perfmodel}
We use a performance model to predict the time to perform one iteration of training with given a configuration,
such as the mini-batch size and the number of nodes,
to understand the quantitative behavior of the framework and validate performance.
We first collect the time to perform (de)convolution, pooling, and batch normalization kernels with various input sizes on a single GPU using \cudnn{}~\cite{2014arXiv1410.0759C},
and then we combine the benchmark results with a communication model to predict layer-wise runtime on multiple GPUs.

The time to perform forward-computation of convolutional or pooling layer $l$ is
\begin{align*}
  {FP}_l &= \max
           \left\{
           {Comp}_l \left( D_l^{main} \right) ,
           \sum_{d=0}^{2} 2 {SR} \left( D^{halo}_{l,d} \right)
           \right\} \\
         &+ {Comp}_l \left( D_l^{halo} \right)
\end{align*}
where ${Comp}_l(D)$ is time to compute layer $l$ on a given domain $D$,
and ${SR}(D)$ is time to perform peer-to-peer send-receive communication between two GPUs
(via NVLink or inter-node InfiniBand depending on the location of the two processes).
The shape of $D_l^{main}$, the domain which can be computed without halo communication,
and $D_{l,d}^{halo}$, the domain which requires halo region to be computed,
are defined by the partitioning of the layer.
We define ${BD}_l$ and ${BF}_l$, the time to perform the backward-data and backward-filter passes on layer $l$, respectively, in a similar manner.

To estimate ${Comp}_l (D)$,
we benchmark each layer type on a single GPU. Unless noted, we use the largest cuDNN workspace possible, and autotune to find the fastest convolution algorithms. We use the median of three trials after warmup. To estimate $SR(D)$, we use Aluminum's ping-pong benchmark and apply linear regression to estimate the time for arbitrary message sizes.

The time for a batch-normalization layer is the sum of the computational time
and time to perform allreduce of the local sum and squared-sum of each channel.

Finally, the total time of the network is
\begin{align*}
  {Cost} &= \sum_l {FP}_l + \max\left\{ \sum_l \left( {BD}_l + {BF}_l \right), \sum_l {AR}_l (\theta_l) \right\},
\end{align*}
where ${AR}_l$ is time to perform allreduce among all of the GPUs and $\theta_l$ is the number of parameters of layer $l$.
To estimate $AL_l$, measure the performance on one node (4 GPUs) to 128 nodes (512 GPUs), with float vectors of 1 to 16 M elements,
and apply linear regression \cite{10.1177/1094342005051521,oyama-bigdata2016} with logarithmic transformations to predict the time for a given message size and the number of GPUs.

We ignore the cost of non-3D part of the 3D CNNs
(e.g., fully-connected and loss layers),
since their costs are negligible compared to other costs,
such as allreduces or convolution.
We also ignore the cost of I/O for loading data samples from the PFS or between processes,
as our optimized pipeline mitigates I/O costs drastically for the two networks we use in this paper.

%% file: 3_model.tex
\section{Extended \cosmoflow{} Model} \label{section:proposal-model}




\begin{table}[t]
  \caption{\cosmoflow{} network architecture.
    $W_i$ is the input spatial width.
    c$N{\rightarrow}$p$N$ are convolution followed by pooling and fc$N$ are fully connected layers.
    We use stride 1 convolution and stride 2 pooling unless noted.
    All layers use ``same'' padding.}
  \label{table:network_architecture}
  \begin{center}

\input{tables/network.tex}
  \end{center}
\end{table}

We now discuss extensions we make to the \cosmoflow{} network, as this is the first attempt to train the network with $64\times$ larger input data than before. We use the \cosmoflow{} model presented in the previous work as our baseline model and extend it to improve its prediction accuracy by exploiting our new hybrid-parallel training capabilities.


\tabref{table:network_architecture} summarizes three models, corresponding to the $128^3$, $256^3$, and $512^3$ voxel training datasets, respectively. For each of the models, we have applied several extensions to the original baseline model.
First, we add a batch normalization layer~\cite{Ioffe:2015:BNA:3045118.3045167} after every convolutional layer. Ravanbakhsh et al. reported that batch normalization was critical in training a similar model~\cite{Ravanbakhsh:2016:ECP:3045390.3045644}. However, in the original \cosmoflow{} model, it was dropped due to the computational cost of batch normalization, especially in a distributed training setting. We present training results in both configurations (\secref{section:evaluation}) and observe that while batch normalization increases memory requirements, it improves final prediction accuracy.
Second, in order to simplify comparison of the three models, we insert additional pooling layers in the $256^3$ and $512^3$ models (the pool6 layer in both models and the pool7 layer in the $512^3$ model).
Finally, we experimentally identified several minor parametric changes that improve prediction accuracy or simplify the implementation of distributed convolution, including removal of biases and use of padding in convolutional layers. We removed biases as we observed significant performance overheads for them in practice.

The remaining details follow the original model:
We use leaky ReLU~\cite{Maas13rectifiernonlinearities} activations (except for the last layer),
dropout with a keep probability of 0.8 after every fully-connected layer,
and adopt the mean squared error as the loss function.
We use the Adam \cite{2014arXiv1412.6980K} optimizer with $\beta_1=0.9$, $\beta_2=0.999$, and $\epsilon=10^{-8}$
and a linear learning-rate decay schedule which leads to 0.01x of the initial rate in 100 epochs.
We perform grid search to tune initial learning rate $\eta^{(0)}$ for each network.


To compare with prior work and study the impact of data volume,
we synthesize two datasets with data volumes of size $128^3$ and $256^3$, by splitting each $512^3$ cube
into 64 and 8 sub-volumes, respectively. This is analogous to the partitioning
method used in the original work. The intuition behind this was that the data
volumes should be sufficiently large to contain galaxy clusters, which are
sensitive cosmological probes.

Training the largest network needs 4 GPUs to store the 52.7 GiB of memory required (\tabref{table:network_architecture}). When batch normalization layers are introduced, memory requirements double, necessitating at least 8 GPUs (2 nodes) per sample.


%% file: tables/network.tex
\newcommand{\veryshortarrow}[1][3pt]{\mathrel{%
   \hbox{\rule[\dimexpr\fontdimen22\textfont2-.2pt\relax]{#1}{.4pt}}%
   \mkern-4mu\hbox{\usefont{U}{lasy}{m}{n}\symbol{41}}}}

\begin{tabular}{c c|c c c}
  \toprule
  \multicolumn{2}{c|}{\textbf{Layer(s)}} & \multicolumn{3}{c}{\textbf{Output width}} \\
  \textbf{Name(s)} & \textbf{Filter} & \textbf{$W_i=128$} & \textbf{$W_i=256$} & \textbf{$W_i=512$} \\
  \midrule
  c1$\rightarrow$p1 & $16 \times \cubic{3}$ & $\cubic{128}{\rightarrow}\cubic{64}$ & $\cubic{256}{\rightarrow}\cubic{128}$ & $\cubic{512}{\rightarrow}\cubic{256}$ \\
  \midrule
  c2$\rightarrow$p2 & $32 \times \cubic{3}$ & $\cubic{64}{\rightarrow}\cubic{32}$ & $\cubic{128}{\rightarrow}\cubic{64} $  & $\cubic{256}{\rightarrow}\cubic{128}$ \\
  \midrule
  c3$\rightarrow$p3 & $64 \times \cubic{3}$ & $\cubic{32}{\rightarrow}\cubic{16}$ & $\cubic{64}{\rightarrow}\cubic{32}$ & $\cubic{128}{\rightarrow}\cubic{64}$ \\
  \midrule
  \multirow{2}{*}{c4$\rightarrow$p4} & $128 \times \cubic{3}$ & \multirow{2}{*}{$\cubic{8}{\rightarrow}\cubic{4}$} & \multirow{2}{*}{$\cubic{16}{\rightarrow}\cubic{8}$} & \multirow{2}{*}{$\cubic{32}{\rightarrow}\cubic{16}$} \\
                                         & (stride of 2) &&& \\
  \midrule
  c5$\rightarrow$p5 & $256 \times \cubic{3}$ & $\cubic{4}{\rightarrow}\cubic{2}$ & $\cubic{8}{\rightarrow}\cubic{4}$ & $\cubic{16}{\rightarrow}\cubic{8}$ \\
  \midrule
  c6$\rightarrow$p6 & $256 \times \cubic{3}$ & $\cubic{2}{\rightarrow}$N/A & $\cubic{4}{\rightarrow}\cubic{2}$ & $\cubic{8}{\rightarrow}\cubic{4}$ \\
  \midrule
  c7$\rightarrow$p7 & $256 \times \cubic{3}$ & $\cubic{2}{\rightarrow}$N/A & $\cubic{2}{\rightarrow}$N/A & $\cubic{4}{\rightarrow}\cubic{2}$ \\
  \midrule
  fc1   & $2048$ & $2048$ & $2048$ & $2048$ \\
  fc2   & $256$  & $256$  & $256$  & $256$  \\
  fc3   & $4$    & $4$    & $4$    & $4$    \\
  \midrule
  \multicolumn{2}{c|}{\fontsize{6pt}{0pt}\selectfont \# conv. ops. [GFlops/sample]} & 55.55 & 443.8 & 3550 \\ 
  \multicolumn{2}{c|}{\fontsize{6pt}{0pt}\selectfont (Forward) [GFlops/sample]} & 18.52 & 147.9 & 1183 \\
  \multicolumn{2}{c|}{Memory [GiB/sample]} & 0.824 & 6.59 & 52.7 \\ 
  \multicolumn{2}{c|}{\# parameters $\left[ 10^6 \right]$} & 9.44 & 9.44 & 9.44 \\ 
  \bottomrule
\end{tabular}

%% file: 4_evaluation.tex
\section{Evaluation} \label{section:evaluation}
In this section, we first evaluate the computational performance of our hybrid-parallel implementation for \cosmoflow{} and the 3D \unet{} in both
strong scaling (fixed global mini-batch size) and weak scaling (fixed mini-batch size on each GPU) regimes.
Then, we demonstrate the importance of increasing the input data resolution of the \cosmoflow{} network to improve its prediction accuracy.
To our knowledge, this work is the first attempt to train the \cosmoflow{} network with the full-resolution universe data instead of partitioning them into small sub-volumes.

\subsection{Evaluation environment} \label{subsection:environment}
We use \lassen{}, a GPU supercomputer at \llnl{} composed of 792 nodes.
Each node has two IBM POWER9 CPUs with 256 GB memory and four NVIDIA V100 GPUs with 16 GB memory and NVLink2.
Each CPU has two GPUs directly connected to it via NVLink, and the two GPUs on each socket are also directly connected via NVLink. The network is dual-rail EDR InfiniBand.

We use GCC 7.3.1, CUDA 10.1, \cudnn{} 7.6.4, NCCL 2.4.2 and IBM Spectrum MPI 10.2.0.11rtm2.
We use auto-tuning to select \cudnn{} convolution algorithms.
We use FP32 for computation throughout the experiments.
We do not use FP16 mixed-precision training (or Tensor Cores),
as the impact of applying low-precision training to \cosmoflow{} has not yet been evaluated.

For the 3D \unet{}, we use the original network architecture proposed in the paper~\cite{CABR16},
but increase the input/output size to $256^3$.
As mentioned in \secref{subsection:background_unet},
the network consumes much more memory than the \cosmoflow{} network for the same input data size,
so we use a smaller size to keep the number of GPUs per sample the same as the \cosmoflow{} experiments.

\subsection{Strong scaling} \label{subsection:strong_scale}

\begin{figure}[t]
  \input{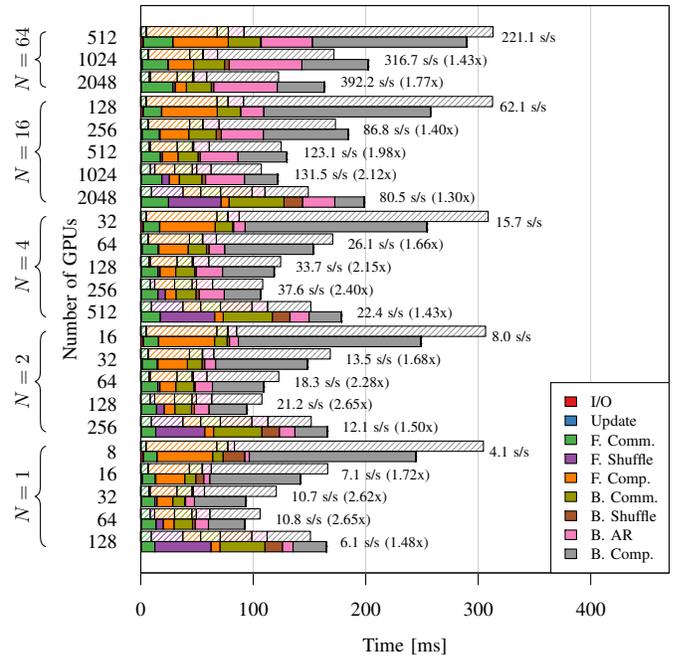}
  \caption{Strong scaling of \cosmoflow{}.
    Shaded bars show time predicted by the performance model.
    ``F.'' and ``B.'' are forward and backward passes, resp. $N$ is the mini-batch size.
    Bars are annotated with throughput (samples/s) and speedup relative to the minimum setting with the same $N$.
  }
  \label{figure:strong_scale_breakdown}
\end{figure}

\begin{figure}[t]
  \input{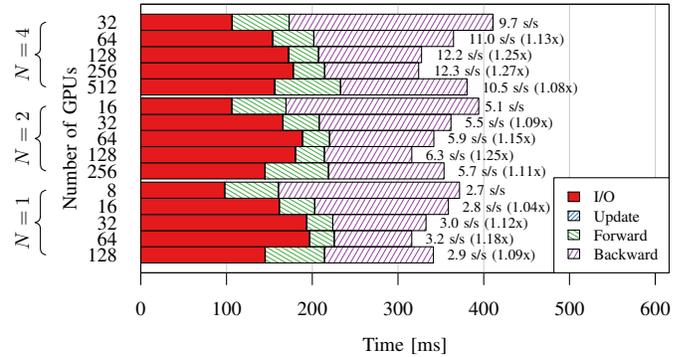}
  \caption{Strong scaling of \cosmoflow{} without spatial-parallel I/O,
    using only distributed caching with \conduit{}.
  }
  \label{figure:strong_scale_breakdown_conduit}
\end{figure}

Training neural networks with strong scaling increases the number of compute
resources brought to bear without perturbing the learning behavior of the model.
In conjunction with hybrid-parallelism this technique allows us to use an
unprecedented number of GPUs per data sample.
\figref{figure:strong_scale_breakdown} shows the strong scaling performance of the \cosmoflow{} network with the \perf{dim512} dataset.
We use global mini-batch sizes ($N$) of 1, 2, 4, 16 and 64, and split the network in the depth dimension.
We run the framework for 4 epochs with a 128-sample subset of the dataset (if the mini-batch size is smaller than 128), or the full dataset, and show the median iteration time except for the first epoch.
We also show predicted times by our performance model, which largely match with the actual measured times, confirming our implementation performed as expected.

As shown in the figure, when the mini-batch size, $N$, is 16 and 64,
we achieve speedups of \perf{strong-scale-cosmoflow-speedup-N16-n32-to-n128} with 512 GPUs (128 nodes) compared to 128 GPUs (32 nodes),
and \perf{strong-scale-cosmoflow-speedup-N64-n128-to-n512} with 2048 GPUs (512 nodes) compared to 512 GPUs (128 nodes), respectively.
We note that for 16 samples the performance gain for going to 1024 GPUs falls off, as the problem becomes over-decomposed.
However, the computational performance in terms of throughput can still be scaled further by increasing the batch size to 64.
As shown in \secref{subsection:training_dim512}, this mini-batch size is a reasonable choice for actual training,
and thus we prove that we successfully scale the training of the CosmoFlow network to thousands of GPUs.
Furthermore, the I/O time is almost invisible in the figure since it is almost completely overlapped with computations in our optimized I/O pipeline.
This makes a significant contrast to conventional I/O methods in terms of strong scaling performance, as their I/O parallelism is limited by the mini-batch size.
In fact, without our spatially-parallel I/O approach, the iteration time does not scale due to the I/O overhead, as indicated in \figref{figure:strong_scale_breakdown_conduit}, which visualizes the impact of the I/O overhead when the spatial-parallel I/O is disabled.
This demonstrates the necessity of strong-scaling I/O along with compute in order to efficiently parallelize training.

\begin{figure}[t]
  \subfloat[$N=4$, \perf{dim512}, \perf{n811}, 32 GPUs]{\input{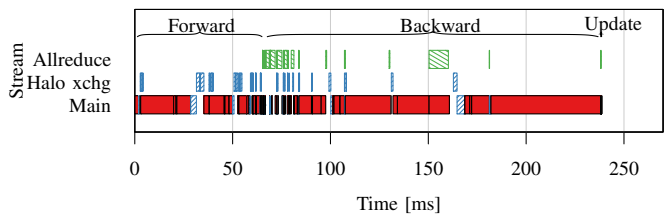}}
  \newline
  \subfloat[$N=4$, \perf{dim512}, \perf{n1611}, 64 GPUs]{\input{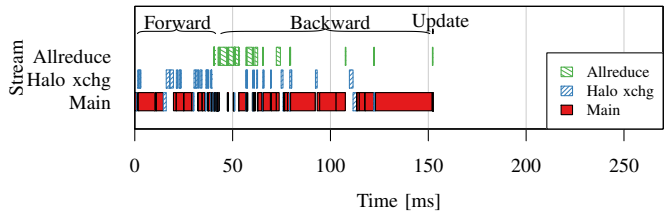}}
  \caption{
    Single-GPU execution timelines for training the \perfx{dim512} dataset with mini-batch size 4.
    Top: 8 GPUs/sample (32 total). Bottom: 16 GPUs/sample (64 total).
    We show one iteration of the root process's GPU of each run.}
  \label{figure:strong_scale_timeline}
\end{figure}

To understand the parallel efficiency of the implementation and identify
potential bottlenecks,
\figref{figure:strong_scale_timeline} shows the GPU execution timeline of a
mini-batch iteration when 32 and 64 GPUs are used to train the $512^3$ model
with a mini-batch size of 4.
A speedup of approximately \perf{strong-scale-cosmoflow-speedup-N4-n8-to-n16} is achieved using $2\times$ the number of GPUs.
The ``Main'' row corresponds to the CUDA stream where compute kernels are launched;
the ``Halo xchg'' row is an asynchronous stream to perform on-device halo exchanges;
and the ``Allreduce'' row corresponds to a stream used by the asynchronous allreduce operations by NCCL.
From the beginning of back propagation, NCCL starts to communicate computed parameter gradients among processes asynchronously to the main computation stream. Since the communication of gradient updates is done asynchronously, this does not block the compute kernels. In both cases, the main streams are nearly fully packed,
indicating the GPU compute units are fully occupied. Similarly, the timelines indicate that the cost of our optimized halo exchanges is almost negligible in these scenarios. Overall, we see that the speedup from the 8-way to 16-way parallelization is mostly determined by the speedups of the individual convolution kernels in the cuDNN library.
In this work, we have exclusively relied on cuDNN for optimized convolution kernels. These results indicate that they may not be well-tuned for non-cube domains, as we only achieved \perf{strong-scale-cosmoflow-speedup-N4-n8-to-n16} speedup going from 8-way to 16-way parallelization.
Identifying the local compute kernels as the bottleneck to better scaling is also corroborated
by our performance model, shown in \figref{figure:strong_scale_breakdown},
which was generated by profiling cuDNN.



\begin{figure}[t]
  \input{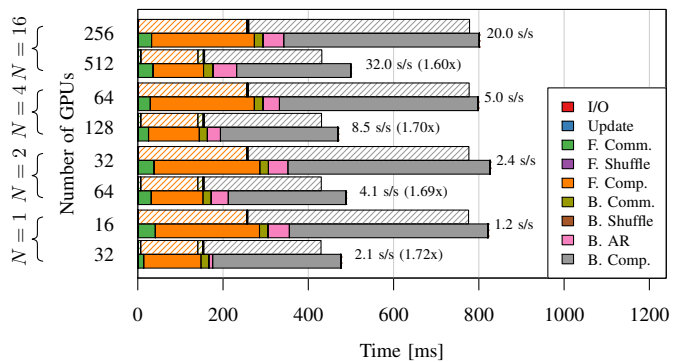}
  \caption{Strong scaling of the 3D \unet{}.
    Shaded bars show iteration time predicted by the performance model.
  }
  \label{figure:strong_scale_breakdown_unet}
\end{figure}

The ability to strong scale the performance of the 3D \unet{} with an input size of
$256^3$ is shown in
\figref{figure:strong_scale_breakdown_unet}.
With this network, we have to use at least 16 GPUs per sample due to the memory requirements.
We achieve good strong scaling performance between 16-way and 32-way partitioning,
such as \perf{strong-scale-unet-speedup-N16-n64-to-n128} on 512 GPUs over 256 GPUs with a mini-batch size of 16.
As shown in \figref{figure:strong_scale_breakdown_unet}, similarly to the \cosmoflow{} network, most of the iteration time is spent in computation,
implying that we achieve near-peak performance, despite the communication overheads of hybrid-parallelism compared to data-parallelism.

\begin{table}[t]
  \caption{Achieved performance of CosmoFlow convolution layers compared
    to peak performance of cuDNN.}
  \label{table:cudnn_performance}
  \begin{center}
    \input{tables/cudnn_performance.tex}
  \end{center}
\end{table}

To better characterize our performance and scaling efficiency, we compare the performance of our distributed convolution layers in the $512^3$ CosmoFlow network to the peak achievable performance of cuDNN in \tabref{table:cudnn_performance}.
The Time and Perf columns give the measured performance of our code (including halo communication, etc.), measured with nvprof.
In the Peak column, we report the TFlop/s achieved by running \emph{only the local cuDNN kernel} for that configuration.
This gives an effective upper bound on the performance we can achieve using cuDNN in our configuration.
Finally, we report the achieved percent of this peak in the Rel column.

We observe that for CosmoFlow, we achieve \perf{performance-model-conv-cosmoflow-depth8-dominant-all-tflops-actual-relative-to-cudnn}\% and \perf{performance-model-conv-cosmoflow-depth32-dominant-all-tflops-actual-relative-to-cudnn}\% of this peak performance for 8- and 32-way partitioning, respectively.
This indicates that the overhead of our distributed convolution is relatively small.
We also observe another benefit of strong scaling: the potential peak performances exhibit super-linear scaling, albeit fairly slightly.
This is due to a larger, aggregated memory space available with a larger number of GPUs, allowing cuDNN to use more efficient convolution algorithms.

We also examined which layer dominates runtime, and for CosmoFlow we find that the conv1 layer accounts for almost half of the entire network runtime.
This is due to the layer processing the largest spatial dimensions.
For 8-way partitioning, we achieve excellent scaling efficiency; for 32-way partitioning, communication overheads limit gains, but still enable overall performance improvements.

While these results show good scaling efficiency with respect to the achievable peak with cuDNN, the TFlops/s achieved is relatively low compared to the theoretical peak of the hardware.
This indicates that there is significant potential for further optimizing 3D convolution kernels.

\subsection{Weak scaling} \label{subsection:weak_scale}

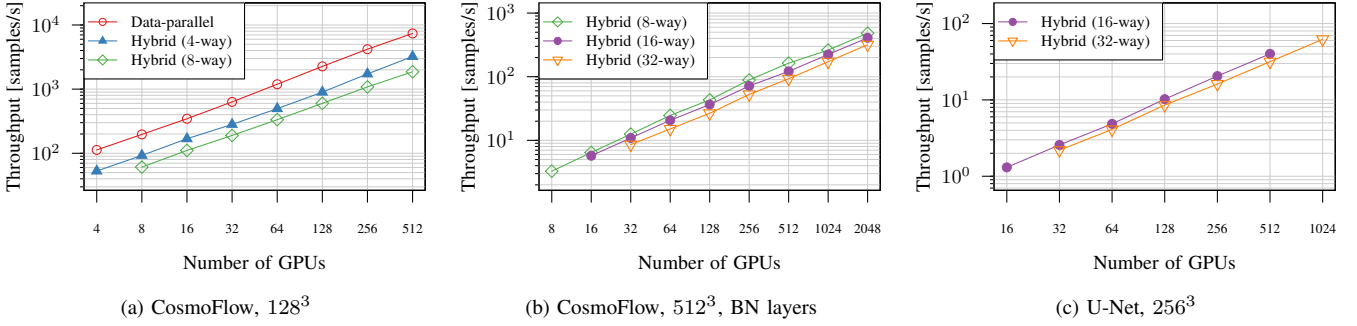
\begin{figure*}[t]
  \subfloat[\cosmoflow{}, \perf{dim128}]{\input{plot_figures/weak_scale_cosmoflow_128.tikz}}
  \subfloat[\cosmoflow{}, \perf{dim512}, BN layers]{\input{plot_figures/weak_scale_cosmoflow_512_bn.tikz}}
  \subfloat[\unet{}, \perf{dim256}]{\input{plot_figures/weak_scale_unet_256_bn.tikz}}
  \caption{
    Weak scaling of the two different 3D CNNs. We increase the global mini-batch size as we increase the number of GPUs.
    In the hybrid results, we partition a single sample by multiple GPUs in its spatial domain.
  }
  \label{figure:weak_scale}
\end{figure*}

The second component of our hybrid-parallelism approach is to exploit data
parallelism in conjunction with spatial parallelism.  This allows us to explore
the impact of weak scaling by increasing the global mini-batch size while tuning
the learning rate.
\figref{figure:weak_scale} shows the weak scaling performance of the two 3D CNNs with different input sizes.
For the \cosmoflow{} network with \perf{dim128} cubes, the data size used in the previous work, we use per-GPU batch sizes of $8$ and increase the global mini-batch size as we increase the number of GPUs.
We evaluate the performance using 4-way and 8-way partitioning for reference.
In the \perf{dim512} case, we only evaluate hybrid parallelization where each data sample is partitioned among 8, 16 or 32 GPUs
as nearly 53 GB of memory is required per sample as shown in \tabref{table:network_architecture}.
While it is smaller than the aggregate capacity of 4 GPUs, we found that it results in an out-of-memory error as additional auxiliary data need to be allocated.
We measure performance as in \secref{subsection:strong_scale}.

In the case of \cosmoflow{} with \perf{dim128} cubes, our implementation achieves nearly linear speedup up to 512 GPUs (128 compute nodes),
in part because of the asynchronous overlapped communication engine of Aluminum and also because of the relatively high compute-to-communication ratio in 3D CNNs.
We achieve a \perf{weak-scale-speedup-dim128-n1-to-n128} speedup on 512 GPUs compared to 4 GPUs with the \perf{dim128} cubes.
In this case, the highest efficiency is achieved with the data-parallel scheme since the hybrid parallelization involves additional communications due to halo exchanges.

With \perf{dim512}, however, hybrid parallelization is required as the model is too large to fit into the device memory of a single GPU.
Thus, we evaluate three configurations, 8-way, 16-way and 32-way, and the global mini-batch size is linearly increased as the number of GPUs is increased,
resulting in \perf{weak-scale-speedup-dim512bn-nx8x1x1-n2-to-n512}, \perf{weak-scale-speedup-dim512bn-nx16x1x1-n4-to-n512}, and \perf{weak-scale-speedup-dim512bn-nx32x1x1-n8-to-n512} of speedup
on 2048 GPUs over 8, 16, 32 GPUs (where the mini-batch size is 1) respectively.
With the 3D \unet{}, we achieve good weak scalability (\perf{weak-scale-speedup-unet-dim256-bn-n8-to-n256} on 1024 GPUs over 32 GPUs with 32-way partitioning) as well.

In all cases, increasing the spatial parallelism results in lower throughput due to the additional communication overhead as well as the decreased compute efficiency of the cuDNN kernel library. However, we note that the hybrid parallelization enables further speedups for a given fixed mini-batch size as it is also shown in \secref{subsection:strong_scale}.

\subsection{\cosmoflow{} model accuracy improvement with $512^3$ universe cubes}
\label{subsection:training_dim512}

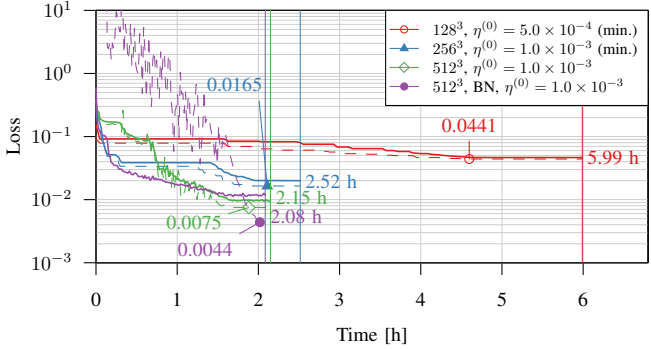
\begin{figure}[t]
  \input{plot_figures/201905_4parE_time.tikz}
  \caption{
    Training/validation losses (solid/dashed lines respectively) and the smallest validation losses (points) of the \cosmoflow{} network with four different configurations.
    For \perf{dim128} and \perf{dim256}, we show the minimum loss values at each point in time for visibility.
  }
  \label{figure:201905_4parE}
\end{figure}

This experiment set out to test if training on entire data samples would improve
the quality of the model learned by the network.
\figref{figure:201905_4parE} shows training results of the \cosmoflow{} network with the full-resolution dataset (\perf{dim512}) and
split versions (\perf{dim128} and \perf{dim256}).
We swept the initial learning rate from $10^{-4}$ to $10^{-2}$ logarithmically and show the results with the best.
We train for 130 epochs with a mini-batch size of 64 in every configuration,
and use the 4-way partitioning (256 GPUs in total) for the networks without batch normalization layers, or 8-way (512 GPUs in total) for networks with batch normalization,
due to the increased memory requirements.
To account for training variance, we show the median result of five trials with
different initial random seeds.

We observe that the test loss decreases significantly as we increase the
dataset size to \perf{dim256-training-test-mse} MSE with $256^3$
and \perf{dim512-training-test-mse} MSE with $512^3$ data.
Adding batch normalization improves this result further,
to \perf{dim512bn-training-test-mse} MSE,
achieving an order-of-magnitude improvement compared to the baseline $128^3$ data.
At the same time, we get \perf{training-speedup-dim128-dim512} of speedup from $128^3$ to $512^3$ with the same number of GPUs and the same mini-batch size.
This result implies that the CNN can be trained with the same computing resources
and dataset size, but with a smaller mini-batch and small overheads (see \secref{subsection:weak_scale}).
This brings an opportunity to keep mini-batch sizes fixed and strong-scale onto
more GPUs for speedup.

\begin{figure}[t]
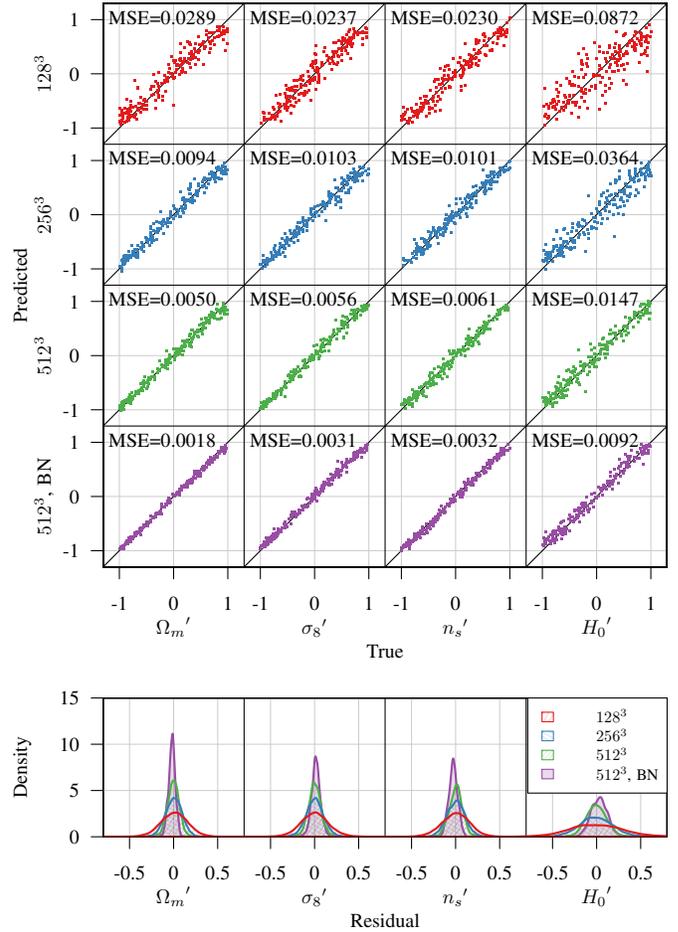

  \subfloat{\input{plot_figures/201905_4parE_residuals.tikz}}
  \\
  \subfloat{\input{plot_figures/201905_4parE_residuals_density.tikz}}
  \caption{True/predicted cosmological parameters (normalized to $[-1,1]$) from four different configurations (top) and the distribution of the residuals (bottom).
    In the top figure, we show 200 randomly chosen data points for visibility.
  }
  \label{figure:201905_4parE_residuals}
\end{figure}

\figref{figure:201905_4parE_residuals} shows the correlation between the predicted
and actual cosmological parameters and the associated residuals for our networks
on each dataset. We clearly demonstrate improvements in the quality of predictions
with increasing data volume; and the benefit of batch normalization. In particular,
we observe that prediction of $H_0$ (the Hubble constant) shows the most improvement
in accuracy with increasing data volume. This makes intuitive sense, as it is
related to the large-scale expansion of the universe.
As cosmological simulations move to sub-percent measurements, being able to test the
quality of the surrogates via a greatly improved Cosmoflow network, with an order of
magnitude improvement in the measurement of the cosmological parameters, is the
only way to quickly validate the quality and precision of the models.

%% file: tables/cudnn_performance.tex
\begin{tabular}{c c |c r r r r}
  \toprule
  \multirow{2}{*}{Depth}
  & \multirow{2}{*}{$N$}
  & \multirow{2}{*}{Layer}
  & Time
  & \multicolumn{1}{c}{Perf.}
  & \multicolumn{1}{c}{Peak}
  & \multicolumn{1}{c}{Rel.}
  \\
  & &
  & [ms]
  & \multicolumn{2}{c}{[TFlop/s]}
  & \multicolumn{1}{c}{[\%]} \\
  \midrule

                                8-way & 64 & All
  & \perf{performance-model-conv-cosmoflow-depth8-dominant-all-time-actual}
  & \perf{performance-model-conv-cosmoflow-depth8-dominant-all-tflops-actual}
  & \perf{performance-model-conv-cosmoflow-depth8-dominant-all-tflops}
  & \perf{performance-model-conv-cosmoflow-depth8-dominant-all-tflops-actual-relative-to-cudnn}
  \\
                                32-way & 64& All
  & \perf{performance-model-conv-cosmoflow-depth32-dominant-all-time-actual}
  & \perf{performance-model-conv-cosmoflow-depth32-dominant-all-tflops-actual}
  & \perf{performance-model-conv-cosmoflow-depth32-dominant-all-tflops}
  & \perf{performance-model-conv-cosmoflow-depth32-dominant-all-tflops-actual-relative-to-cudnn}
  \\

  \midrule

                                8-way & 64
  & \perf{performance-model-conv-cosmoflow-depth8-dominant-layer}
  & \perf{performance-model-conv-cosmoflow-depth8-dominant-time-actual}
  & \perf{performance-model-conv-cosmoflow-depth8-dominant-tflops-actual}
  & \perf{performance-model-conv-cosmoflow-depth8-dominant-tflops}
  & \perf{performance-model-conv-cosmoflow-depth8-dominant-tflops-actual-relative-to-cudnn}
  \\
                                32-way & 64
  & \perf{performance-model-conv-cosmoflow-depth32-dominant-layer}
  & \perf{performance-model-conv-cosmoflow-depth32-dominant-time-actual}
  & \perf{performance-model-conv-cosmoflow-depth32-dominant-tflops-actual}
  & \perf{performance-model-conv-cosmoflow-depth32-dominant-tflops}
  & \perf{performance-model-conv-cosmoflow-depth32-dominant-tflops-actual-relative-to-cudnn}
  \\

  \bottomrule
\end{tabular}

%% file: plot_figures/weak_scale_cosmoflow_128.tikz
\begin{tikzpicture}[x=1pt,y=1pt]
\definecolor{fillColor}{RGB}{255,255,255}
\path[use as bounding box,fill=fillColor,fill opacity=0.00] (0,0) rectangle (168.63,110.57);
\begin{scope}
\path[clip] (  0.00,  0.00) rectangle (168.63,110.57);
\definecolor{drawColor}{RGB}{0,0,0}

\path[draw=drawColor,line width= 0.4pt,line join=round,line cap=round] ( 36.00, 36.00) --
	(165.03, 36.00) --
	(165.03,106.97) --
	( 36.00,106.97) --
	( 36.00, 36.00);
\end{scope}
\begin{scope}
\path[clip] (  0.00,  0.00) rectangle (168.63,110.57);
\definecolor{drawColor}{RGB}{0,0,0}

\node[text=drawColor,anchor=base,inner sep=0pt, outer sep=0pt, scale=  0.75] at (100.51,  6.30) {Number of GPUs};
\end{scope}
\begin{scope}
\path[clip] (  0.00,  0.00) rectangle (168.63,110.57);
\definecolor{drawColor}{RGB}{0,0,0}

\path[draw=drawColor,line width= 0.4pt,line join=round,line cap=round] ( 40.78, 36.00) -- ( 40.78, 36.00);

\path[draw=drawColor,line width= 0.4pt,line join=round,line cap=round] ( 40.78, 36.00) -- ( 40.78, 31.50);

\node[text=drawColor,anchor=base,inner sep=0pt, outer sep=0pt, scale=  0.51] at ( 40.78, 19.80) {4};

\path[draw=drawColor,line width= 0.4pt,line join=round,line cap=round] ( 57.85, 36.00) -- ( 57.85, 36.00);

\path[draw=drawColor,line width= 0.4pt,line join=round,line cap=round] ( 57.85, 36.00) -- ( 57.85, 31.50);

\node[text=drawColor,anchor=base,inner sep=0pt, outer sep=0pt, scale=  0.51] at ( 57.85, 19.80) {8};

\path[draw=drawColor,line width= 0.4pt,line join=round,line cap=round] ( 74.91, 36.00) -- ( 74.91, 36.00);

\path[draw=drawColor,line width= 0.4pt,line join=round,line cap=round] ( 74.91, 36.00) -- ( 74.91, 31.50);

\node[text=drawColor,anchor=base,inner sep=0pt, outer sep=0pt, scale=  0.51] at ( 74.91, 19.80) {16};

\path[draw=drawColor,line width= 0.4pt,line join=round,line cap=round] ( 91.98, 36.00) -- ( 91.98, 36.00);

\path[draw=drawColor,line width= 0.4pt,line join=round,line cap=round] ( 91.98, 36.00) -- ( 91.98, 31.50);

\node[text=drawColor,anchor=base,inner sep=0pt, outer sep=0pt, scale=  0.51] at ( 91.98, 19.80) {32};

\path[draw=drawColor,line width= 0.4pt,line join=round,line cap=round] (109.05, 36.00) -- (109.05, 36.00);

\path[draw=drawColor,line width= 0.4pt,line join=round,line cap=round] (109.05, 36.00) -- (109.05, 31.50);

\node[text=drawColor,anchor=base,inner sep=0pt, outer sep=0pt, scale=  0.51] at (109.05, 19.80) {64};

\path[draw=drawColor,line width= 0.4pt,line join=round,line cap=round] (126.12, 36.00) -- (126.12, 36.00);

\path[draw=drawColor,line width= 0.4pt,line join=round,line cap=round] (126.12, 36.00) -- (126.12, 31.50);

\node[text=drawColor,anchor=base,inner sep=0pt, outer sep=0pt, scale=  0.51] at (126.12, 19.80) {128};

\path[draw=drawColor,line width= 0.4pt,line join=round,line cap=round] (143.18, 36.00) -- (143.18, 36.00);

\path[draw=drawColor,line width= 0.4pt,line join=round,line cap=round] (143.18, 36.00) -- (143.18, 31.50);

\node[text=drawColor,anchor=base,inner sep=0pt, outer sep=0pt, scale=  0.51] at (143.18, 19.80) {256};

\path[draw=drawColor,line width= 0.4pt,line join=round,line cap=round] (160.25, 36.00) -- (160.25, 36.00);

\path[draw=drawColor,line width= 0.4pt,line join=round,line cap=round] (160.25, 36.00) -- (160.25, 31.50);

\node[text=drawColor,anchor=base,inner sep=0pt, outer sep=0pt, scale=  0.51] at (160.25, 19.80) {512};

\path[draw=drawColor,line width= 0.4pt,line join=round,line cap=round] ( 36.00, 36.00) -- ( 36.00,106.97);

\path[draw=drawColor,line width= 0.4pt,line join=round,line cap=round] ( 36.00, 37.34) -- ( 34.58, 37.34);

\path[draw=drawColor,line width= 0.4pt,line join=round,line cap=round] ( 36.00, 40.37) -- ( 34.58, 40.37);

\path[draw=drawColor,line width= 0.4pt,line join=round,line cap=round] ( 36.00, 42.72) -- ( 34.58, 42.72);

\path[draw=drawColor,line width= 0.4pt,line join=round,line cap=round] ( 36.00, 44.65) -- ( 34.58, 44.65);

\path[draw=drawColor,line width= 0.4pt,line join=round,line cap=round] ( 36.00, 46.27) -- ( 34.58, 46.27);

\path[draw=drawColor,line width= 0.4pt,line join=round,line cap=round] ( 36.00, 47.68) -- ( 34.58, 47.68);

\path[draw=drawColor,line width= 0.4pt,line join=round,line cap=round] ( 36.00, 48.92) -- ( 34.58, 48.92);

\path[draw=drawColor,line width= 0.4pt,line join=round,line cap=round] ( 36.00, 50.03) -- ( 34.58, 50.03);

\path[draw=drawColor,line width= 0.4pt,line join=round,line cap=round] ( 36.00, 57.34) -- ( 34.58, 57.34);

\path[draw=drawColor,line width= 0.4pt,line join=round,line cap=round] ( 36.00, 61.62) -- ( 34.58, 61.62);

\path[draw=drawColor,line width= 0.4pt,line join=round,line cap=round] ( 36.00, 64.65) -- ( 34.58, 64.65);

\path[draw=drawColor,line width= 0.4pt,line join=round,line cap=round] ( 36.00, 67.01) -- ( 34.58, 67.01);

\path[draw=drawColor,line width= 0.4pt,line join=round,line cap=round] ( 36.00, 68.93) -- ( 34.58, 68.93);

\path[draw=drawColor,line width= 0.4pt,line join=round,line cap=round] ( 36.00, 70.56) -- ( 34.58, 70.56);

\path[draw=drawColor,line width= 0.4pt,line join=round,line cap=round] ( 36.00, 71.96) -- ( 34.58, 71.96);

\path[draw=drawColor,line width= 0.4pt,line join=round,line cap=round] ( 36.00, 73.21) -- ( 34.58, 73.21);

\path[draw=drawColor,line width= 0.4pt,line join=round,line cap=round] ( 36.00, 74.32) -- ( 34.58, 74.32);

\path[draw=drawColor,line width= 0.4pt,line join=round,line cap=round] ( 36.00, 81.63) -- ( 34.58, 81.63);

\path[draw=drawColor,line width= 0.4pt,line join=round,line cap=round] ( 36.00, 85.90) -- ( 34.58, 85.90);

\path[draw=drawColor,line width= 0.4pt,line join=round,line cap=round] ( 36.00, 88.94) -- ( 34.58, 88.94);

\path[draw=drawColor,line width= 0.4pt,line join=round,line cap=round] ( 36.00, 91.29) -- ( 34.58, 91.29);

\path[draw=drawColor,line width= 0.4pt,line join=round,line cap=round] ( 36.00, 93.21) -- ( 34.58, 93.21);

\path[draw=drawColor,line width= 0.4pt,line join=round,line cap=round] ( 36.00, 94.84) -- ( 34.58, 94.84);

\path[draw=drawColor,line width= 0.4pt,line join=round,line cap=round] ( 36.00, 96.25) -- ( 34.58, 96.25);

\path[draw=drawColor,line width= 0.4pt,line join=round,line cap=round] ( 36.00, 97.49) -- ( 34.58, 97.49);

\path[draw=drawColor,line width= 0.4pt,line join=round,line cap=round] ( 36.00, 98.60) -- ( 34.58, 98.60);

\path[draw=drawColor,line width= 0.4pt,line join=round,line cap=round] ( 36.00,105.91) -- ( 34.58,105.91);

\path[draw=drawColor,line width= 0.4pt,line join=round,line cap=round] ( 36.00, 36.00) -- ( 36.00,106.97);

\path[draw=drawColor,line width= 0.4pt,line join=round,line cap=round] ( 36.00, 50.03) -- ( 31.50, 50.03);

\path[draw=drawColor,line width= 0.4pt,line join=round,line cap=round] ( 36.00, 74.32) -- ( 31.50, 74.32);

\path[draw=drawColor,line width= 0.4pt,line join=round,line cap=round] ( 36.00, 98.60) -- ( 31.50, 98.60);

\node[text=drawColor,anchor=base east,inner sep=0pt, outer sep=0pt, scale=  0.75] at ( 27.00, 47.45) {$10^2$};

\node[text=drawColor,anchor=base east,inner sep=0pt, outer sep=0pt, scale=  0.75] at ( 27.00, 71.74) {$10^3$};

\node[text=drawColor,anchor=base east,inner sep=0pt, outer sep=0pt, scale=  0.75] at ( 27.00, 96.02) {$10^4$};
\end{scope}
\begin{scope}
\path[clip] (  0.00,  0.00) rectangle (168.63,110.57);
\definecolor{drawColor}{RGB}{0,0,0}

\node[text=drawColor,rotate= 90.00,anchor=base,inner sep=0pt, outer sep=0pt, scale=  0.75] at ( 11.70, 71.49) {Throughput [samples/s]};
\end{scope}
\begin{scope}
\path[clip] (  0.00,  0.00) rectangle (168.63,110.57);
\definecolor{drawColor}{gray}{0.80}

\path[draw=drawColor,line width= 0.1pt,line join=round,line cap=round] ( 40.78, 36.00) -- (160.25, 36.00);

\path[draw=drawColor,line width= 0.1pt,line join=round,line cap=round] ( 40.78, 36.00) -- ( 40.78,106.97);

\path[draw=drawColor,line width= 0.1pt,line join=round,line cap=round] ( 57.85, 36.00) -- ( 57.85,106.97);

\path[draw=drawColor,line width= 0.1pt,line join=round,line cap=round] ( 74.91, 36.00) -- ( 74.91,106.97);

\path[draw=drawColor,line width= 0.1pt,line join=round,line cap=round] ( 91.98, 36.00) -- ( 91.98,106.97);

\path[draw=drawColor,line width= 0.1pt,line join=round,line cap=round] (109.05, 36.00) -- (109.05,106.97);

\path[draw=drawColor,line width= 0.1pt,line join=round,line cap=round] (126.12, 36.00) -- (126.12,106.97);

\path[draw=drawColor,line width= 0.1pt,line join=round,line cap=round] (143.18, 36.00) -- (143.18,106.97);

\path[draw=drawColor,line width= 0.1pt,line join=round,line cap=round] (160.25, 36.00) -- (160.25,106.97);
\definecolor{drawColor}{RGB}{0,0,0}

\path[draw=drawColor,line width= 0.4pt,line join=round,line cap=round] ( 36.00, 36.00) --
	(165.03, 36.00) --
	(165.03,106.97) --
	( 36.00,106.97) --
	( 36.00, 36.00);
\definecolor{drawColor}{gray}{0.80}

\path[draw=drawColor,line width= 0.1pt,line join=round,line cap=round] ( 36.00, 36.00) -- ( 36.00,106.97);

\path[draw=drawColor,line width= 0.1pt,line join=round,line cap=round] ( 36.00, 37.34) -- (165.03, 37.34);

\path[draw=drawColor,line width= 0.1pt,line join=round,line cap=round] ( 36.00, 40.37) -- (165.03, 40.37);

\path[draw=drawColor,line width= 0.1pt,line join=round,line cap=round] ( 36.00, 42.72) -- (165.03, 42.72);

\path[draw=drawColor,line width= 0.1pt,line join=round,line cap=round] ( 36.00, 44.65) -- (165.03, 44.65);

\path[draw=drawColor,line width= 0.1pt,line join=round,line cap=round] ( 36.00, 46.27) -- (165.03, 46.27);

\path[draw=drawColor,line width= 0.1pt,line join=round,line cap=round] ( 36.00, 47.68) -- (165.03, 47.68);

\path[draw=drawColor,line width= 0.1pt,line join=round,line cap=round] ( 36.00, 48.92) -- (165.03, 48.92);

\path[draw=drawColor,line width= 0.1pt,line join=round,line cap=round] ( 36.00, 50.03) -- (165.03, 50.03);

\path[draw=drawColor,line width= 0.1pt,line join=round,line cap=round] ( 36.00, 57.34) -- (165.03, 57.34);

\path[draw=drawColor,line width= 0.1pt,line join=round,line cap=round] ( 36.00, 61.62) -- (165.03, 61.62);

\path[draw=drawColor,line width= 0.1pt,line join=round,line cap=round] ( 36.00, 64.65) -- (165.03, 64.65);

\path[draw=drawColor,line width= 0.1pt,line join=round,line cap=round] ( 36.00, 67.01) -- (165.03, 67.01);

\path[draw=drawColor,line width= 0.1pt,line join=round,line cap=round] ( 36.00, 68.93) -- (165.03, 68.93);

\path[draw=drawColor,line width= 0.1pt,line join=round,line cap=round] ( 36.00, 70.56) -- (165.03, 70.56);

\path[draw=drawColor,line width= 0.1pt,line join=round,line cap=round] ( 36.00, 71.96) -- (165.03, 71.96);

\path[draw=drawColor,line width= 0.1pt,line join=round,line cap=round] ( 36.00, 73.21) -- (165.03, 73.21);

\path[draw=drawColor,line width= 0.1pt,line join=round,line cap=round] ( 36.00, 74.32) -- (165.03, 74.32);

\path[draw=drawColor,line width= 0.1pt,line join=round,line cap=round] ( 36.00, 81.63) -- (165.03, 81.63);

\path[draw=drawColor,line width= 0.1pt,line join=round,line cap=round] ( 36.00, 85.90) -- (165.03, 85.90);

\path[draw=drawColor,line width= 0.1pt,line join=round,line cap=round] ( 36.00, 88.94) -- (165.03, 88.94);

\path[draw=drawColor,line width= 0.1pt,line join=round,line cap=round] ( 36.00, 91.29) -- (165.03, 91.29);

\path[draw=drawColor,line width= 0.1pt,line join=round,line cap=round] ( 36.00, 93.21) -- (165.03, 93.21);

\path[draw=drawColor,line width= 0.1pt,line join=round,line cap=round] ( 36.00, 94.84) -- (165.03, 94.84);

\path[draw=drawColor,line width= 0.1pt,line join=round,line cap=round] ( 36.00, 96.25) -- (165.03, 96.25);

\path[draw=drawColor,line width= 0.1pt,line join=round,line cap=round] ( 36.00, 97.49) -- (165.03, 97.49);

\path[draw=drawColor,line width= 0.1pt,line join=round,line cap=round] ( 36.00, 98.60) -- (165.03, 98.60);

\path[draw=drawColor,line width= 0.1pt,line join=round,line cap=round] ( 36.00,105.91) -- (165.03,105.91);
\definecolor{drawColor}{RGB}{0,0,0}

\path[draw=drawColor,line width= 0.4pt,line join=round,line cap=round] ( 36.00, 36.00) --
	(165.03, 36.00) --
	(165.03,106.97) --
	( 36.00,106.97) --
	( 36.00, 36.00);
\end{scope}
\begin{scope}
\path[clip] ( 36.00, 36.00) rectangle (165.03,106.97);
\definecolor{drawColor}{RGB}{228,26,28}

\path[draw=drawColor,line width= 0.4pt,line join=round,line cap=round] ( 40.78, 51.30) --
	( 57.85, 57.18) --
	( 74.91, 63.10) --
	( 91.98, 69.48) --
	(109.05, 76.17) --
	(126.12, 82.93) --
	(143.18, 89.44) --
	(160.25, 95.39);

\path[draw=drawColor,line width= 0.4pt,line join=round,line cap=round] ( 40.78, 51.30) circle (  1.69);

\path[draw=drawColor,line width= 0.4pt,line join=round,line cap=round] ( 57.85, 57.18) circle (  1.69);

\path[draw=drawColor,line width= 0.4pt,line join=round,line cap=round] ( 74.91, 63.10) circle (  1.69);

\path[draw=drawColor,line width= 0.4pt,line join=round,line cap=round] ( 91.98, 69.48) circle (  1.69);

\path[draw=drawColor,line width= 0.4pt,line join=round,line cap=round] (109.05, 76.17) circle (  1.69);

\path[draw=drawColor,line width= 0.4pt,line join=round,line cap=round] (126.12, 82.93) circle (  1.69);

\path[draw=drawColor,line width= 0.4pt,line join=round,line cap=round] (143.18, 89.44) circle (  1.69);

\path[draw=drawColor,line width= 0.4pt,line join=round,line cap=round] (160.25, 95.39) circle (  1.69);
\end{scope}
\begin{scope}
\path[clip] ( 36.00, 36.00) rectangle (165.03,106.97);
\definecolor{drawColor}{RGB}{55,126,184}

\path[draw=drawColor,line width= 0.4pt,line join=round,line cap=round] ( 40.78, 43.31) --
	( 57.85, 49.22) --
	( 74.91, 55.54) --
	( 91.98, 60.96) --
	(109.05, 66.93) --
	(126.12, 73.17) --
	(143.18, 80.08) --
	(160.25, 86.67);
\definecolor{fillColor}{RGB}{55,126,184}

\path[fill=fillColor] ( 40.78, 45.93) --
	( 43.05, 42.00) --
	( 38.51, 42.00) --
	cycle;

\path[fill=fillColor] ( 57.85, 51.85) --
	( 60.12, 47.91) --
	( 55.57, 47.91) --
	cycle;

\path[fill=fillColor] ( 74.91, 58.16) --
	( 77.19, 54.22) --
	( 72.64, 54.22) --
	cycle;

\path[fill=fillColor] ( 91.98, 63.58) --
	( 94.25, 59.64) --
	( 89.71, 59.64) --
	cycle;

\path[fill=fillColor] (109.05, 69.55) --
	(111.32, 65.61) --
	(106.78, 65.61) --
	cycle;

\path[fill=fillColor] (126.12, 75.80) --
	(128.39, 71.86) --
	(123.84, 71.86) --
	cycle;

\path[fill=fillColor] (143.18, 82.70) --
	(145.46, 78.76) --
	(140.91, 78.76) --
	cycle;

\path[fill=fillColor] (160.25, 89.30) --
	(162.52, 85.36) --
	(157.98, 85.36) --
	cycle;
\end{scope}
\begin{scope}
\path[clip] ( 36.00, 36.00) rectangle (165.03,106.97);
\definecolor{drawColor}{RGB}{77,175,74}

\path[draw=drawColor,line width= 0.4pt,line join=round,line cap=round] ( 57.85, 44.80) --
	( 74.91, 51.11) --
	( 91.98, 56.78) --
	(109.05, 62.75) --
	(126.12, 68.93) --
	(143.18, 75.16) --
	(160.25, 80.89);

\path[draw=drawColor,line width= 0.4pt,line join=round,line cap=round] ( 55.46, 44.80) --
	( 57.85, 47.19) --
	( 60.23, 44.80) --
	( 57.85, 42.41) --
	( 55.46, 44.80);

\path[draw=drawColor,line width= 0.4pt,line join=round,line cap=round] ( 72.53, 51.11) --
	( 74.91, 53.50) --
	( 77.30, 51.11) --
	( 74.91, 48.73) --
	( 72.53, 51.11);

\path[draw=drawColor,line width= 0.4pt,line join=round,line cap=round] ( 89.59, 56.78) --
	( 91.98, 59.17) --
	( 94.37, 56.78) --
	( 91.98, 54.39) --
	( 89.59, 56.78);

\path[draw=drawColor,line width= 0.4pt,line join=round,line cap=round] (106.66, 62.75) --
	(109.05, 65.13) --
	(111.44, 62.75) --
	(109.05, 60.36) --
	(106.66, 62.75);

\path[draw=drawColor,line width= 0.4pt,line join=round,line cap=round] (123.73, 68.93) --
	(126.12, 71.32) --
	(128.50, 68.93) --
	(126.12, 66.55) --
	(123.73, 68.93);

\path[draw=drawColor,line width= 0.4pt,line join=round,line cap=round] (140.80, 75.16) --
	(143.18, 77.54) --
	(145.57, 75.16) --
	(143.18, 72.77) --
	(140.80, 75.16);

\path[draw=drawColor,line width= 0.4pt,line join=round,line cap=round] (157.86, 80.89) --
	(160.25, 83.28) --
	(162.64, 80.89) --
	(160.25, 78.51) --
	(157.86, 80.89);
\end{scope}
\begin{scope}
\path[clip] ( 36.00, 36.00) rectangle (165.03,106.97);
\definecolor{drawColor}{RGB}{0,0,0}
\definecolor{fillColor}{RGB}{255,255,255}

\path[draw=drawColor,line width= 0.4pt,line join=round,line cap=round,fill=fillColor] ( 36.00,106.97) rectangle ( 96.69, 78.17);
\definecolor{drawColor}{RGB}{228,26,28}

\path[draw=drawColor,line width= 0.4pt,line join=round,line cap=round] ( 37.62, 99.77) -- ( 48.42, 99.77);
\definecolor{drawColor}{RGB}{55,126,184}

\path[draw=drawColor,line width= 0.4pt,line join=round,line cap=round] ( 37.62, 92.57) -- ( 48.42, 92.57);
\definecolor{drawColor}{RGB}{77,175,74}

\path[draw=drawColor,line width= 0.4pt,line join=round,line cap=round] ( 37.62, 85.37) -- ( 48.42, 85.37);
\definecolor{drawColor}{RGB}{228,26,28}

\path[draw=drawColor,line width= 0.4pt,line join=round,line cap=round] ( 43.02, 99.77) circle (  1.35);
\definecolor{fillColor}{RGB}{55,126,184}

\path[fill=fillColor] ( 43.02, 94.67) --
	( 44.84, 91.52) --
	( 41.20, 91.52) --
	cycle;
\definecolor{drawColor}{RGB}{77,175,74}

\path[draw=drawColor,line width= 0.4pt,line join=round,line cap=round] ( 41.11, 85.37) --
	( 43.02, 87.28) --
	( 44.93, 85.37) --
	( 43.02, 83.46) --
	( 41.11, 85.37);
\definecolor{drawColor}{RGB}{0,0,0}

\node[text=drawColor,anchor=base west,inner sep=0pt, outer sep=0pt, scale=  0.60] at ( 53.82, 97.71) {Data-parallel};

\node[text=drawColor,anchor=base west,inner sep=0pt, outer sep=0pt, scale=  0.60] at ( 53.82, 90.51) {Hybrid ($4$-way)};

\node[text=drawColor,anchor=base west,inner sep=0pt, outer sep=0pt, scale=  0.60] at ( 53.82, 83.31) {Hybrid ($8$-way)};
\end{scope}
\end{tikzpicture}

%% file: plot_figures/weak_scale_cosmoflow_512_bn.tikz
\begin{tikzpicture}[x=1pt,y=1pt]
\definecolor{fillColor}{RGB}{255,255,255}
\path[use as bounding box,fill=fillColor,fill opacity=0.00] (0,0) rectangle (168.63,110.57);
\begin{scope}
\path[clip] (  0.00,  0.00) rectangle (168.63,110.57);
\definecolor{drawColor}{RGB}{0,0,0}

\path[draw=drawColor,line width= 0.4pt,line join=round,line cap=round] ( 36.00, 36.00) --
	(165.03, 36.00) --
	(165.03,106.97) --
	( 36.00,106.97) --
	( 36.00, 36.00);
\end{scope}
\begin{scope}
\path[clip] (  0.00,  0.00) rectangle (168.63,110.57);
\definecolor{drawColor}{RGB}{0,0,0}

\node[text=drawColor,anchor=base,inner sep=0pt, outer sep=0pt, scale=  0.75] at (100.51,  6.30) {Number of GPUs};
\end{scope}
\begin{scope}
\path[clip] (  0.00,  0.00) rectangle (168.63,110.57);
\definecolor{drawColor}{RGB}{0,0,0}

\path[draw=drawColor,line width= 0.4pt,line join=round,line cap=round] ( 40.78, 36.00) -- ( 40.78, 36.00);

\path[draw=drawColor,line width= 0.4pt,line join=round,line cap=round] ( 40.78, 36.00) -- ( 40.78, 31.50);

\node[text=drawColor,anchor=base,inner sep=0pt, outer sep=0pt, scale=  0.51] at ( 40.78, 19.80) {8};

\path[draw=drawColor,line width= 0.4pt,line join=round,line cap=round] ( 55.71, 36.00) -- ( 55.71, 36.00);

\path[draw=drawColor,line width= 0.4pt,line join=round,line cap=round] ( 55.71, 36.00) -- ( 55.71, 31.50);

\node[text=drawColor,anchor=base,inner sep=0pt, outer sep=0pt, scale=  0.51] at ( 55.71, 19.80) {16};

\path[draw=drawColor,line width= 0.4pt,line join=round,line cap=round] ( 70.65, 36.00) -- ( 70.65, 36.00);

\path[draw=drawColor,line width= 0.4pt,line join=round,line cap=round] ( 70.65, 36.00) -- ( 70.65, 31.50);

\node[text=drawColor,anchor=base,inner sep=0pt, outer sep=0pt, scale=  0.51] at ( 70.65, 19.80) {32};

\path[draw=drawColor,line width= 0.4pt,line join=round,line cap=round] ( 85.58, 36.00) -- ( 85.58, 36.00);

\path[draw=drawColor,line width= 0.4pt,line join=round,line cap=round] ( 85.58, 36.00) -- ( 85.58, 31.50);

\node[text=drawColor,anchor=base,inner sep=0pt, outer sep=0pt, scale=  0.51] at ( 85.58, 19.80) {64};

\path[draw=drawColor,line width= 0.4pt,line join=round,line cap=round] (100.51, 36.00) -- (100.51, 36.00);

\path[draw=drawColor,line width= 0.4pt,line join=round,line cap=round] (100.51, 36.00) -- (100.51, 31.50);

\node[text=drawColor,anchor=base,inner sep=0pt, outer sep=0pt, scale=  0.51] at (100.51, 19.80) {128};

\path[draw=drawColor,line width= 0.4pt,line join=round,line cap=round] (115.45, 36.00) -- (115.45, 36.00);

\path[draw=drawColor,line width= 0.4pt,line join=round,line cap=round] (115.45, 36.00) -- (115.45, 31.50);

\node[text=drawColor,anchor=base,inner sep=0pt, outer sep=0pt, scale=  0.51] at (115.45, 19.80) {256};

\path[draw=drawColor,line width= 0.4pt,line join=round,line cap=round] (130.38, 36.00) -- (130.38, 36.00);

\path[draw=drawColor,line width= 0.4pt,line join=round,line cap=round] (130.38, 36.00) -- (130.38, 31.50);

\node[text=drawColor,anchor=base,inner sep=0pt, outer sep=0pt, scale=  0.51] at (130.38, 19.80) {512};

\path[draw=drawColor,line width= 0.4pt,line join=round,line cap=round] (145.32, 36.00) -- (145.32, 36.00);

\path[draw=drawColor,line width= 0.4pt,line join=round,line cap=round] (145.32, 36.00) -- (145.32, 31.50);

\node[text=drawColor,anchor=base,inner sep=0pt, outer sep=0pt, scale=  0.51] at (145.32, 19.80) {1024};

\path[draw=drawColor,line width= 0.4pt,line join=round,line cap=round] (160.25, 36.00) -- (160.25, 36.00);

\path[draw=drawColor,line width= 0.4pt,line join=round,line cap=round] (160.25, 36.00) -- (160.25, 31.50);

\node[text=drawColor,anchor=base,inner sep=0pt, outer sep=0pt, scale=  0.51] at (160.25, 19.80) {2048};

\path[draw=drawColor,line width= 0.4pt,line join=round,line cap=round] ( 36.00, 36.00) -- ( 36.00,106.97);

\path[draw=drawColor,line width= 0.4pt,line join=round,line cap=round] ( 36.00, 38.09) -- ( 34.58, 38.09);

\path[draw=drawColor,line width= 0.4pt,line join=round,line cap=round] ( 36.00, 42.34) -- ( 34.58, 42.34);

\path[draw=drawColor,line width= 0.4pt,line join=round,line cap=round] ( 36.00, 45.34) -- ( 34.58, 45.34);

\path[draw=drawColor,line width= 0.4pt,line join=round,line cap=round] ( 36.00, 47.68) -- ( 34.58, 47.68);

\path[draw=drawColor,line width= 0.4pt,line join=round,line cap=round] ( 36.00, 49.59) -- ( 34.58, 49.59);

\path[draw=drawColor,line width= 0.4pt,line join=round,line cap=round] ( 36.00, 51.20) -- ( 34.58, 51.20);

\path[draw=drawColor,line width= 0.4pt,line join=round,line cap=round] ( 36.00, 52.60) -- ( 34.58, 52.60);

\path[draw=drawColor,line width= 0.4pt,line join=round,line cap=round] ( 36.00, 53.83) -- ( 34.58, 53.83);

\path[draw=drawColor,line width= 0.4pt,line join=round,line cap=round] ( 36.00, 54.93) -- ( 34.58, 54.93);

\path[draw=drawColor,line width= 0.4pt,line join=round,line cap=round] ( 36.00, 62.18) -- ( 34.58, 62.18);

\path[draw=drawColor,line width= 0.4pt,line join=round,line cap=round] ( 36.00, 66.42) -- ( 34.58, 66.42);

\path[draw=drawColor,line width= 0.4pt,line join=round,line cap=round] ( 36.00, 69.43) -- ( 34.58, 69.43);

\path[draw=drawColor,line width= 0.4pt,line join=round,line cap=round] ( 36.00, 71.77) -- ( 34.58, 71.77);

\path[draw=drawColor,line width= 0.4pt,line join=round,line cap=round] ( 36.00, 73.67) -- ( 34.58, 73.67);

\path[draw=drawColor,line width= 0.4pt,line join=round,line cap=round] ( 36.00, 75.29) -- ( 34.58, 75.29);

\path[draw=drawColor,line width= 0.4pt,line join=round,line cap=round] ( 36.00, 76.68) -- ( 34.58, 76.68);

\path[draw=drawColor,line width= 0.4pt,line join=round,line cap=round] ( 36.00, 77.92) -- ( 34.58, 77.92);

\path[draw=drawColor,line width= 0.4pt,line join=round,line cap=round] ( 36.00, 79.02) -- ( 34.58, 79.02);

\path[draw=drawColor,line width= 0.4pt,line join=round,line cap=round] ( 36.00, 86.27) -- ( 34.58, 86.27);

\path[draw=drawColor,line width= 0.4pt,line join=round,line cap=round] ( 36.00, 90.51) -- ( 34.58, 90.51);

\path[draw=drawColor,line width= 0.4pt,line join=round,line cap=round] ( 36.00, 93.52) -- ( 34.58, 93.52);

\path[draw=drawColor,line width= 0.4pt,line join=round,line cap=round] ( 36.00, 95.86) -- ( 34.58, 95.86);

\path[draw=drawColor,line width= 0.4pt,line join=round,line cap=round] ( 36.00, 97.76) -- ( 34.58, 97.76);

\path[draw=drawColor,line width= 0.4pt,line join=round,line cap=round] ( 36.00, 99.38) -- ( 34.58, 99.38);

\path[draw=drawColor,line width= 0.4pt,line join=round,line cap=round] ( 36.00,100.77) -- ( 34.58,100.77);

\path[draw=drawColor,line width= 0.4pt,line join=round,line cap=round] ( 36.00,102.00) -- ( 34.58,102.00);

\path[draw=drawColor,line width= 0.4pt,line join=round,line cap=round] ( 36.00,103.11) -- ( 34.58,103.11);

\path[draw=drawColor,line width= 0.4pt,line join=round,line cap=round] ( 36.00, 36.00) -- ( 36.00,106.97);

\path[draw=drawColor,line width= 0.4pt,line join=round,line cap=round] ( 36.00, 54.93) -- ( 31.50, 54.93);

\path[draw=drawColor,line width= 0.4pt,line join=round,line cap=round] ( 36.00, 79.02) -- ( 31.50, 79.02);

\path[draw=drawColor,line width= 0.4pt,line join=round,line cap=round] ( 36.00,103.11) -- ( 31.50,103.11);

\node[text=drawColor,anchor=base east,inner sep=0pt, outer sep=0pt, scale=  0.75] at ( 27.00, 52.35) {$10^1$};

\node[text=drawColor,anchor=base east,inner sep=0pt, outer sep=0pt, scale=  0.75] at ( 27.00, 76.44) {$10^2$};

\node[text=drawColor,anchor=base east,inner sep=0pt, outer sep=0pt, scale=  0.75] at ( 27.00,100.52) {$10^3$};
\end{scope}
\begin{scope}
\path[clip] (  0.00,  0.00) rectangle (168.63,110.57);
\definecolor{drawColor}{RGB}{0,0,0}

\node[text=drawColor,rotate= 90.00,anchor=base,inner sep=0pt, outer sep=0pt, scale=  0.75] at ( 11.70, 71.49) {Throughput [samples/s]};
\end{scope}
\begin{scope}
\path[clip] (  0.00,  0.00) rectangle (168.63,110.57);
\definecolor{drawColor}{gray}{0.80}

\path[draw=drawColor,line width= 0.1pt,line join=round,line cap=round] ( 40.78, 36.00) -- (160.25, 36.00);

\path[draw=drawColor,line width= 0.1pt,line join=round,line cap=round] ( 40.78, 36.00) -- ( 40.78,106.97);

\path[draw=drawColor,line width= 0.1pt,line join=round,line cap=round] ( 55.71, 36.00) -- ( 55.71,106.97);

\path[draw=drawColor,line width= 0.1pt,line join=round,line cap=round] ( 70.65, 36.00) -- ( 70.65,106.97);

\path[draw=drawColor,line width= 0.1pt,line join=round,line cap=round] ( 85.58, 36.00) -- ( 85.58,106.97);

\path[draw=drawColor,line width= 0.1pt,line join=round,line cap=round] (100.51, 36.00) -- (100.51,106.97);

\path[draw=drawColor,line width= 0.1pt,line join=round,line cap=round] (115.45, 36.00) -- (115.45,106.97);

\path[draw=drawColor,line width= 0.1pt,line join=round,line cap=round] (130.38, 36.00) -- (130.38,106.97);

\path[draw=drawColor,line width= 0.1pt,line join=round,line cap=round] (145.32, 36.00) -- (145.32,106.97);

\path[draw=drawColor,line width= 0.1pt,line join=round,line cap=round] (160.25, 36.00) -- (160.25,106.97);
\definecolor{drawColor}{RGB}{0,0,0}

\path[draw=drawColor,line width= 0.4pt,line join=round,line cap=round] ( 36.00, 36.00) --
	(165.03, 36.00) --
	(165.03,106.97) --
	( 36.00,106.97) --
	( 36.00, 36.00);
\definecolor{drawColor}{gray}{0.80}

\path[draw=drawColor,line width= 0.1pt,line join=round,line cap=round] ( 36.00, 36.00) -- ( 36.00,106.97);

\path[draw=drawColor,line width= 0.1pt,line join=round,line cap=round] ( 36.00, 38.09) -- (165.03, 38.09);

\path[draw=drawColor,line width= 0.1pt,line join=round,line cap=round] ( 36.00, 42.34) -- (165.03, 42.34);

\path[draw=drawColor,line width= 0.1pt,line join=round,line cap=round] ( 36.00, 45.34) -- (165.03, 45.34);

\path[draw=drawColor,line width= 0.1pt,line join=round,line cap=round] ( 36.00, 47.68) -- (165.03, 47.68);

\path[draw=drawColor,line width= 0.1pt,line join=round,line cap=round] ( 36.00, 49.59) -- (165.03, 49.59);

\path[draw=drawColor,line width= 0.1pt,line join=round,line cap=round] ( 36.00, 51.20) -- (165.03, 51.20);

\path[draw=drawColor,line width= 0.1pt,line join=round,line cap=round] ( 36.00, 52.60) -- (165.03, 52.60);

\path[draw=drawColor,line width= 0.1pt,line join=round,line cap=round] ( 36.00, 53.83) -- (165.03, 53.83);

\path[draw=drawColor,line width= 0.1pt,line join=round,line cap=round] ( 36.00, 54.93) -- (165.03, 54.93);

\path[draw=drawColor,line width= 0.1pt,line join=round,line cap=round] ( 36.00, 62.18) -- (165.03, 62.18);

\path[draw=drawColor,line width= 0.1pt,line join=round,line cap=round] ( 36.00, 66.42) -- (165.03, 66.42);

\path[draw=drawColor,line width= 0.1pt,line join=round,line cap=round] ( 36.00, 69.43) -- (165.03, 69.43);

\path[draw=drawColor,line width= 0.1pt,line join=round,line cap=round] ( 36.00, 71.77) -- (165.03, 71.77);

\path[draw=drawColor,line width= 0.1pt,line join=round,line cap=round] ( 36.00, 73.67) -- (165.03, 73.67);

\path[draw=drawColor,line width= 0.1pt,line join=round,line cap=round] ( 36.00, 75.29) -- (165.03, 75.29);

\path[draw=drawColor,line width= 0.1pt,line join=round,line cap=round] ( 36.00, 76.68) -- (165.03, 76.68);

\path[draw=drawColor,line width= 0.1pt,line join=round,line cap=round] ( 36.00, 77.92) -- (165.03, 77.92);

\path[draw=drawColor,line width= 0.1pt,line join=round,line cap=round] ( 36.00, 79.02) -- (165.03, 79.02);

\path[draw=drawColor,line width= 0.1pt,line join=round,line cap=round] ( 36.00, 86.27) -- (165.03, 86.27);

\path[draw=drawColor,line width= 0.1pt,line join=round,line cap=round] ( 36.00, 90.51) -- (165.03, 90.51);

\path[draw=drawColor,line width= 0.1pt,line join=round,line cap=round] ( 36.00, 93.52) -- (165.03, 93.52);

\path[draw=drawColor,line width= 0.1pt,line join=round,line cap=round] ( 36.00, 95.86) -- (165.03, 95.86);

\path[draw=drawColor,line width= 0.1pt,line join=round,line cap=round] ( 36.00, 97.76) -- (165.03, 97.76);

\path[draw=drawColor,line width= 0.1pt,line join=round,line cap=round] ( 36.00, 99.38) -- (165.03, 99.38);

\path[draw=drawColor,line width= 0.1pt,line join=round,line cap=round] ( 36.00,100.77) -- (165.03,100.77);

\path[draw=drawColor,line width= 0.1pt,line join=round,line cap=round] ( 36.00,102.00) -- (165.03,102.00);

\path[draw=drawColor,line width= 0.1pt,line join=round,line cap=round] ( 36.00,103.11) -- (165.03,103.11);
\definecolor{drawColor}{RGB}{0,0,0}

\path[draw=drawColor,line width= 0.4pt,line join=round,line cap=round] ( 36.00, 36.00) --
	(165.03, 36.00) --
	(165.03,106.97) --
	( 36.00,106.97) --
	( 36.00, 36.00);
\end{scope}
\begin{scope}
\path[clip] ( 36.00, 36.00) rectangle (165.03,106.97);
\definecolor{drawColor}{RGB}{77,175,74}

\path[draw=drawColor,line width= 0.4pt,line join=round,line cap=round] ( 40.78, 43.25) --
	( 55.71, 50.33) --
	( 70.65, 57.28) --
	( 85.58, 64.36) --
	(100.51, 70.25) --
	(115.45, 77.78) --
	(130.38, 84.16) --
	(145.32, 89.10) --
	(160.25, 95.48);

\path[draw=drawColor,line width= 0.4pt,line join=round,line cap=round] ( 38.39, 43.25) --
	( 40.78, 45.64) --
	( 43.17, 43.25) --
	( 40.78, 40.86) --
	( 38.39, 43.25);

\path[draw=drawColor,line width= 0.4pt,line join=round,line cap=round] ( 53.33, 50.33) --
	( 55.71, 52.72) --
	( 58.10, 50.33) --
	( 55.71, 47.94) --
	( 53.33, 50.33);

\path[draw=drawColor,line width= 0.4pt,line join=round,line cap=round] ( 68.26, 57.28) --
	( 70.65, 59.67) --
	( 73.03, 57.28) --
	( 70.65, 54.90) --
	( 68.26, 57.28);

\path[draw=drawColor,line width= 0.4pt,line join=round,line cap=round] ( 83.19, 64.36) --
	( 85.58, 66.75) --
	( 87.97, 64.36) --
	( 85.58, 61.97) --
	( 83.19, 64.36);

\path[draw=drawColor,line width= 0.4pt,line join=round,line cap=round] ( 98.13, 70.25) --
	(100.51, 72.64) --
	(102.90, 70.25) --
	(100.51, 67.86) --
	( 98.13, 70.25);

\path[draw=drawColor,line width= 0.4pt,line join=round,line cap=round] (113.06, 77.78) --
	(115.45, 80.17) --
	(117.84, 77.78) --
	(115.45, 75.40) --
	(113.06, 77.78);

\path[draw=drawColor,line width= 0.4pt,line join=round,line cap=round] (128.00, 84.16) --
	(130.38, 86.55) --
	(132.77, 84.16) --
	(130.38, 81.78) --
	(128.00, 84.16);

\path[draw=drawColor,line width= 0.4pt,line join=round,line cap=round] (142.93, 89.10) --
	(145.32, 91.48) --
	(147.70, 89.10) --
	(145.32, 86.71) --
	(142.93, 89.10);

\path[draw=drawColor,line width= 0.4pt,line join=round,line cap=round] (157.86, 95.48) --
	(160.25, 97.87) --
	(162.64, 95.48) --
	(160.25, 93.09) --
	(157.86, 95.48);
\end{scope}
\begin{scope}
\path[clip] ( 36.00, 36.00) rectangle (165.03,106.97);
\definecolor{drawColor}{RGB}{152,78,163}

\path[draw=drawColor,line width= 0.4pt,line join=round,line cap=round] ( 55.71, 49.10) --
	( 70.65, 55.82) --
	( 85.58, 62.51) --
	(100.51, 68.48) --
	(115.45, 75.51) --
	(130.38, 81.04) --
	(145.32, 87.41) --
	(160.25, 93.74);
\definecolor{fillColor}{RGB}{152,78,163}

\path[draw=drawColor,line width= 0.4pt,line join=round,line cap=round,fill=fillColor] ( 55.71, 49.10) circle (  1.69);

\path[draw=drawColor,line width= 0.4pt,line join=round,line cap=round,fill=fillColor] ( 70.65, 55.82) circle (  1.69);

\path[draw=drawColor,line width= 0.4pt,line join=round,line cap=round,fill=fillColor] ( 85.58, 62.51) circle (  1.69);

\path[draw=drawColor,line width= 0.4pt,line join=round,line cap=round,fill=fillColor] (100.51, 68.48) circle (  1.69);

\path[draw=drawColor,line width= 0.4pt,line join=round,line cap=round,fill=fillColor] (115.45, 75.51) circle (  1.69);

\path[draw=drawColor,line width= 0.4pt,line join=round,line cap=round,fill=fillColor] (130.38, 81.04) circle (  1.69);

\path[draw=drawColor,line width= 0.4pt,line join=round,line cap=round,fill=fillColor] (145.32, 87.41) circle (  1.69);

\path[draw=drawColor,line width= 0.4pt,line join=round,line cap=round,fill=fillColor] (160.25, 93.74) circle (  1.69);
\end{scope}
\begin{scope}
\path[clip] ( 36.00, 36.00) rectangle (165.03,106.97);
\definecolor{drawColor}{RGB}{255,127,0}

\path[draw=drawColor,line width= 0.4pt,line join=round,line cap=round] ( 70.65, 53.32) --
	( 85.58, 59.25) --
	(100.51, 65.17) --
	(115.45, 72.31) --
	(130.38, 78.24) --
	(145.32, 84.69) --
	(160.25, 91.15);

\path[draw=drawColor,line width= 0.4pt,line join=round,line cap=round] ( 70.65, 50.69) --
	( 72.92, 54.63) --
	( 68.37, 54.63) --
	( 70.65, 50.69);

\path[draw=drawColor,line width= 0.4pt,line join=round,line cap=round] ( 85.58, 56.62) --
	( 87.85, 60.56) --
	( 83.31, 60.56) --
	( 85.58, 56.62);

\path[draw=drawColor,line width= 0.4pt,line join=round,line cap=round] (100.51, 62.54) --
	(102.79, 66.48) --
	( 98.24, 66.48) --
	(100.51, 62.54);

\path[draw=drawColor,line width= 0.4pt,line join=round,line cap=round] (115.45, 69.68) --
	(117.72, 73.62) --
	(113.18, 73.62) --
	(115.45, 69.68);

\path[draw=drawColor,line width= 0.4pt,line join=round,line cap=round] (130.38, 75.61) --
	(132.66, 79.55) --
	(128.11, 79.55) --
	(130.38, 75.61);

\path[draw=drawColor,line width= 0.4pt,line join=round,line cap=round] (145.32, 82.07) --
	(147.59, 86.00) --
	(143.04, 86.00) --
	(145.32, 82.07);

\path[draw=drawColor,line width= 0.4pt,line join=round,line cap=round] (160.25, 88.52) --
	(162.52, 92.46) --
	(157.98, 92.46) --
	(160.25, 88.52);
\end{scope}
\begin{scope}
\path[clip] ( 36.00, 36.00) rectangle (165.03,106.97);
\definecolor{drawColor}{RGB}{0,0,0}
\definecolor{fillColor}{RGB}{255,255,255}

\path[draw=drawColor,line width= 0.4pt,line join=round,line cap=round,fill=fillColor] ( 36.00,106.97) rectangle ( 99.69, 78.17);
\definecolor{drawColor}{RGB}{77,175,74}

\path[draw=drawColor,line width= 0.4pt,line join=round,line cap=round] ( 37.62, 99.77) -- ( 48.42, 99.77);
\definecolor{drawColor}{RGB}{152,78,163}

\path[draw=drawColor,line width= 0.4pt,line join=round,line cap=round] ( 37.62, 92.57) -- ( 48.42, 92.57);
\definecolor{drawColor}{RGB}{255,127,0}

\path[draw=drawColor,line width= 0.4pt,line join=round,line cap=round] ( 37.62, 85.37) -- ( 48.42, 85.37);
\definecolor{drawColor}{RGB}{77,175,74}

\path[draw=drawColor,line width= 0.4pt,line join=round,line cap=round] ( 41.11, 99.77) --
	( 43.02,101.68) --
	( 44.93, 99.77) --
	( 43.02, 97.86) --
	( 41.11, 99.77);
\definecolor{drawColor}{RGB}{152,78,163}
\definecolor{fillColor}{RGB}{152,78,163}

\path[draw=drawColor,line width= 0.4pt,line join=round,line cap=round,fill=fillColor] ( 43.02, 92.57) circle (  1.35);
\definecolor{drawColor}{RGB}{255,127,0}

\path[draw=drawColor,line width= 0.4pt,line join=round,line cap=round] ( 43.02, 83.27) --
	( 44.84, 86.42) --
	( 41.20, 86.42) --
	( 43.02, 83.27);
\definecolor{drawColor}{RGB}{0,0,0}

\node[text=drawColor,anchor=base west,inner sep=0pt, outer sep=0pt, scale=  0.60] at ( 53.82, 97.71) {Hybrid ($8$-way)};

\node[text=drawColor,anchor=base west,inner sep=0pt, outer sep=0pt, scale=  0.60] at ( 53.82, 90.51) {Hybrid ($16$-way)};

\node[text=drawColor,anchor=base west,inner sep=0pt, outer sep=0pt, scale=  0.60] at ( 53.82, 83.31) {Hybrid ($32$-way)};
\end{scope}
\end{tikzpicture}

%% file: plot_figures/weak_scale_unet_256_bn.tikz
\begin{tikzpicture}[x=1pt,y=1pt]
\definecolor{fillColor}{RGB}{255,255,255}
\path[use as bounding box,fill=fillColor,fill opacity=0.00] (0,0) rectangle (168.63,110.57);
\begin{scope}
\path[clip] (  0.00,  0.00) rectangle (168.63,110.57);
\definecolor{drawColor}{RGB}{0,0,0}

\path[draw=drawColor,line width= 0.4pt,line join=round,line cap=round] ( 36.00, 36.00) --
	(165.03, 36.00) --
	(165.03,106.97) --
	( 36.00,106.97) --
	( 36.00, 36.00);
\end{scope}
\begin{scope}
\path[clip] (  0.00,  0.00) rectangle (168.63,110.57);
\definecolor{drawColor}{RGB}{0,0,0}

\node[text=drawColor,anchor=base,inner sep=0pt, outer sep=0pt, scale=  0.75] at (100.51,  6.30) {Number of GPUs};
\end{scope}
\begin{scope}
\path[clip] (  0.00,  0.00) rectangle (168.63,110.57);
\definecolor{drawColor}{RGB}{0,0,0}

\path[draw=drawColor,line width= 0.4pt,line join=round,line cap=round] ( 40.78, 36.00) -- ( 40.78, 36.00);

\path[draw=drawColor,line width= 0.4pt,line join=round,line cap=round] ( 40.78, 36.00) -- ( 40.78, 31.50);

\node[text=drawColor,anchor=base,inner sep=0pt, outer sep=0pt, scale=  0.51] at ( 40.78, 19.80) {16};

\path[draw=drawColor,line width= 0.4pt,line join=round,line cap=round] ( 60.69, 36.00) -- ( 60.69, 36.00);

\path[draw=drawColor,line width= 0.4pt,line join=round,line cap=round] ( 60.69, 36.00) -- ( 60.69, 31.50);

\node[text=drawColor,anchor=base,inner sep=0pt, outer sep=0pt, scale=  0.51] at ( 60.69, 19.80) {32};

\path[draw=drawColor,line width= 0.4pt,line join=round,line cap=round] ( 80.60, 36.00) -- ( 80.60, 36.00);

\path[draw=drawColor,line width= 0.4pt,line join=round,line cap=round] ( 80.60, 36.00) -- ( 80.60, 31.50);

\node[text=drawColor,anchor=base,inner sep=0pt, outer sep=0pt, scale=  0.51] at ( 80.60, 19.80) {64};

\path[draw=drawColor,line width= 0.4pt,line join=round,line cap=round] (100.51, 36.00) -- (100.51, 36.00);

\path[draw=drawColor,line width= 0.4pt,line join=round,line cap=round] (100.51, 36.00) -- (100.51, 31.50);

\node[text=drawColor,anchor=base,inner sep=0pt, outer sep=0pt, scale=  0.51] at (100.51, 19.80) {128};

\path[draw=drawColor,line width= 0.4pt,line join=round,line cap=round] (120.43, 36.00) -- (120.43, 36.00);

\path[draw=drawColor,line width= 0.4pt,line join=round,line cap=round] (120.43, 36.00) -- (120.43, 31.50);

\node[text=drawColor,anchor=base,inner sep=0pt, outer sep=0pt, scale=  0.51] at (120.43, 19.80) {256};

\path[draw=drawColor,line width= 0.4pt,line join=round,line cap=round] (140.34, 36.00) -- (140.34, 36.00);

\path[draw=drawColor,line width= 0.4pt,line join=round,line cap=round] (140.34, 36.00) -- (140.34, 31.50);

\node[text=drawColor,anchor=base,inner sep=0pt, outer sep=0pt, scale=  0.51] at (140.34, 19.80) {512};

\path[draw=drawColor,line width= 0.4pt,line join=round,line cap=round] (160.25, 36.00) -- (160.25, 36.00);

\path[draw=drawColor,line width= 0.4pt,line join=round,line cap=round] (160.25, 36.00) -- (160.25, 31.50);

\node[text=drawColor,anchor=base,inner sep=0pt, outer sep=0pt, scale=  0.51] at (160.25, 19.80) {1024};

\path[draw=drawColor,line width= 0.4pt,line join=round,line cap=round] ( 36.00, 36.00) -- ( 36.00,106.97);

\path[draw=drawColor,line width= 0.4pt,line join=round,line cap=round] ( 36.00, 36.86) -- ( 34.58, 36.86);

\path[draw=drawColor,line width= 0.4pt,line join=round,line cap=round] ( 36.00, 38.54) -- ( 34.58, 38.54);

\path[draw=drawColor,line width= 0.4pt,line join=round,line cap=round] ( 36.00, 40.02) -- ( 34.58, 40.02);

\path[draw=drawColor,line width= 0.4pt,line join=round,line cap=round] ( 36.00, 41.34) -- ( 34.58, 41.34);

\path[draw=drawColor,line width= 0.4pt,line join=round,line cap=round] ( 36.00, 50.04) -- ( 34.58, 50.04);

\path[draw=drawColor,line width= 0.4pt,line join=round,line cap=round] ( 36.00, 55.13) -- ( 34.58, 55.13);

\path[draw=drawColor,line width= 0.4pt,line join=round,line cap=round] ( 36.00, 58.74) -- ( 34.58, 58.74);

\path[draw=drawColor,line width= 0.4pt,line join=round,line cap=round] ( 36.00, 61.54) -- ( 34.58, 61.54);

\path[draw=drawColor,line width= 0.4pt,line join=round,line cap=round] ( 36.00, 63.83) -- ( 34.58, 63.83);

\path[draw=drawColor,line width= 0.4pt,line join=round,line cap=round] ( 36.00, 65.76) -- ( 34.58, 65.76);

\path[draw=drawColor,line width= 0.4pt,line join=round,line cap=round] ( 36.00, 67.44) -- ( 34.58, 67.44);

\path[draw=drawColor,line width= 0.4pt,line join=round,line cap=round] ( 36.00, 68.92) -- ( 34.58, 68.92);

\path[draw=drawColor,line width= 0.4pt,line join=round,line cap=round] ( 36.00, 70.24) -- ( 34.58, 70.24);

\path[draw=drawColor,line width= 0.4pt,line join=round,line cap=round] ( 36.00, 78.94) -- ( 34.58, 78.94);

\path[draw=drawColor,line width= 0.4pt,line join=round,line cap=round] ( 36.00, 84.03) -- ( 34.58, 84.03);

\path[draw=drawColor,line width= 0.4pt,line join=round,line cap=round] ( 36.00, 87.64) -- ( 34.58, 87.64);

\path[draw=drawColor,line width= 0.4pt,line join=round,line cap=round] ( 36.00, 90.44) -- ( 34.58, 90.44);

\path[draw=drawColor,line width= 0.4pt,line join=round,line cap=round] ( 36.00, 92.73) -- ( 34.58, 92.73);

\path[draw=drawColor,line width= 0.4pt,line join=round,line cap=round] ( 36.00, 94.66) -- ( 34.58, 94.66);

\path[draw=drawColor,line width= 0.4pt,line join=round,line cap=round] ( 36.00, 96.34) -- ( 34.58, 96.34);

\path[draw=drawColor,line width= 0.4pt,line join=round,line cap=round] ( 36.00, 97.82) -- ( 34.58, 97.82);

\path[draw=drawColor,line width= 0.4pt,line join=round,line cap=round] ( 36.00, 99.14) -- ( 34.58, 99.14);

\path[draw=drawColor,line width= 0.4pt,line join=round,line cap=round] ( 36.00, 36.00) -- ( 36.00,106.97);

\path[draw=drawColor,line width= 0.4pt,line join=round,line cap=round] ( 36.00, 41.34) -- ( 31.50, 41.34);

\path[draw=drawColor,line width= 0.4pt,line join=round,line cap=round] ( 36.00, 70.24) -- ( 31.50, 70.24);

\path[draw=drawColor,line width= 0.4pt,line join=round,line cap=round] ( 36.00, 99.14) -- ( 31.50, 99.14);

\node[text=drawColor,anchor=base east,inner sep=0pt, outer sep=0pt, scale=  0.75] at ( 27.00, 38.76) {$10^0$};

\node[text=drawColor,anchor=base east,inner sep=0pt, outer sep=0pt, scale=  0.75] at ( 27.00, 67.66) {$10^1$};

\node[text=drawColor,anchor=base east,inner sep=0pt, outer sep=0pt, scale=  0.75] at ( 27.00, 96.56) {$10^2$};
\end{scope}
\begin{scope}
\path[clip] (  0.00,  0.00) rectangle (168.63,110.57);
\definecolor{drawColor}{RGB}{0,0,0}

\node[text=drawColor,rotate= 90.00,anchor=base,inner sep=0pt, outer sep=0pt, scale=  0.75] at ( 11.70, 71.49) {Throughput [samples/s]};
\end{scope}
\begin{scope}
\path[clip] (  0.00,  0.00) rectangle (168.63,110.57);
\definecolor{drawColor}{gray}{0.80}

\path[draw=drawColor,line width= 0.1pt,line join=round,line cap=round] ( 40.78, 36.00) -- (160.25, 36.00);

\path[draw=drawColor,line width= 0.1pt,line join=round,line cap=round] ( 40.78, 36.00) -- ( 40.78,106.97);

\path[draw=drawColor,line width= 0.1pt,line join=round,line cap=round] ( 60.69, 36.00) -- ( 60.69,106.97);

\path[draw=drawColor,line width= 0.1pt,line join=round,line cap=round] ( 80.60, 36.00) -- ( 80.60,106.97);

\path[draw=drawColor,line width= 0.1pt,line join=round,line cap=round] (100.51, 36.00) -- (100.51,106.97);

\path[draw=drawColor,line width= 0.1pt,line join=round,line cap=round] (120.43, 36.00) -- (120.43,106.97);

\path[draw=drawColor,line width= 0.1pt,line join=round,line cap=round] (140.34, 36.00) -- (140.34,106.97);

\path[draw=drawColor,line width= 0.1pt,line join=round,line cap=round] (160.25, 36.00) -- (160.25,106.97);
\definecolor{drawColor}{RGB}{0,0,0}

\path[draw=drawColor,line width= 0.4pt,line join=round,line cap=round] ( 36.00, 36.00) --
	(165.03, 36.00) --
	(165.03,106.97) --
	( 36.00,106.97) --
	( 36.00, 36.00);
\definecolor{drawColor}{gray}{0.80}

\path[draw=drawColor,line width= 0.1pt,line join=round,line cap=round] ( 36.00, 36.00) -- ( 36.00,106.97);

\path[draw=drawColor,line width= 0.1pt,line join=round,line cap=round] ( 36.00, 36.86) -- (165.03, 36.86);

\path[draw=drawColor,line width= 0.1pt,line join=round,line cap=round] ( 36.00, 38.54) -- (165.03, 38.54);

\path[draw=drawColor,line width= 0.1pt,line join=round,line cap=round] ( 36.00, 40.02) -- (165.03, 40.02);

\path[draw=drawColor,line width= 0.1pt,line join=round,line cap=round] ( 36.00, 41.34) -- (165.03, 41.34);

\path[draw=drawColor,line width= 0.1pt,line join=round,line cap=round] ( 36.00, 50.04) -- (165.03, 50.04);

\path[draw=drawColor,line width= 0.1pt,line join=round,line cap=round] ( 36.00, 55.13) -- (165.03, 55.13);

\path[draw=drawColor,line width= 0.1pt,line join=round,line cap=round] ( 36.00, 58.74) -- (165.03, 58.74);

\path[draw=drawColor,line width= 0.1pt,line join=round,line cap=round] ( 36.00, 61.54) -- (165.03, 61.54);

\path[draw=drawColor,line width= 0.1pt,line join=round,line cap=round] ( 36.00, 63.83) -- (165.03, 63.83);

\path[draw=drawColor,line width= 0.1pt,line join=round,line cap=round] ( 36.00, 65.76) -- (165.03, 65.76);

\path[draw=drawColor,line width= 0.1pt,line join=round,line cap=round] ( 36.00, 67.44) -- (165.03, 67.44);

\path[draw=drawColor,line width= 0.1pt,line join=round,line cap=round] ( 36.00, 68.92) -- (165.03, 68.92);

\path[draw=drawColor,line width= 0.1pt,line join=round,line cap=round] ( 36.00, 70.24) -- (165.03, 70.24);

\path[draw=drawColor,line width= 0.1pt,line join=round,line cap=round] ( 36.00, 78.94) -- (165.03, 78.94);

\path[draw=drawColor,line width= 0.1pt,line join=round,line cap=round] ( 36.00, 84.03) -- (165.03, 84.03);

\path[draw=drawColor,line width= 0.1pt,line join=round,line cap=round] ( 36.00, 87.64) -- (165.03, 87.64);

\path[draw=drawColor,line width= 0.1pt,line join=round,line cap=round] ( 36.00, 90.44) -- (165.03, 90.44);

\path[draw=drawColor,line width= 0.1pt,line join=round,line cap=round] ( 36.00, 92.73) -- (165.03, 92.73);

\path[draw=drawColor,line width= 0.1pt,line join=round,line cap=round] ( 36.00, 94.66) -- (165.03, 94.66);

\path[draw=drawColor,line width= 0.1pt,line join=round,line cap=round] ( 36.00, 96.34) -- (165.03, 96.34);

\path[draw=drawColor,line width= 0.1pt,line join=round,line cap=round] ( 36.00, 97.82) -- (165.03, 97.82);

\path[draw=drawColor,line width= 0.1pt,line join=round,line cap=round] ( 36.00, 99.14) -- (165.03, 99.14);
\definecolor{drawColor}{RGB}{0,0,0}

\path[draw=drawColor,line width= 0.4pt,line join=round,line cap=round] ( 36.00, 36.00) --
	(165.03, 36.00) --
	(165.03,106.97) --
	( 36.00,106.97) --
	( 36.00, 36.00);
\end{scope}
\begin{scope}
\path[clip] ( 36.00, 36.00) rectangle (165.03,106.97);
\definecolor{drawColor}{RGB}{152,78,163}

\path[draw=drawColor,line width= 0.4pt,line join=round,line cap=round] ( 40.78, 44.70) --
	( 60.69, 53.19) --
	( 80.60, 61.16) --
	(100.51, 70.53) --
	(120.43, 79.26) --
	(140.34, 87.68);
\definecolor{fillColor}{RGB}{152,78,163}

\path[draw=drawColor,line width= 0.4pt,line join=round,line cap=round,fill=fillColor] ( 40.78, 44.70) circle (  1.69);

\path[draw=drawColor,line width= 0.4pt,line join=round,line cap=round,fill=fillColor] ( 60.69, 53.19) circle (  1.69);

\path[draw=drawColor,line width= 0.4pt,line join=round,line cap=round,fill=fillColor] ( 80.60, 61.16) circle (  1.69);

\path[draw=drawColor,line width= 0.4pt,line join=round,line cap=round,fill=fillColor] (100.51, 70.53) circle (  1.69);

\path[draw=drawColor,line width= 0.4pt,line join=round,line cap=round,fill=fillColor] (120.43, 79.26) circle (  1.69);

\path[draw=drawColor,line width= 0.4pt,line join=round,line cap=round,fill=fillColor] (140.34, 87.68) circle (  1.69);
\end{scope}
\begin{scope}
\path[clip] ( 36.00, 36.00) rectangle (165.03,106.97);
\definecolor{drawColor}{RGB}{255,127,0}

\path[draw=drawColor,line width= 0.4pt,line join=round,line cap=round] ( 60.69, 51.19) --
	( 80.60, 59.08) --
	(100.51, 68.28) --
	(120.43, 76.24) --
	(140.34, 84.71) --
	(160.25, 93.18);

\path[draw=drawColor,line width= 0.4pt,line join=round,line cap=round] ( 60.69, 48.57) --
	( 62.96, 52.51) --
	( 58.42, 52.51) --
	( 60.69, 48.57);

\path[draw=drawColor,line width= 0.4pt,line join=round,line cap=round] ( 80.60, 56.46) --
	( 82.88, 60.39) --
	( 78.33, 60.39) --
	( 80.60, 56.46);

\path[draw=drawColor,line width= 0.4pt,line join=round,line cap=round] (100.51, 65.65) --
	(102.79, 69.59) --
	( 98.24, 69.59) --
	(100.51, 65.65);

\path[draw=drawColor,line width= 0.4pt,line join=round,line cap=round] (120.43, 73.61) --
	(122.70, 77.55) --
	(118.15, 77.55) --
	(120.43, 73.61);

\path[draw=drawColor,line width= 0.4pt,line join=round,line cap=round] (140.34, 82.09) --
	(142.61, 86.02) --
	(138.07, 86.02) --
	(140.34, 82.09);

\path[draw=drawColor,line width= 0.4pt,line join=round,line cap=round] (160.25, 90.56) --
	(162.52, 94.50) --
	(157.98, 94.50) --
	(160.25, 90.56);
\end{scope}
\begin{scope}
\path[clip] ( 36.00, 36.00) rectangle (165.03,106.97);
\definecolor{drawColor}{RGB}{0,0,0}
\definecolor{fillColor}{RGB}{255,255,255}

\path[draw=drawColor,line width= 0.4pt,line join=round,line cap=round,fill=fillColor] ( 36.00,106.97) rectangle ( 99.69, 85.37);
\definecolor{drawColor}{RGB}{152,78,163}

\path[draw=drawColor,line width= 0.4pt,line join=round,line cap=round] ( 37.62, 99.77) -- ( 48.42, 99.77);
\definecolor{drawColor}{RGB}{255,127,0}

\path[draw=drawColor,line width= 0.4pt,line join=round,line cap=round] ( 37.62, 92.57) -- ( 48.42, 92.57);
\definecolor{drawColor}{RGB}{152,78,163}
\definecolor{fillColor}{RGB}{152,78,163}

\path[draw=drawColor,line width= 0.4pt,line join=round,line cap=round,fill=fillColor] ( 43.02, 99.77) circle (  1.35);
\definecolor{drawColor}{RGB}{255,127,0}

\path[draw=drawColor,line width= 0.4pt,line join=round,line cap=round] ( 43.02, 90.47) --
	( 44.84, 93.62) --
	( 41.20, 93.62) --
	( 43.02, 90.47);
\definecolor{drawColor}{RGB}{0,0,0}

\node[text=drawColor,anchor=base west,inner sep=0pt, outer sep=0pt, scale=  0.60] at ( 53.82, 97.71) {Hybrid ($16$-way)};

\node[text=drawColor,anchor=base west,inner sep=0pt, outer sep=0pt, scale=  0.60] at ( 53.82, 90.51) {Hybrid ($32$-way)};
\end{scope}
\end{tikzpicture}

%% file: plot_figures/201905_4parE_time.tikz
\begin{tikzpicture}[x=1pt,y=1pt]
\definecolor{fillColor}{RGB}{255,255,255}
\path[use as bounding box,fill=fillColor,fill opacity=0.00] (0,0) rectangle (252.94,135.14);
\begin{scope}
\path[clip] (  0.00,  0.00) rectangle (252.94,135.14);
\definecolor{drawColor}{RGB}{0,0,0}

\path[draw=drawColor,line width= 0.4pt,line join=round,line cap=round] ( 40.50, 36.00) -- (224.78, 36.00);

\path[draw=drawColor,line width= 0.4pt,line join=round,line cap=round] ( 40.50, 36.00) -- ( 40.50, 31.50);

\path[draw=drawColor,line width= 0.4pt,line join=round,line cap=round] ( 71.21, 36.00) -- ( 71.21, 31.50);

\path[draw=drawColor,line width= 0.4pt,line join=round,line cap=round] (101.92, 36.00) -- (101.92, 31.50);

\path[draw=drawColor,line width= 0.4pt,line join=round,line cap=round] (132.64, 36.00) -- (132.64, 31.50);

\path[draw=drawColor,line width= 0.4pt,line join=round,line cap=round] (163.35, 36.00) -- (163.35, 31.50);

\path[draw=drawColor,line width= 0.4pt,line join=round,line cap=round] (194.06, 36.00) -- (194.06, 31.50);

\path[draw=drawColor,line width= 0.4pt,line join=round,line cap=round] (224.78, 36.00) -- (224.78, 31.50);

\node[text=drawColor,anchor=base,inner sep=0pt, outer sep=0pt, scale=  0.75] at ( 40.50, 19.80) {0};

\node[text=drawColor,anchor=base,inner sep=0pt, outer sep=0pt, scale=  0.75] at ( 71.21, 19.80) {1};

\node[text=drawColor,anchor=base,inner sep=0pt, outer sep=0pt, scale=  0.75] at (101.93, 19.80) {2};

\node[text=drawColor,anchor=base,inner sep=0pt, outer sep=0pt, scale=  0.75] at (132.64, 19.80) {3};

\node[text=drawColor,anchor=base,inner sep=0pt, outer sep=0pt, scale=  0.75] at (163.35, 19.80) {4};

\node[text=drawColor,anchor=base,inner sep=0pt, outer sep=0pt, scale=  0.75] at (194.06, 19.80) {5};

\node[text=drawColor,anchor=base,inner sep=0pt, outer sep=0pt, scale=  0.75] at (224.78, 19.80) {6};

\path[draw=drawColor,line width= 0.4pt,line join=round,line cap=round] ( 40.50, 36.00) --
	(249.34, 36.00) --
	(249.34,131.54) --
	( 40.50,131.54) --
	( 40.50, 36.00);
\end{scope}
\begin{scope}
\path[clip] (  0.00,  0.00) rectangle (252.94,135.14);
\definecolor{drawColor}{RGB}{0,0,0}

\node[text=drawColor,anchor=base,inner sep=0pt, outer sep=0pt, scale=  0.75] at (144.92,  6.30) {Time [h]};

\node[text=drawColor,rotate= 90.00,anchor=base,inner sep=0pt, outer sep=0pt, scale=  0.75] at ( 11.70, 83.77) {Loss};
\end{scope}
\begin{scope}
\path[clip] (  0.00,  0.00) rectangle (252.94,135.14);
\definecolor{drawColor}{gray}{0.80}

\path[draw=drawColor,line width= 0.1pt,line join=round,line cap=round] ( 40.50, 36.00) -- (224.78, 36.00);

\path[draw=drawColor,line width= 0.1pt,line join=round,line cap=round] ( 40.50, 36.00) -- ( 40.50,131.54);

\path[draw=drawColor,line width= 0.1pt,line join=round,line cap=round] ( 71.21, 36.00) -- ( 71.21,131.54);

\path[draw=drawColor,line width= 0.1pt,line join=round,line cap=round] (101.92, 36.00) -- (101.92,131.54);

\path[draw=drawColor,line width= 0.1pt,line join=round,line cap=round] (132.64, 36.00) -- (132.64,131.54);

\path[draw=drawColor,line width= 0.1pt,line join=round,line cap=round] (163.35, 36.00) -- (163.35,131.54);

\path[draw=drawColor,line width= 0.1pt,line join=round,line cap=round] (194.06, 36.00) -- (194.06,131.54);

\path[draw=drawColor,line width= 0.1pt,line join=round,line cap=round] (224.78, 36.00) -- (224.78,131.54);
\definecolor{drawColor}{RGB}{0,0,0}

\path[draw=drawColor,line width= 0.4pt,line join=round,line cap=round] ( 40.50, 36.00) --
	(249.34, 36.00) --
	(249.34,131.54) --
	( 40.50,131.54) --
	( 40.50, 36.00);

\path[draw=drawColor,line width= 0.4pt,line join=round,line cap=round] ( 40.50, 36.00) -- ( 40.50,131.54);

\path[draw=drawColor,line width= 0.4pt,line join=round,line cap=round] ( 40.50, 36.00) -- ( 36.00, 36.00);

\path[draw=drawColor,line width= 0.4pt,line join=round,line cap=round] ( 40.50, 59.89) -- ( 36.00, 59.89);

\path[draw=drawColor,line width= 0.4pt,line join=round,line cap=round] ( 40.50, 83.77) -- ( 36.00, 83.77);

\path[draw=drawColor,line width= 0.4pt,line join=round,line cap=round] ( 40.50,107.66) -- ( 36.00,107.66);

\path[draw=drawColor,line width= 0.4pt,line join=round,line cap=round] ( 40.50,131.54) -- ( 36.00,131.54);

\node[text=drawColor,anchor=base east,inner sep=0pt, outer sep=0pt, scale=  0.75] at ( 31.50, 33.42) {$10^{-3}$};

\node[text=drawColor,anchor=base east,inner sep=0pt, outer sep=0pt, scale=  0.75] at ( 31.50, 57.30) {$10^{-2}$};

\node[text=drawColor,anchor=base east,inner sep=0pt, outer sep=0pt, scale=  0.75] at ( 31.50, 81.19) {$10^{-1}$};

\node[text=drawColor,anchor=base east,inner sep=0pt, outer sep=0pt, scale=  0.75] at ( 31.50,105.08) {$10^{0}$};

\node[text=drawColor,anchor=base east,inner sep=0pt, outer sep=0pt, scale=  0.75] at ( 31.50,128.96) {$10^{1}$};
\definecolor{drawColor}{gray}{0.80}

\path[draw=drawColor,line width= 0.1pt,line join=round,line cap=round] ( 40.50, 36.00) -- ( 40.50,131.54);

\path[draw=drawColor,line width= 0.1pt,line join=round,line cap=round] ( 40.50, 36.00) -- (249.34, 36.00);

\path[draw=drawColor,line width= 0.1pt,line join=round,line cap=round] ( 40.50, 43.19) -- (249.34, 43.19);

\path[draw=drawColor,line width= 0.1pt,line join=round,line cap=round] ( 40.50, 47.40) -- (249.34, 47.40);

\path[draw=drawColor,line width= 0.1pt,line join=round,line cap=round] ( 40.50, 50.38) -- (249.34, 50.38);

\path[draw=drawColor,line width= 0.1pt,line join=round,line cap=round] ( 40.50, 52.70) -- (249.34, 52.70);

\path[draw=drawColor,line width= 0.1pt,line join=round,line cap=round] ( 40.50, 54.59) -- (249.34, 54.59);

\path[draw=drawColor,line width= 0.1pt,line join=round,line cap=round] ( 40.50, 56.19) -- (249.34, 56.19);

\path[draw=drawColor,line width= 0.1pt,line join=round,line cap=round] ( 40.50, 57.57) -- (249.34, 57.57);

\path[draw=drawColor,line width= 0.1pt,line join=round,line cap=round] ( 40.50, 58.79) -- (249.34, 58.79);

\path[draw=drawColor,line width= 0.1pt,line join=round,line cap=round] ( 40.50, 59.89) -- (249.34, 59.89);

\path[draw=drawColor,line width= 0.1pt,line join=round,line cap=round] ( 40.50, 67.08) -- (249.34, 67.08);

\path[draw=drawColor,line width= 0.1pt,line join=round,line cap=round] ( 40.50, 71.28) -- (249.34, 71.28);

\path[draw=drawColor,line width= 0.1pt,line join=round,line cap=round] ( 40.50, 74.27) -- (249.34, 74.27);

\path[draw=drawColor,line width= 0.1pt,line join=round,line cap=round] ( 40.50, 76.58) -- (249.34, 76.58);

\path[draw=drawColor,line width= 0.1pt,line join=round,line cap=round] ( 40.50, 78.47) -- (249.34, 78.47);

\path[draw=drawColor,line width= 0.1pt,line join=round,line cap=round] ( 40.50, 80.07) -- (249.34, 80.07);

\path[draw=drawColor,line width= 0.1pt,line join=round,line cap=round] ( 40.50, 81.46) -- (249.34, 81.46);

\path[draw=drawColor,line width= 0.1pt,line join=round,line cap=round] ( 40.50, 82.68) -- (249.34, 82.68);

\path[draw=drawColor,line width= 0.1pt,line join=round,line cap=round] ( 40.50, 83.77) -- (249.34, 83.77);

\path[draw=drawColor,line width= 0.1pt,line join=round,line cap=round] ( 40.50, 90.96) -- (249.34, 90.96);

\path[draw=drawColor,line width= 0.1pt,line join=round,line cap=round] ( 40.50, 95.17) -- (249.34, 95.17);

\path[draw=drawColor,line width= 0.1pt,line join=round,line cap=round] ( 40.50, 98.15) -- (249.34, 98.15);

\path[draw=drawColor,line width= 0.1pt,line join=round,line cap=round] ( 40.50,100.47) -- (249.34,100.47);

\path[draw=drawColor,line width= 0.1pt,line join=round,line cap=round] ( 40.50,102.36) -- (249.34,102.36);

\path[draw=drawColor,line width= 0.1pt,line join=round,line cap=round] ( 40.50,103.96) -- (249.34,103.96);

\path[draw=drawColor,line width= 0.1pt,line join=round,line cap=round] ( 40.50,105.34) -- (249.34,105.34);

\path[draw=drawColor,line width= 0.1pt,line join=round,line cap=round] ( 40.50,106.57) -- (249.34,106.57);

\path[draw=drawColor,line width= 0.1pt,line join=round,line cap=round] ( 40.50,107.66) -- (249.34,107.66);

\path[draw=drawColor,line width= 0.1pt,line join=round,line cap=round] ( 40.50,114.85) -- (249.34,114.85);

\path[draw=drawColor,line width= 0.1pt,line join=round,line cap=round] ( 40.50,119.06) -- (249.34,119.06);

\path[draw=drawColor,line width= 0.1pt,line join=round,line cap=round] ( 40.50,122.04) -- (249.34,122.04);

\path[draw=drawColor,line width= 0.1pt,line join=round,line cap=round] ( 40.50,124.35) -- (249.34,124.35);

\path[draw=drawColor,line width= 0.1pt,line join=round,line cap=round] ( 40.50,126.25) -- (249.34,126.25);

\path[draw=drawColor,line width= 0.1pt,line join=round,line cap=round] ( 40.50,127.84) -- (249.34,127.84);

\path[draw=drawColor,line width= 0.1pt,line join=round,line cap=round] ( 40.50,129.23) -- (249.34,129.23);

\path[draw=drawColor,line width= 0.1pt,line join=round,line cap=round] ( 40.50,130.45) -- (249.34,130.45);

\path[draw=drawColor,line width= 0.1pt,line join=round,line cap=round] ( 40.50,131.54) -- (249.34,131.54);
\definecolor{drawColor}{RGB}{0,0,0}

\path[draw=drawColor,line width= 0.4pt,line join=round,line cap=round] ( 40.50, 36.00) --
	(249.34, 36.00) --
	(249.34,131.54) --
	( 40.50,131.54) --
	( 40.50, 36.00);
\end{scope}
\begin{scope}
\path[clip] ( 40.50, 36.00) rectangle (249.34,131.54);
\definecolor{drawColor}{RGB}{228,26,28}

\path[draw=drawColor,line width= 0.6pt,line join=round,line cap=round] ( 40.50, 88.24) --
	( 41.93, 82.89) --
	( 43.37, 82.89) --
	( 44.80, 82.89) --
	( 46.22, 82.89) --
	( 47.65, 82.89) --
	( 49.07, 82.89) --
	( 50.50, 82.89) --
	( 51.92, 82.89) --
	( 53.35, 82.89) --
	( 54.78, 82.89) --
	( 56.20, 82.89) --
	( 57.63, 82.89) --
	( 59.06, 82.89) --
	( 60.48, 82.89) --
	( 61.91, 82.89) --
	( 63.34, 82.89) --
	( 64.77, 82.89) --
	( 66.19, 82.89) --
	( 67.62, 82.89) --
	( 69.05, 82.89) --
	( 70.47, 82.89) --
	( 71.90, 82.89) --
	( 73.32, 82.89) --
	( 74.75, 82.89) --
	( 76.18, 82.89) --
	( 77.60, 82.89) --
	( 79.03, 82.89) --
	( 80.46, 82.89) --
	( 81.88, 82.89) --
	( 83.31, 82.89) --
	( 84.74, 82.89) --
	( 86.16, 82.89) --
	( 87.59, 82.89) --
	( 89.01, 82.87) --
	( 90.44, 82.01) --
	( 91.86, 82.01) --
	( 93.29, 82.01) --
	( 94.72, 82.01) --
	( 96.14, 82.01) --
	( 97.57, 82.01) --
	( 98.99, 82.01) --
	(100.42, 82.01) --
	(101.84, 82.01) --
	(103.27, 81.83) --
	(104.69, 81.83) --
	(106.12, 81.83) --
	(107.54, 81.83) --
	(108.97, 81.83) --
	(110.40, 81.83) --
	(111.82, 81.83) --
	(113.25, 81.83) --
	(114.68, 81.83) --
	(116.10, 81.83) --
	(117.53, 81.54) --
	(118.95, 80.86) --
	(120.38, 80.86) --
	(121.80, 80.86) --
	(123.23, 80.86) --
	(124.66, 80.86) --
	(126.08, 80.86) --
	(127.51, 80.86) --
	(128.93, 80.86) --
	(130.36, 80.86) --
	(131.79, 79.82) --
	(133.21, 79.82) --
	(134.64, 79.82) --
	(136.06, 79.82) --
	(137.49, 79.20) --
	(138.91, 79.20) --
	(140.34, 79.20) --
	(141.77, 79.20) --
	(143.19, 79.20) --
	(144.62, 79.20) --
	(146.04, 79.20) --
	(147.47, 79.20) --
	(148.89, 79.20) --
	(150.32, 78.47) --
	(151.74, 78.47) --
	(153.17, 78.47) --
	(154.60, 78.13) --
	(156.02, 78.13) --
	(157.44, 78.13) --
	(158.87, 78.13) --
	(160.30, 77.58) --
	(161.72, 77.46) --
	(163.15, 77.33) --
	(164.57, 77.32) --
	(166.00, 77.32) --
	(167.42, 77.04) --
	(168.85, 76.91) --
	(170.28, 76.87) --
	(171.70, 76.64) --
	(173.13, 76.55) --
	(174.55, 76.41) --
	(175.98, 76.34) --
	(177.40, 76.19) --
	(178.83, 76.15) --
	(180.26, 76.00) --
	(181.68, 75.91) --
	(183.11, 75.87) --
	(184.53, 75.85) --
	(185.96, 75.85) --
	(187.39, 75.85) --
	(188.81, 75.83) --
	(190.24, 75.83) --
	(191.66, 75.83) --
	(193.09, 75.83) --
	(194.51, 75.83) --
	(195.94, 75.83) --
	(197.36, 75.83) --
	(198.79, 75.83) --
	(200.22, 75.83) --
	(201.64, 75.83) --
	(203.07, 75.83) --
	(204.49, 75.83) --
	(205.92, 75.83) --
	(207.34, 75.83) --
	(208.77, 75.83) --
	(210.20, 75.83) --
	(211.62, 75.83) --
	(213.05, 75.83) --
	(214.47, 75.83) --
	(215.90, 75.83) --
	(217.32, 75.83) --
	(218.75, 75.83) --
	(220.18, 75.83) --
	(221.60, 75.83) --
	(223.03, 75.83) --
	(224.45, 75.83);
\end{scope}
\begin{scope}
\path[clip] ( 40.50, 36.00) rectangle (249.34,131.54);
\definecolor{drawColor}{RGB}{228,26,28}

\path[draw=drawColor,line width= 0.2pt,line join=round,line cap=round] (224.45, 36.00) -- (224.45,131.54);

\node[text=drawColor,anchor=base west,inner sep=0pt, outer sep=0pt, scale=  0.75] at (226.45, 73.25) {5.99 h};
\end{scope}
\begin{scope}
\path[clip] ( 40.50, 36.00) rectangle (249.34,131.54);
\definecolor{drawColor}{RGB}{228,26,28}

\path[draw=drawColor,line width= 0.4pt,dash pattern=on 4pt off 4pt ,line join=round,line cap=round] ( 40.50, 82.85) --
	( 41.93, 81.27) --
	( 43.37, 81.27) --
	( 44.80, 81.27) --
	( 46.22, 81.27) --
	( 47.65, 81.27) --
	( 49.07, 81.27) --
	( 50.50, 81.27) --
	( 51.92, 81.27) --
	( 53.35, 81.27) --
	( 54.78, 81.27) --
	( 56.20, 81.27) --
	( 57.63, 81.27) --
	( 59.06, 81.27) --
	( 60.48, 81.27) --
	( 61.91, 81.27) --
	( 63.34, 81.27) --
	( 64.77, 81.27) --
	( 66.19, 81.27) --
	( 67.62, 81.27) --
	( 69.05, 81.27) --
	( 70.47, 81.27) --
	( 71.90, 81.27) --
	( 73.32, 81.27) --
	( 74.75, 81.27) --
	( 76.18, 81.27) --
	( 77.60, 81.27) --
	( 79.03, 81.27) --
	( 80.46, 81.27) --
	( 81.88, 81.27) --
	( 83.31, 81.27) --
	( 84.74, 81.27) --
	( 86.16, 81.27) --
	( 87.59, 81.27) --
	( 89.01, 81.27) --
	( 90.44, 80.02) --
	( 91.86, 80.02) --
	( 93.29, 80.02) --
	( 94.72, 80.02) --
	( 96.14, 80.02) --
	( 97.57, 80.02) --
	( 98.99, 80.02) --
	(100.42, 80.02) --
	(101.84, 80.02) --
	(103.27, 79.13) --
	(104.69, 79.13) --
	(106.12, 79.13) --
	(107.54, 79.13) --
	(108.97, 79.13) --
	(110.40, 78.64) --
	(111.82, 78.64) --
	(113.25, 78.64) --
	(114.68, 78.64) --
	(116.10, 78.64) --
	(117.53, 78.64) --
	(118.95, 78.64) --
	(120.38, 78.64) --
	(121.80, 78.64) --
	(123.23, 78.64) --
	(124.66, 78.20) --
	(126.08, 78.20) --
	(127.51, 78.20) --
	(128.93, 77.91) --
	(130.36, 77.91) --
	(131.79, 77.91) --
	(133.21, 77.29) --
	(134.64, 77.29) --
	(136.06, 77.29) --
	(137.49, 77.29) --
	(138.91, 77.29) --
	(140.34, 77.29) --
	(141.77, 77.29) --
	(143.19, 77.09) --
	(144.62, 77.09) --
	(146.04, 77.09) --
	(147.47, 77.09) --
	(148.89, 77.09) --
	(150.32, 76.73) --
	(151.74, 76.73) --
	(153.17, 76.58) --
	(154.60, 76.58) --
	(156.02, 76.58) --
	(157.44, 76.58) --
	(158.87, 76.58) --
	(160.30, 76.58) --
	(161.72, 75.85) --
	(163.15, 75.85) --
	(164.57, 75.85) --
	(166.00, 75.85) --
	(167.42, 75.83) --
	(168.85, 75.83) --
	(170.28, 75.83) --
	(171.70, 75.68) --
	(173.13, 75.58) --
	(174.55, 75.48) --
	(175.98, 75.48) --
	(177.40, 75.44) --
	(178.83, 75.39) --
	(180.26, 75.39) --
	(181.68, 75.27) --
	(183.11, 75.27) --
	(184.53, 75.27) --
	(185.96, 75.27) --
	(187.39, 75.27) --
	(188.81, 75.27) --
	(190.24, 75.27) --
	(191.66, 75.27) --
	(193.09, 75.27) --
	(194.51, 75.27) --
	(195.94, 75.27) --
	(197.36, 75.27) --
	(198.79, 75.27) --
	(200.22, 75.27) --
	(201.64, 75.27) --
	(203.07, 75.27) --
	(204.49, 75.27) --
	(205.92, 75.27) --
	(207.34, 75.27) --
	(208.77, 75.27) --
	(210.20, 75.27) --
	(211.62, 75.27) --
	(213.05, 75.27) --
	(214.47, 75.27) --
	(215.90, 75.27) --
	(217.32, 75.27) --
	(218.75, 75.27) --
	(220.18, 75.27) --
	(221.60, 75.27) --
	(223.03, 75.27) --
	(224.45, 75.27);
\end{scope}
\begin{scope}
\path[clip] ( 40.50, 36.00) rectangle (249.34,131.54);
\definecolor{drawColor}{RGB}{228,26,28}

\path[draw=drawColor,line width= 0.4pt,line join=round,line cap=round] (181.68, 75.27) circle (  1.69);
\end{scope}
\begin{scope}
\path[clip] ( 40.50, 36.00) rectangle (249.34,131.54);
\definecolor{drawColor}{RGB}{228,26,28}

\path[draw=drawColor,line width= 0.4pt,line join=round,line cap=round] (181.68, 75.27) --
	(181.68, 81.97);

\node[text=drawColor,anchor=base,inner sep=0pt, outer sep=0pt, scale=  0.75] at (181.68, 85.60) {0.0441};
\end{scope}
\begin{scope}
\path[clip] ( 40.50, 36.00) rectangle (249.34,131.54);
\definecolor{drawColor}{RGB}{55,126,184}

\path[draw=drawColor,line width= 0.6pt,line join=round,line cap=round] ( 40.50, 94.93) --
	( 41.09, 91.08) --
	( 41.82, 89.37) --
	( 42.43, 87.16) --
	( 43.03, 82.59) --
	( 43.63, 80.21) --
	( 44.23, 78.29) --
	( 44.83, 76.90) --
	( 45.42, 76.90) --
	( 46.03, 76.90) --
	( 46.62, 76.90) --
	( 47.22, 76.90) --
	( 47.82, 76.90) --
	( 48.43, 76.41) --
	( 49.03, 75.99) --
	( 49.62, 74.74) --
	( 50.22, 73.89) --
	( 50.82, 73.89) --
	( 51.41, 73.89) --
	( 52.01, 73.89) --
	( 52.61, 73.89) --
	( 53.21, 73.89) --
	( 53.81, 73.89) --
	( 54.40, 73.89) --
	( 55.01, 73.89) --
	( 55.60, 73.89) --
	( 56.20, 73.89) --
	( 56.80, 73.89) --
	( 57.40, 73.89) --
	( 57.99, 73.89) --
	( 58.59, 73.89) --
	( 59.19, 73.89) --
	( 59.79, 73.89) --
	( 60.39, 73.89) --
	( 60.99, 73.89) --
	( 61.59, 73.89) --
	( 62.19, 73.89) --
	( 62.78, 73.89) --
	( 63.38, 73.89) --
	( 63.98, 73.89) --
	( 64.57, 73.89) --
	( 65.17, 73.89) --
	( 65.76, 73.89) --
	( 66.36, 73.89) --
	( 66.97, 73.89) --
	( 67.57, 73.89) --
	( 68.17, 73.89) --
	( 68.77, 73.89) --
	( 69.37, 73.89) --
	( 69.96, 73.89) --
	( 70.56, 73.89) --
	( 71.16, 73.89) --
	( 71.77, 73.89) --
	( 72.37, 73.89) --
	( 72.97, 73.89) --
	( 73.57, 73.89) --
	( 74.17, 73.89) --
	( 74.78, 73.89) --
	( 75.38, 73.89) --
	( 75.98, 73.89) --
	( 76.57, 73.89) --
	( 77.17, 73.89) --
	( 77.77, 73.89) --
	( 78.37, 73.89) --
	( 78.97, 73.89) --
	( 79.57, 73.89) --
	( 80.16, 73.89) --
	( 80.77, 73.89) --
	( 81.37, 73.86) --
	( 81.97, 73.86) --
	( 82.57, 73.86) --
	( 83.17, 73.86) --
	( 83.76, 73.86) --
	( 84.36, 73.58) --
	( 84.96, 73.30) --
	( 85.55, 72.97) --
	( 86.15, 72.52) --
	( 86.74, 71.67) --
	( 87.35, 71.54) --
	( 87.96, 71.44) --
	( 88.56, 70.66) --
	( 89.16, 70.21) --
	( 89.76, 70.21) --
	( 90.36, 69.97) --
	( 90.96, 69.73) --
	( 91.56, 69.22) --
	( 92.16, 69.22) --
	( 92.76, 68.96) --
	( 93.36, 68.93) --
	( 93.95, 68.63) --
	( 94.55, 68.63) --
	( 95.14, 68.63) --
	( 95.73, 68.63) --
	( 96.34, 68.48) --
	( 96.94, 68.28) --
	( 97.53, 67.99) --
	( 98.13, 67.92) --
	( 98.73, 67.70) --
	( 99.32, 67.46) --
	( 99.91, 67.28) --
	(100.51, 67.18) --
	(101.10, 67.13) --
	(101.70, 67.13) --
	(102.30, 67.13) --
	(102.89, 67.13) --
	(103.49, 67.13) --
	(104.09, 67.13) --
	(104.68, 67.13) --
	(105.28, 67.13) --
	(105.88, 67.12) --
	(106.48, 67.12) --
	(107.07, 67.12) --
	(107.66, 67.12) --
	(108.26, 67.12) --
	(108.85, 67.12) --
	(109.44, 67.12) --
	(110.03, 67.11) --
	(110.62, 67.11) --
	(111.22, 67.11) --
	(111.81, 67.11) --
	(112.40, 67.10) --
	(113.00, 67.10) --
	(113.60, 67.10) --
	(114.19, 67.10) --
	(114.78, 67.10) --
	(115.38, 67.10) --
	(115.98, 67.10) --
	(116.57, 67.10) --
	(117.17, 67.08) --
	(117.76, 67.08);
\end{scope}
\begin{scope}
\path[clip] ( 40.50, 36.00) rectangle (249.34,131.54);
\definecolor{drawColor}{RGB}{55,126,184}

\path[draw=drawColor,line width= 0.2pt,line join=round,line cap=round] (117.76, 36.00) -- (117.76,131.54);

\node[text=drawColor,anchor=base west,inner sep=0pt, outer sep=0pt, scale=  0.75] at (119.76, 64.49) {2.52 h};
\end{scope}
\begin{scope}
\path[clip] ( 40.50, 36.00) rectangle (249.34,131.54);
\definecolor{drawColor}{RGB}{55,126,184}

\path[draw=drawColor,line width= 0.4pt,dash pattern=on 4pt off 4pt ,line join=round,line cap=round] ( 40.50, 93.34) --
	( 41.09, 88.78) --
	( 41.82, 88.60) --
	( 42.43, 84.12) --
	( 43.03, 79.34) --
	( 43.63, 79.34) --
	( 44.23, 78.03) --
	( 44.83, 76.20) --
	( 45.42, 76.20) --
	( 46.03, 76.20) --
	( 46.62, 76.20) --
	( 47.22, 76.20) --
	( 47.82, 76.20) --
	( 48.43, 76.20) --
	( 49.03, 73.84) --
	( 49.62, 72.51) --
	( 50.22, 72.51) --
	( 50.82, 72.51) --
	( 51.41, 72.51) --
	( 52.01, 72.51) --
	( 52.61, 72.51) --
	( 53.21, 72.51) --
	( 53.81, 72.51) --
	( 54.40, 72.51) --
	( 55.01, 72.51) --
	( 55.60, 72.51) --
	( 56.20, 72.51) --
	( 56.80, 72.51) --
	( 57.40, 72.51) --
	( 57.99, 72.51) --
	( 58.59, 72.51) --
	( 59.19, 72.51) --
	( 59.79, 72.51) --
	( 60.39, 72.51) --
	( 60.99, 72.51) --
	( 61.59, 72.51) --
	( 62.19, 72.51) --
	( 62.78, 72.51) --
	( 63.38, 72.51) --
	( 63.98, 72.51) --
	( 64.57, 72.51) --
	( 65.17, 72.51) --
	( 65.76, 72.51) --
	( 66.36, 72.51) --
	( 66.97, 72.51) --
	( 67.57, 72.51) --
	( 68.17, 72.51) --
	( 68.77, 72.51) --
	( 69.37, 72.51) --
	( 69.96, 72.51) --
	( 70.56, 72.51) --
	( 71.16, 72.51) --
	( 71.77, 72.51) --
	( 72.37, 72.51) --
	( 72.97, 72.51) --
	( 73.57, 72.51) --
	( 74.17, 72.51) --
	( 74.78, 72.51) --
	( 75.38, 72.51) --
	( 75.98, 72.51) --
	( 76.57, 72.51) --
	( 77.17, 72.51) --
	( 77.77, 72.51) --
	( 78.37, 72.51) --
	( 78.97, 72.51) --
	( 79.57, 71.21) --
	( 80.16, 71.21) --
	( 80.77, 71.21) --
	( 81.37, 70.07) --
	( 81.97, 70.07) --
	( 82.57, 70.07) --
	( 83.17, 70.07) --
	( 83.76, 70.07) --
	( 84.36, 70.07) --
	( 84.96, 70.07) --
	( 85.55, 69.40) --
	( 86.15, 68.90) --
	( 86.74, 68.90) --
	( 87.35, 68.90) --
	( 87.96, 68.90) --
	( 88.56, 67.19) --
	( 89.16, 67.19) --
	( 89.76, 67.19) --
	( 90.36, 67.15) --
	( 90.96, 67.15) --
	( 91.56, 66.62) --
	( 92.16, 66.23) --
	( 92.76, 65.63) --
	( 93.36, 65.63) --
	( 93.95, 65.63) --
	( 94.55, 65.63) --
	( 95.14, 65.63) --
	( 95.73, 65.63) --
	( 96.34, 65.63) --
	( 96.94, 65.56) --
	( 97.53, 65.48) --
	( 98.13, 65.48) --
	( 98.73, 65.48) --
	( 99.32, 65.36) --
	( 99.91, 65.06) --
	(100.51, 65.06) --
	(101.10, 65.06) --
	(101.70, 65.06) --
	(102.30, 65.06) --
	(102.89, 65.06) --
	(103.49, 65.06) --
	(104.09, 65.06) --
	(104.68, 65.06) --
	(105.28, 65.06) --
	(105.88, 65.06) --
	(106.48, 65.06) --
	(107.07, 65.06) --
	(107.66, 65.06) --
	(108.26, 65.06) --
	(108.85, 65.06) --
	(109.44, 65.06) --
	(110.03, 65.06) --
	(110.62, 65.06) --
	(111.22, 65.06) --
	(111.81, 65.06) --
	(112.40, 65.06) --
	(113.00, 65.06) --
	(113.60, 65.06) --
	(114.19, 65.06) --
	(114.78, 65.06) --
	(115.38, 65.06) --
	(115.98, 65.06) --
	(116.57, 65.06) --
	(117.17, 65.06) --
	(117.76, 65.06);
\end{scope}
\begin{scope}
\path[clip] ( 40.50, 36.00) rectangle (249.34,131.54);
\definecolor{fillColor}{RGB}{55,126,184}

\path[fill=fillColor] (105.28, 67.68) --
	(107.56, 63.74) --
	(103.01, 63.74) --
	cycle;
\end{scope}
\begin{scope}
\path[clip] ( 40.50, 36.00) rectangle (249.34,131.54);
\definecolor{drawColor}{RGB}{55,126,184}

\path[draw=drawColor,line width= 0.4pt,line join=round,line cap=round] (105.28, 65.06) --
	(102.21, 97.24);

\node[text=drawColor,anchor=base east,inner sep=0pt, outer sep=0pt, scale=  0.75] at (103.20,101.21) {0.0165};
\end{scope}
\begin{scope}
\path[clip] ( 40.50, 36.00) rectangle (249.34,131.54);
\definecolor{drawColor}{RGB}{77,175,74}

\path[draw=drawColor,line width= 0.6pt,line join=round,line cap=round] ( 40.50, 97.48) --
	( 41.01, 93.68) --
	( 41.53, 93.04) --
	( 42.04, 90.43) --
	( 42.55, 90.11) --
	( 43.06, 89.78) --
	( 43.57, 89.60) --
	( 44.07, 89.39) --
	( 44.58, 89.44) --
	( 45.09, 89.27) --
	( 45.60, 89.34) --
	( 46.11, 89.31) --
	( 46.62, 89.23) --
	( 47.13, 89.34) --
	( 47.63, 89.22) --
	( 48.14, 89.25) --
	( 48.65, 89.04) --
	( 49.16, 89.11) --
	( 49.67, 88.92) --
	( 50.18, 87.80) --
	( 50.69, 86.73) --
	( 51.20, 86.83) --
	( 51.72, 85.06) --
	( 52.23, 84.97) --
	( 52.74, 84.74) --
	( 53.25, 84.07) --
	( 53.76, 84.33) --
	( 54.27, 83.90) --
	( 54.78, 83.69) --
	( 55.29, 83.32) --
	( 55.79, 83.19) --
	( 56.30, 83.19) --
	( 56.81, 82.67) --
	( 57.31, 82.77) --
	( 57.81, 82.95) --
	( 58.33, 82.41) --
	( 58.84, 82.11) --
	( 59.34, 82.24) --
	( 59.86, 82.05) --
	( 60.37, 83.09) --
	( 60.87, 81.83) --
	( 61.38, 79.27) --
	( 61.89, 81.70) --
	( 62.41, 78.88) --
	( 62.92, 76.99) --
	( 63.43, 74.61) --
	( 63.95, 73.54) --
	( 64.46, 71.21) --
	( 64.97, 71.50) --
	( 65.48, 71.48) --
	( 65.99, 71.99) --
	( 66.50, 70.24) --
	( 67.00, 71.69) --
	( 67.51, 72.79) --
	( 68.02, 68.68) --
	( 68.53, 69.49) --
	( 69.04, 69.93) --
	( 69.55, 69.56) --
	( 70.06, 68.52) --
	( 70.57, 68.82) --
	( 71.08, 67.59) --
	( 71.59, 67.98) --
	( 72.10, 67.27) --
	( 72.61, 66.80) --
	( 73.13, 67.02) --
	( 73.64, 67.58) --
	( 74.16, 65.98) --
	( 74.66, 66.72) --
	( 75.18, 66.63) --
	( 75.68, 65.67) --
	( 76.20, 65.46) --
	( 76.70, 66.31) --
	( 77.21, 65.93) --
	( 77.72, 66.71) --
	( 78.23, 64.81) --
	( 78.73, 64.36) --
	( 79.24, 64.80) --
	( 79.75, 65.47) --
	( 80.25, 64.68) --
	( 80.78, 65.13) --
	( 81.29, 64.08) --
	( 81.79, 63.88) --
	( 82.30, 63.53) --
	( 82.81, 63.25) --
	( 83.33, 63.68) --
	( 83.85, 63.26) --
	( 84.36, 62.29) --
	( 84.88, 62.88) --
	( 85.39, 62.20) --
	( 85.91, 62.21) --
	( 86.42, 62.18) --
	( 86.94, 61.10) --
	( 87.45, 61.12) --
	( 87.96, 61.02) --
	( 88.48, 61.02) --
	( 89.00, 60.66) --
	( 89.51, 60.63) --
	( 90.03, 60.41) --
	( 90.54, 59.93) --
	( 91.06, 59.74) --
	( 91.57, 59.80) --
	( 92.09, 59.42) --
	( 92.60, 59.72) --
	( 93.11, 59.58) --
	( 93.63, 59.48) --
	( 94.14, 59.42) --
	( 94.66, 59.53) --
	( 95.17, 59.49) --
	( 95.70, 59.59) --
	( 96.21, 59.87) --
	( 96.72, 59.77) --
	( 97.23, 59.73) --
	( 97.74, 59.63) --
	( 98.26, 59.78) --
	( 98.77, 59.66) --
	( 99.28, 59.72) --
	( 99.81, 59.45) --
	(100.32, 59.39) --
	(100.84, 59.63) --
	(101.36, 59.66) --
	(101.87, 59.55) --
	(102.39, 59.49) --
	(102.90, 59.75) --
	(103.42, 59.83) --
	(103.93, 59.27) --
	(104.45, 59.52) --
	(104.96, 59.79) --
	(105.47, 59.43) --
	(105.99, 59.36) --
	(106.50, 59.30);
\end{scope}
\begin{scope}
\path[clip] ( 40.50, 36.00) rectangle (249.34,131.54);
\definecolor{drawColor}{RGB}{77,175,74}

\path[draw=drawColor,line width= 0.2pt,line join=round,line cap=round] (106.50, 36.00) -- (106.50,131.54);

\node[text=drawColor,anchor=base west,inner sep=0pt, outer sep=0pt, scale=  0.75] at (108.50, 58.27) {2.15 h};
\end{scope}
\begin{scope}
\path[clip] ( 40.50, 36.00) rectangle (249.34,131.54);
\definecolor{drawColor}{RGB}{77,175,74}

\path[draw=drawColor,line width= 0.4pt,dash pattern=on 4pt off 4pt ,line join=round,line cap=round] ( 40.50, 93.15) --
	( 41.01, 93.29) --
	( 41.53, 91.04) --
	( 42.04, 88.75) --
	( 42.55, 88.55) --
	( 43.06, 88.30) --
	( 43.57, 88.76) --
	( 44.07, 88.38) --
	( 44.58, 88.24) --
	( 45.09, 88.31) --
	( 45.60, 88.45) --
	( 46.11, 88.37) --
	( 46.62, 88.84) --
	( 47.13, 88.42) --
	( 47.63, 88.35) --
	( 48.14, 88.36) --
	( 48.65, 88.25) --
	( 49.16, 88.60) --
	( 49.67, 88.10) --
	( 50.18, 85.43) --
	( 50.69, 93.72) --
	( 51.20, 84.50) --
	( 51.72, 87.42) --
	( 52.23, 84.30) --
	( 52.74, 82.82) --
	( 53.25, 84.43) --
	( 53.76, 82.93) --
	( 54.27, 83.04) --
	( 54.78, 81.70) --
	( 55.29, 82.27) --
	( 55.79, 81.68) --
	( 56.30, 82.11) --
	( 56.81, 81.85) --
	( 57.31, 81.34) --
	( 57.81, 80.76) --
	( 58.33, 81.94) --
	( 58.84, 80.82) --
	( 59.34, 81.11) --
	( 59.86, 84.80) --
	( 60.37, 82.70) --
	( 60.87, 80.16) --
	( 61.38, 78.32) --
	( 61.89, 81.55) --
	( 62.41, 79.40) --
	( 62.92, 74.49) --
	( 63.43, 75.77) --
	( 63.95, 69.69) --
	( 64.46, 67.19) --
	( 64.97, 67.75) --
	( 65.48, 70.93) --
	( 65.99, 68.27) --
	( 66.50, 75.83) --
	( 67.00, 67.23) --
	( 67.51, 66.90) --
	( 68.02, 66.29) --
	( 68.53, 70.90) --
	( 69.04, 64.61) --
	( 69.55, 71.52) --
	( 70.06, 63.70) --
	( 70.57, 68.13) --
	( 71.08, 66.25) --
	( 71.59, 70.92) --
	( 72.10, 68.94) --
	( 72.61, 61.21) --
	( 73.13, 63.28) --
	( 73.64, 62.08) --
	( 74.16, 63.48) --
	( 74.66, 67.01) --
	( 75.18, 68.09) --
	( 75.68, 63.13) --
	( 76.20, 64.98) --
	( 76.70, 65.11) --
	( 77.21, 64.57) --
	( 77.72, 62.72) --
	( 78.23, 65.90) --
	( 78.73, 59.92) --
	( 79.24, 63.10) --
	( 79.75, 62.46) --
	( 80.25, 61.09) --
	( 80.78, 61.73) --
	( 81.29, 63.45) --
	( 81.79, 59.66) --
	( 82.30, 61.18) --
	( 82.81, 60.58) --
	( 83.33, 60.05) --
	( 83.85, 62.00) --
	( 84.36, 59.07) --
	( 84.88, 58.73) --
	( 85.39, 58.14) --
	( 85.91, 59.45) --
	( 86.42, 58.86) --
	( 86.94, 58.16) --
	( 87.45, 57.63) --
	( 87.96, 57.59) --
	( 88.48, 57.70) --
	( 89.00, 57.88) --
	( 89.51, 57.76) --
	( 90.03, 57.12) --
	( 90.54, 57.21) --
	( 91.06, 56.98) --
	( 91.57, 56.96) --
	( 92.09, 56.97) --
	( 92.60, 56.96) --
	( 93.11, 56.94) --
	( 93.63, 56.92) --
	( 94.14, 56.93) --
	( 94.66, 56.93) --
	( 95.17, 56.93) --
	( 95.70, 56.92) --
	( 96.21, 56.93) --
	( 96.72, 56.92) --
	( 97.23, 56.91) --
	( 97.74, 56.90) --
	( 98.26, 56.90) --
	( 98.77, 56.91) --
	( 99.28, 56.93) --
	( 99.81, 56.92) --
	(100.32, 56.94) --
	(100.84, 56.94) --
	(101.36, 56.98) --
	(101.87, 56.92) --
	(102.39, 56.93) --
	(102.90, 56.91) --
	(103.42, 56.91) --
	(103.93, 56.91) --
	(104.45, 56.91) --
	(104.96, 56.93) --
	(105.47, 56.95) --
	(105.99, 56.93) --
	(106.50, 56.91);
\end{scope}
\begin{scope}
\path[clip] ( 40.50, 36.00) rectangle (249.34,131.54);
\definecolor{drawColor}{RGB}{77,175,74}

\path[draw=drawColor,line width= 0.4pt,line join=round,line cap=round] ( 95.87, 56.90) --
	( 98.26, 59.29) --
	(100.64, 56.90) --
	( 98.26, 54.51) --
	( 95.87, 56.90);
\end{scope}
\begin{scope}
\path[clip] ( 40.50, 36.00) rectangle (249.34,131.54);
\definecolor{drawColor}{RGB}{77,175,74}

\path[draw=drawColor,line width= 0.4pt,line join=round,line cap=round] ( 98.26, 56.90) --
	( 90.58, 54.58);

\node[text=drawColor,anchor=base east,inner sep=0pt, outer sep=0pt, scale=  0.75] at ( 87.84, 48.63) {0.0075};
\end{scope}
\begin{scope}
\path[clip] ( 40.50, 36.00) rectangle (249.34,131.54);
\definecolor{drawColor}{RGB}{152,78,163}

\path[draw=drawColor,line width= 0.6pt,line join=round,line cap=round] ( 40.50,102.01) --
	( 40.98, 89.88) --
	( 41.45, 86.97) --
	( 41.92, 85.81) --
	( 42.40, 84.90) --
	( 42.87, 84.47) --
	( 43.35, 84.14) --
	( 43.83, 84.10) --
	( 44.31, 83.41) --
	( 44.79, 82.31) --
	( 45.28, 77.61) --
	( 45.76, 76.30) --
	( 46.24, 73.98) --
	( 46.72, 73.72) --
	( 47.20, 73.00) --
	( 47.68, 72.81) --
	( 48.16, 72.85) --
	( 48.66, 72.60) --
	( 49.15, 71.37) --
	( 49.64, 71.41) --
	( 50.14, 71.49) --
	( 50.63, 70.68) --
	( 51.12, 70.52) --
	( 51.61, 70.27) --
	( 52.10, 70.13) --
	( 52.59, 69.95) --
	( 53.08, 70.65) --
	( 53.58, 69.98) --
	( 54.07, 69.37) --
	( 54.56, 69.33) --
	( 55.05, 69.91) --
	( 55.54, 69.94) --
	( 56.03, 69.42) --
	( 56.53, 68.59) --
	( 57.03, 68.63) --
	( 57.53, 68.62) --
	( 58.02, 68.87) --
	( 58.52, 67.94) --
	( 59.03, 69.01) --
	( 59.52, 67.90) --
	( 60.02, 68.26) --
	( 60.51, 68.44) --
	( 61.00, 67.19) --
	( 61.49, 67.78) --
	( 61.98, 68.80) --
	( 62.48, 67.50) --
	( 62.98, 68.06) --
	( 63.47, 67.50) --
	( 63.96, 67.70) --
	( 64.45, 67.42) --
	( 64.94, 67.55) --
	( 65.42, 67.44) --
	( 65.92, 66.99) --
	( 66.41, 66.83) --
	( 66.91, 66.45) --
	( 67.41, 66.55) --
	( 67.90, 66.36) --
	( 68.40, 66.41) --
	( 68.91, 65.96) --
	( 69.41, 66.11) --
	( 69.91, 65.94) --
	( 70.40, 65.46) --
	( 70.89, 65.75) --
	( 71.39, 65.35) --
	( 71.89, 65.40) --
	( 72.39, 65.77) --
	( 72.88, 65.60) --
	( 73.38, 65.42) --
	( 73.88, 65.42) --
	( 74.38, 65.03) --
	( 74.87, 65.27) --
	( 75.38, 64.83) --
	( 75.89, 65.65) --
	( 76.39, 64.17) --
	( 76.89, 64.37) --
	( 77.39, 64.53) --
	( 77.89, 64.67) --
	( 78.39, 64.43) --
	( 78.89, 63.63) --
	( 79.39, 63.67) --
	( 79.88, 63.87) --
	( 80.38, 64.04) --
	( 80.88, 63.50) --
	( 81.38, 63.58) --
	( 81.89, 64.00) --
	( 82.38, 63.75) --
	( 82.88, 63.30) --
	( 83.39, 62.99) --
	( 83.88, 62.90) --
	( 84.38, 63.13) --
	( 84.89, 62.61) --
	( 85.40, 62.73) --
	( 85.91, 62.24) --
	( 86.40, 61.97) --
	( 86.90, 62.45) --
	( 87.41, 62.39) --
	( 87.92, 62.11) --
	( 88.42, 62.42) --
	( 88.92, 61.79) --
	( 89.42, 61.35) --
	( 89.94, 61.38) --
	( 90.44, 61.76) --
	( 90.94, 61.70) --
	( 91.43, 61.81) --
	( 91.93, 61.44) --
	( 92.43, 61.11) --
	( 92.94, 61.78) --
	( 93.44, 61.54) --
	( 93.95, 61.63) --
	( 94.46, 61.34) --
	( 94.97, 61.77) --
	( 95.48, 61.33) --
	( 95.99, 62.26) --
	( 96.48, 61.74) --
	( 96.97, 61.21) --
	( 97.48, 61.15) --
	( 97.98, 61.49) --
	( 98.48, 61.75) --
	( 98.98, 61.36) --
	( 99.48, 61.68) --
	( 99.98, 61.57) --
	(100.49, 61.68) --
	(100.99, 61.82) --
	(101.50, 61.60) --
	(102.00, 61.30) --
	(102.50, 61.64) --
	(103.01, 61.26) --
	(103.51, 61.78) --
	(104.02, 62.29) --
	(104.52, 61.57);
\end{scope}
\begin{scope}
\path[clip] ( 40.50, 36.00) rectangle (249.34,131.54);
\definecolor{drawColor}{RGB}{152,78,163}

\path[draw=drawColor,line width= 0.2pt,line join=round,line cap=round] (104.52, 36.00) -- (104.52,131.54);

\node[text=drawColor,anchor=base west,inner sep=0pt, outer sep=0pt, scale=  0.75] at (106.52, 51.24) {2.08 h};
\end{scope}
\begin{scope}
\path[clip] ( 40.50, 36.00) rectangle (249.34,131.54);
\definecolor{drawColor}{RGB}{152,78,163}

\path[draw=drawColor,line width= 0.4pt,dash pattern=on 4pt off 4pt ,line join=round,line cap=round] ( 44.53,135.14) --
	( 44.79,117.63) --
	( 45.28,117.04) --
	( 45.76,119.92) --
	( 46.24,113.08) --
	( 46.72,119.18) --
	( 47.20,117.15) --
	( 47.68,121.97) --
	( 48.16,131.03) --
	( 48.66,128.79) --
	( 49.15,124.97) --
	( 49.64,123.34) --
	( 50.14,132.77) --
	( 50.63,132.47) --
	( 51.12,115.09) --
	( 51.61,126.27) --
	( 52.10,121.55) --
	( 52.59,112.62) --
	( 53.08,116.14) --
	( 53.58,126.53) --
	( 54.07,110.28) --
	( 54.56,114.00) --
	( 55.05,124.76) --
	( 55.54,110.91) --
	( 56.03,108.49) --
	( 56.53,107.58) --
	( 57.03,122.71) --
	( 57.53,120.64) --
	( 58.02,100.05) --
	( 58.52,125.14) --
	( 59.03,115.92) --
	( 59.52, 98.36) --
	( 60.02,109.00) --
	( 60.51,121.93) --
	( 61.00,100.95) --
	( 61.49, 95.12) --
	( 61.98,112.71) --
	( 62.48,116.32) --
	( 62.98,108.89) --
	( 63.47,112.71) --
	( 63.96, 99.84) --
	( 64.45, 98.08) --
	( 64.94,115.83) --
	( 65.42,105.30) --
	( 65.92,115.55) --
	( 66.41,115.06) --
	( 66.91, 99.90) --
	( 67.41,101.88) --
	( 67.90,114.92) --
	( 68.40,114.86) --
	( 68.91, 97.52) --
	( 69.41, 92.81) --
	( 69.91, 78.19) --
	( 70.40, 93.41) --
	( 70.89, 95.13) --
	( 71.39, 99.61) --
	( 71.89, 83.48) --
	( 72.39,108.14) --
	( 72.88, 99.86) --
	( 73.38,100.08) --
	( 73.88, 92.64) --
	( 74.38,100.60) --
	( 74.87, 86.09) --
	( 75.38, 99.39) --
	( 75.89, 96.27) --
	( 76.39, 83.60) --
	( 76.89, 78.14) --
	( 77.39, 92.24) --
	( 77.89, 95.61) --
	( 78.39,103.50) --
	( 78.89,100.49) --
	( 79.39, 87.38) --
	( 79.88, 73.35) --
	( 80.38, 80.42) --
	( 80.88, 87.26) --
	( 81.38,101.46) --
	( 81.89, 88.98) --
	( 82.38, 90.32) --
	( 82.88, 89.26) --
	( 83.39, 90.32) --
	( 83.88, 86.60) --
	( 84.38, 84.00) --
	( 84.89, 82.99) --
	( 85.40, 75.30) --
	( 85.91, 82.19) --
	( 86.40, 72.96) --
	( 86.90, 87.96) --
	( 87.41, 79.35) --
	( 87.92, 78.09) --
	( 88.42, 81.86) --
	( 88.92, 80.88) --
	( 89.42, 84.70) --
	( 89.94, 82.45) --
	( 90.44, 80.38) --
	( 90.94, 78.26) --
	( 91.43, 76.15) --
	( 91.93, 74.05) --
	( 92.43, 72.17) --
	( 92.94, 70.58) --
	( 93.44, 68.43) --
	( 93.95, 66.96) --
	( 94.46, 66.02) --
	( 94.97, 63.90) --
	( 95.48, 62.80) --
	( 95.99, 61.56) --
	( 96.48, 60.13) --
	( 96.97, 58.46) --
	( 97.48, 57.83) --
	( 97.98, 56.66) --
	( 98.48, 55.54) --
	( 98.98, 55.35) --
	( 99.48, 54.41) --
	( 99.98, 54.15) --
	(100.49, 53.80) --
	(100.99, 52.83) --
	(101.50, 52.73) --
	(102.00, 51.78) --
	(102.50, 51.34) --
	(103.01, 51.64) --
	(103.51, 51.90) --
	(104.02, 51.84) --
	(104.52, 51.55);
\end{scope}
\begin{scope}
\path[clip] ( 40.50, 36.00) rectangle (249.34,131.54);
\definecolor{drawColor}{RGB}{152,78,163}
\definecolor{fillColor}{RGB}{152,78,163}

\path[draw=drawColor,line width= 0.4pt,line join=round,line cap=round,fill=fillColor] (102.50, 51.34) circle (  1.69);
\end{scope}
\begin{scope}
\path[clip] ( 40.50, 36.00) rectangle (249.34,131.54);
\definecolor{drawColor}{RGB}{152,78,163}

\path[draw=drawColor,line width= 0.4pt,line join=round,line cap=round] (102.50, 51.34) --
	( 94.82, 47.00);

\node[text=drawColor,anchor=base east,inner sep=0pt, outer sep=0pt, scale=  0.75] at ( 92.08, 38.42) {0.0044};
\definecolor{drawColor}{RGB}{0,0,0}
\definecolor{fillColor}{RGB}{255,255,255}

\path[draw=drawColor,line width= 0.4pt,line join=round,line cap=round,fill=fillColor] (150.06,131.54) rectangle (249.34, 95.54);
\definecolor{drawColor}{RGB}{228,26,28}

\path[draw=drawColor,line width= 0.4pt,line join=round,line cap=round] (151.68,124.34) -- (162.48,124.34);
\definecolor{drawColor}{RGB}{55,126,184}

\path[draw=drawColor,line width= 0.4pt,line join=round,line cap=round] (151.68,117.14) -- (162.48,117.14);
\definecolor{drawColor}{RGB}{77,175,74}

\path[draw=drawColor,line width= 0.4pt,line join=round,line cap=round] (151.68,109.94) -- (162.48,109.94);
\definecolor{drawColor}{RGB}{152,78,163}

\path[draw=drawColor,line width= 0.4pt,line join=round,line cap=round] (151.68,102.74) -- (162.48,102.74);
\definecolor{drawColor}{RGB}{228,26,28}

\path[draw=drawColor,line width= 0.4pt,line join=round,line cap=round] (157.08,124.34) circle (  1.35);
\definecolor{fillColor}{RGB}{55,126,184}

\path[fill=fillColor] (157.08,119.24) --
	(158.90,116.10) --
	(155.26,116.10) --
	cycle;
\definecolor{drawColor}{RGB}{77,175,74}

\path[draw=drawColor,line width= 0.4pt,line join=round,line cap=round] (155.17,109.94) --
	(157.08,111.85) --
	(158.99,109.94) --
	(157.08,108.04) --
	(155.17,109.94);
\definecolor{drawColor}{RGB}{152,78,163}
\definecolor{fillColor}{RGB}{152,78,163}

\path[draw=drawColor,line width= 0.4pt,line join=round,line cap=round,fill=fillColor] (157.08,102.74) circle (  1.35);
\definecolor{drawColor}{RGB}{0,0,0}

\node[text=drawColor,anchor=base west,inner sep=0pt, outer sep=0pt, scale=  0.60] at (167.88,122.28) {$128^3$, $\eta^{(0)} = 5.0 \times 10^{-4}$ (min.)};

\node[text=drawColor,anchor=base west,inner sep=0pt, outer sep=0pt, scale=  0.60] at (167.88,115.08) {$256^3$, $\eta^{(0)} = 1.0 \times 10^{-3}$ (min.)};

\node[text=drawColor,anchor=base west,inner sep=0pt, outer sep=0pt, scale=  0.60] at (167.88,107.88) {$512^3$, $\eta^{(0)} = 1.0 \times 10^{-3}$};

\node[text=drawColor,anchor=base west,inner sep=0pt, outer sep=0pt, scale=  0.60] at (167.88,100.68) {$512^3$, BN, $\eta^{(0)} = 1.0 \times 10^{-3}$};
\end{scope}
\end{tikzpicture}

%% file: 5_related.tex
\section{Related work} \label{section:related}

Scalable training and model-parallelism has a long history in deep learning;
Ben-Nun \& Hoefler~\cite{ben2019demystifying} provide a comprehensive overview.
We discuss the most relevant. Early work on models such as AlexNet incorporated
model-parallelism using grouped convolution or partitioning fully-connected
layers~\cite{Krizhevsky2012a,krizhevsky2014one}. Coates et al.~\cite{coates2013deep}
applied spatial partitioning to locally-connected layers. Gholami et al.~\cite{gholami2017integrated}
consider spatial parallelism, but provide only simulated results.
FlexFlow~\cite{jia2019beyond} presents an automated system for identifying model- and hybrid-parallelism.
Frameworks such as DistBelief~\cite{dean2012large}, Project Adam~\cite{chilimbi2014project},
Mesh-TensorFlow~\cite{shazeer2018mesh}, and TF-Replicator~\cite{buchlovsky2019tf}
also support limited forms of model-parallelism, but do not do spatial partitioning.
The Distconv library~\cite{distconv} provides the basis for our 3D hybrid parallelism, but is limited to 2D CNNs. Extending this to efficiently support 3D CNNs requires significant novel work (Section~\ref{subsection:proposal-scaling-extending}). Further, this work does not consider I/O, which is a key bottleneck for scaling 3D CNNs, nor does it demonstrate improved learning. Similarly, channel and filter parallelism~\cite{dryden2019channel} provides another method for partitioning data, primarily targeting \emph{wide} CNNs. It also only considers only 2D data and neglects I/O. Further, channel/filter parallelism requires communication of entire activation tensors, instead of only halo regions, so the communication overheads will be significantly greater for the large, 3D data we consider.
In general, these prior approaches have not considered the extreme strong scaling
regime required for training 3D CNNs. Further, with CosmoFlow, we have demonstrated
that training on full-resolution input data actually produces better results.


Many approaches, such as pipelining~\cite{chen2012pipelined,li2018pipe,huang2019gpipe,narayanan2019pipedream}
and microbatching~\cite{oyama2018accelerating} are orthogonal to our 3D spatial
partitioning. Others directly target memory pressure during training, but perform additional computation, including
gradient accumulation~\cite{ito2017ooc_cudnn}, out-of-core algorithms~\cite{rhu2016vdnn,meng2017training,wang2018superneurons},
and recomputation~\cite{chen2016training,jain2019checkmate}.

I/O performance has also been a recent focus, particularly for data sets with many
or large samples. This has typically involved optimizing I/O pipelines with data
staging and asynchronous I/O~\cite{Mathuriya:2018:CUD:3291656.3291743,kurth2018exascale,pumma2017towards,chien2018characterizing,zhu2018entropy,chowdhury2019characterization,jacobs2019}. We build upon these to
handle the case where I/O for even a single sample is a bottleneck by
partitioning and scaling I/O spatially, adapting collective I/O techniques developed for large-scale parallel scientific workloads~\cite{750599, 1592942, hdf5_tuning}.

3D CNNs have been widely used for 3D volume datasets,
including medical imagery~\cite{2016arXiv160604797M,CABR16,ohou2019high} and video action
recognition~\cite{Tran:2015:LSF:2919332.2919929,2017arXiv170507750C}.
These typically extend a 2D CNN architecture by replacing the 2D convolutions with
3D operations~\cite{CABR16}. While 3D convolutional layers enjoy similar learning
properties to 2D ones and are more parameter-efficient than fully-connected
layers for extracting spatial features, the memory requirements have made them
challenging to use~\cite{2016arXiv160604797M,2017arXiv170507750C,Tran:2015:LSF:2919332.2919929}.
In particular, this has limited many works to small mini-batches, low-resolution
data, or CNN architecture tradeoffs~\cite{2016arXiv160604797M,2017arXiv170507750C}.
We demonstrate that tackling these problems enables further improvements
in prediction accuracy.

%% file: 6_conclusion.tex

\section{Conclusions} \label{section:conclusion}


Parallel training of DNNs is now considered a common practice rather than an art thanks to the wide availability of parallelized software stacks for deep learning.
However, as shown in this paper, addressing the computational requirements in applying deep learning to scientific problems on 3D data sets necessitates finer-grained scalable parallel algorithms both in compute and I/O.
%
%
We demonstrated the ability to spatially partition the
training over many GPU-accelerated HPC nodes, enabling the traditional strong
scaling that other HPC applications enjoy: accelerated time to solution without
a compromise in the quality of the learned model.
Further, we have demonstrated this with two networks that differ significantly in task, architecture, and performance characteristics, so we expect our work to be broadly applicable.
As a result, we have created a
scalable framework that can tackle the 3D CNNs that are
rapidly emerging at the forefront of scientific machine learning.
%

Another hypothesis of this work was that learning on full-resolution
data would allow models
to learn better representations of long-range features present in the data.
Our work with \cosmoflow{} demonstrated the strength of this hypothesis, where
the extended \cosmoflow{} model achieved an order-of-magnitude improvement in prediction quality while significantly reducing training time by exploiting a larger-scale system. %

%% file: acknowledgement.tex
\section*{Acknowledgments}
Prepared by LLNL under Contract DE-AC52-07NA27344 (LLNL-JRNL-812691).
This research was supported by JSPS KAKENHI Grant Number JP18J22858, Japan
and by the Exascale Computing Project (17-SC-20-SC).
Experiments were performed at the Livermore Computing facility.
This research used resources of the National Energy Research Scientific Computing Center (NERSC), a U.S. Department of Energy Office of Science User Facility operated under Contract No. DE-AC02-05CH11231.